\documentclass[fleqn,usenatbib]{mnras}

\usepackage{newtxtext,newtxmath}
\usepackage[T1]{fontenc}
\usepackage{ae,aecompl}
\usepackage[dvipsnames]{xcolor}

\usepackage{graphicx}	
\usepackage{multirow}
\usepackage{multicol}
\usepackage{makecell}
\usepackage{array}

\newcommand{\ujy}{\mu\mathrm{Jy}}
\newcommand{\km}{\mbox{km}}

\newcommand{\kev}{\mbox{keV}}

\newcommand{\msun}{\mbox{$M_\odot$}}

\newcommand{\hst}{\emph{HST}}
\newcommand{\chandra}{\emph{Chandra}}

\newcommand{\ergs}{\mbox{${\rm erg}~{\rm s}^{-1}$}}

\newcommand{\B}{\mbox{$B_{435}$}}
\newcommand{\R}{\mbox{$R_{625}$}}
\newcommand{\br}{\mbox{$\B\!-\!\R$}}
\newcommand{\ha}{\mbox{H$\alpha$}}
\newcommand{\hr}{\mbox{$\ha\!-\!\R$}}
\newcommand{\UV}{\mbox{$UV_{275}$}}
\newcommand{\U}{\mbox{$U_{336}$}}
\newcommand{\uvu}{\mbox{$\UV\!-\!\U$}}
\newcommand{\uvb}{\mbox{$\UV\!-\!\B$}}
\newcommand{\ub}{\mbox{$\U\!-\!\B$}}
\newcommand\nodata{ ~$\cdots$~ }%
\newcommand{\nd}{\nodata}
\newcommand{\rerr}{\mbox{$r_{\rm err}$}}

\newcommand{\ignore}[1]{}
\newcommand{\no}{N_\mathrm{obs}}
\newcommand{\np}{N_\mathrm{pred}}

\newcommand{\cory}[1]{{\color{black}{#1}}}
\newcommand{\coryrv}[1]{{\color{black}{#1}}}
\newcommand{\craig}[1]{{\color{black}{#1}}}
\newcommand{\mspf}[1]{{\color{black}{#1}}}
\newcommand{\revised}[1]{{\color{black}{#1}}}
\newcommand{\hp}[1]{{\color{black}{#1}}}

\title[Faint \chandra\ Sources in NGC 6752]{A Deep Search for Faint \chandra\ X-ray Sources, Radio Sources, and Optical Counterparts in NGC 6752}

\author[H.N. Cohn et al.]{%
Haldan N. Cohn,$^{1}$\thanks{E-mail: cohn@iu.edu (HNC)}
Phyllis M. Lugger,$^{1}$
Yue Zhao,$^{2}$
Vlad Tudor$^{3}$,
Craig O. Heinke,$^{2}$\newauthor
Adrienne M. Cool,$^{4}$
Jay Anderson,$^{5}$
Jay Strader,$^{6}$ and
James C.~A.~Miller-Jones$^{7}$\\
$^{1}$Department of Astronomy, Indiana University, 727 E. Third St.,
Bloomington, IN 47405, USA\\
$^{2}$Department of Physics, University of Alberta, Edmonton, AB T6G
2G7, Canada\\
$^{3}$International Centre for Radio Astronomy Research -- Curtin University, GPO Box U1987, Perth, WA 6845, Australia\\
$^{4}$Department of Physics and Astronomy, San Francisco State
University, 1600 Holloway Ave., San Francisco, CA 94132, USA\\
$^{5}$Space Telescope Science Institute, 3700 San Martin Dr.,
Baltimore, MD 21218, USA\\
$^{6}$ Center for Data Intensive and Time Domain Astronomy, Department of Physics and Astronomy, Michigan State University, East Lansing MI 48824, USA\\
$^{7}$ International Centre for Radio Astronomy Research, Curtin University, GPO Box U1987, Perth, WA 6845, Australia}

\date{Accepted XXX. Received YYY; in original form ZZZ}

\pubyear{2021}

\begin{document}
\label{firstpage}
\pagerange{\pageref{firstpage}--\pageref{lastpage}}
\maketitle

\begin{abstract}

We report the results of a deep search for faint \chandra\ X-ray sources, radio sources, and optical counterparts in the nearby, core-collapsed globular cluster, NGC 6752. We combined new and archival \chandra\ imaging to detect 51 X-ray sources (12 of which are new) within the 1\farcm9 half-light radius. Three radio sources in deep ATCA 5 and 9 GHz radio images match with \chandra\ sources. We have searched for optical identifications for the expanded \chandra\ source list using deep \emph{Hubble Space Telescope} photometry in \B, \R, \ha, \UV, and \U. Among the entire sample of 51 \chandra\ sources, we identify \revised{18}
cataclysmic variables (CVs), 9 chromospherically active binaries (ABs), 3 red giants (RGs), 3 galaxies (GLXs), and 6 active galactic nuclei (AGNs). \revised{Three of the sources are associated with millisecond pulsars (MSPs).} \revised{As in our previous study of NGC 6752, we find that the brightest CVs appear to be more centrally concentrated than are the faintest CVs, although the effect is no longer statistically significant as a consequence of the inclusion in the faint group of two intermediate brightness CVs. This possible difference in the radial distributions of the bright and faint CV groups appears to indicate that mass segregation has separated them. We note that photometric incompleteness in the crowded central region of the cluster may also play a role. Both groups of CVs have an inferred mass above that of the main-sequence turnoff stars. We discuss the implications for the masses of the CV components.} 

\end{abstract}




\section{Introduction}

Globular clusters are of particular dynamical interest because of the phenomena of core collapse and binary burning. Stellar interactions in dense globular clusters tend to drive the central core towards a collapse to a state of extremely small radius and high density. Primordial binary populations may delay the onset of this core collapse by serving as dynamical energy sources, but eventually become depleted by various destruction mechanisms \citep[e.g.][]{Verbunt14}, allowing core collapse to proceed \citep[e.g.][and references therein]{Fregeau03}. Several studies suggest that black-hole binaries, rather than main-sequence (MS) binaries, are critical for delaying core collapse \citep[e.g.][]{Breen13,Wang16,Kremer19}. \revised{In addition, an intermediate-mass black hole (IMBH) in a cluster core will act as a strong central energy source able to delay, or even prevent, core collapse \citep[e.g.][]{Baumgardt04,Gill08,Lutzgendorf13}.} In any case, after an initial collapse, the cluster core will undergo gravothermal oscillations of expansion and contraction, as first demonstrated by \citet{Sugimoto83}, during which the core radius remains quite small.  About 20--25 globular clusters, including NGC 6397 and NGC 6752, have very compact ($r_c \lesssim 10\arcsec$), high-density cores that appear to be post-collapse. The post-collapse oscillations of these cores should produce episodic bursts of strongly enhanced dynamical binary formation and ejection during the densest phases, as noted by \citet{Lugger07}. Their reasoning is based on the scaling of the encounter rate $\Gamma$, which is given by the integral of $\rho^2/v$ over the cluster volume, where $\rho$ is spatial mass density and $v$ is the velocity dispersion \citep{Verbunt87,Bahramian13}. The encounter rate can be approximated by the simplified expression $\Gamma \propto \rho_0^2 r_c^3/v_0$ \citep{Pooley03}, where $\rho_0$ is central density, $r_c$ is the core radius, and $v_0$ is the central velocity dispersion. This results in $\Gamma \propto r_c^{-1.4}$ for a simple homologous model for core collapse, in which $\rho_0 \propto r_c^{-2.2}$ and $v_0 \propto r_c^{-0.05}$ \citep{Cohn80}. 

Simulations of clusters undergoing core collapse oscillations indicate oscillation timescales of $10^{7-9}$ years, and that most produced binaries are ultimately ejected from the cluster. \citet{Beccari19} have recently reported evidence that the core-collapsed cluster M15 has undergone two discrete core-collapse events, dating to 2 and 5.5 Gyr ago, which each produced a coeval blue straggler star (BSS) sequence. \revised{It is important to note that there is a delay of a few hundred Myr to a few Gyr between the formation of a detached binary consisting of a white dwarf (WD) and a MS star and the evolution of this CV progenitor into a semi-detached binary, i.e.\ a CV. Thus, the observation of an apparently young group of CVs does not necessarily indicate a recent core-collapse event.}

Cataclysmic variables (CVs) \revised{and strong candidates} have been 
identified in a number of globular clusters \citep[e.g.][]{Cool95,Pooley02,Edmonds03a,Cohn10,Cool13,Lugger17,RiveraSandoval18}. 
\revised{Identification of an X-ray source's optical counterpart as a likely CV generally has been claimed using one or more of blue colors, H$\alpha$ excess, optical variability, proper motion consistent with the cluster, and/or spectroscopy, though some objects with several of these features have been discovered to be something else \citep[e.g. X9 in 47 Tuc,][]{Miller-Jones15}.}
The production of CVs in globular clusters is complex, as some CVs are dynamically produced through exchange interactions with primordial binaries and possibly tidal captures, while others are of primordial origin \citep{Statler87,Davies97,Pooley06,Shara06,Ivanova06,Hong17,Belloni19}.
\revised{Dynamical production of CVs in dense clusters (and quiescent low-mass X-ray binaries, qLMXBs, in all clusters) is supported by the correlation between the stellar encounter rates and numbers of CVs \citep{Johnston96,Heinke03,Pooley03,Pooley06,Bahramian13}.  However, lower-density clusters are theoretically expected to have more primordially-formed CVs than dynamically formed CVs \citep{Davies97,Belloni19}, and indeed there is strong observational support for this \citep{Kong06,Haggard09,Cheng18,Belloni19}. The strength of the correlation between stellar encounter rate and X-ray sources (which varies depending on the sample) has been recently discussed by \citet{Cheng18} and \citet{Heinke20}.}

The numbers of CVs in a cluster are often large enough that their spatial distribution and luminosity function can be studied (in contrast to qLMXBs), giving information on their ages (through their luminosities), and their dynamical state (through their spatial distribution). Bright CVs should be younger \revised{on average}, as 
CVs 
\revised{often} 
start with (relatively) massive companions, and the companion's optical luminosity and $\dot{M}$ 
both decay with time as the companion is whittled away. As the mass of the secondary is reduced and the system luminosity decreases, the orbit tightens and the orbital period shortens. \revised{A fascinating recent discovery is a significant paucity of X-ray luminous magnetic CVs in globular clusters \citep{Bahramian20}, which \citet{Belloni21} argue is due to the large age of the white dwarfs when the systems begin mass transfer.}

Our previous studies of NGC 6397 \citep{Cohn10}, $\omega$ Cen \citep{Cool13}, NGC 6752 \citep{Lugger17}, and 47 Tuc \citep{RiveraSandoval18} have revealed significant populations of X-ray and optically faint CVs, which we expect to have periods generally below 2 hours, comparing to predictions from binary evolution by \citet{Ivanova06} \revised{or \citet{Belloni19}}. 
These simulations predict twice \revised{(or more)} as many detectable CVs below the $2-3$ hr period gap as above it. 
\revised{Observations of local CVs summarised by \citet{Pala20} find $3-8$ times more CVs below the period gap, and suggest that $M_R\sim9$ roughly divides CVs at the period gap (3 of 37 CVs below the gap are brighter, none above the gap are fainter), with 10 CVs brighter than $M_R=9$ and 29 CVs in $9<M_R<12$ (18 CVs in $9<M_R<11$).
\citet{RiveraSandoval18} identified 9 CVs above $M_B\sim9$ vs. 30 candidate CVs between $9<M_B<12$ in a $U_{300}-B_{390}$ CMD of 47 Tuc, though their $B-R$ CMD of 47 Tuc only found 9 CVs below $M_R\sim9$ (the latter CMD was only complete to $M_R\sim10$). \citet{Cool13} identified 8 CVs above $M_R=9$, and 11 between $9<M_R<12$, in $\omega$ Cen, but using shallower {\it Chandra} observations reaching only $L_X=1.0\times10^{30}$ erg/s which would have missed most of the faint 47 Tuc or NGC 6397 CVs (new deeper {\it Chandra} observations of \citealt{Henleywillis18} should allow the identification of additional faint CVs).}
By contrast, NGC 6397 (with the deepest \hst\ and \chandra\ data, \revised{reaching $L_X=1\times10^{29}$ erg/s and $M_R\sim13$}) has almost the same number of CVs between $M_R$=9 and $M_R$=12 
as at brighter $M_V$ values (6 likely above the period gap, vs.\ 7 likely below).  We think this may be due to the dynamical evolution of NGC 6397; either by the destruction or ejection of old binaries through dynamical encounters, or by a recent burst of dynamical CV formation through an extreme-density phase of core collapse, inducing interactions. \revised{However, these comparisons depend on the security of the CV identifications, and on the observational sensitivity limits.}

NGC 6752 is a nearby ($d = 4\,\mbox{kpc}$), low-extinction, high-interaction-rate cluster. It is the only core-collapsed cluster, besides NGC 6397, that is both nearby and low extinction, and thus a good comparison for testing models of how core collapse affects interacting binaries. \revised{\citet{Bailyn96} used \hst\ WFPC2 imaging of the core of NGC 6752 to identify two candidate CVs there from strong \ha\ emission, periodic variability, and in one case, a UV excess.} \citet{Pooley02} used the 30 ks of Cycle 1 \chandra\ data to find 19 X-ray sources, suggest optical counterparts for 12, and perform spectral analyses for relatively bright sources (CX1-9). \citet{Thomson12} suggested two new optical and ultraviolet counterparts to these X-ray sources. \citet{Forestell14} used 67 ks of {\it Chandra} data to construct a deeper X-ray catalog of 39 sources, and analyze the spectra of the 5 X-ray faint MSPs (discovered by \citealt{DAmico01,DAmico02}).
In our previous study of NGC 6752 \citep[][ hereafter L17]{Lugger17}, we searched for counterparts to 
the 39 X-ray sources detected by 
\citet{Forestell14}. 
In the present study, we obtained an additional 277\,ks of \chandra\ ACIS-S exposure, giving an expanded list of 51 sources when combined with the previous exposure.   Since the Cycle 18 observation provides a deeper view and thus more counts, we can extend the analyses to include more faint sources. In this work, we performed detailed spectral analyses for all 51 sources within the half-light radius (including three MSPs) and two other MSPs that are outside. 
In the following sections, we present the X-ray analysis, the radio analysis, and the optical/UV analysis, followed by our conclusions. 

\section{X-ray Observations and Analyses}
\subsection{{\it Chandra} observations \& reprocessing}
In addition to a total of 67\,ks of previous ACIS-S observations (OBSID 948 and 6612; PI: Lewin and Heinke, respectively), we incorporated our Cycle 18 ACIS-S observations (OBSID:19013, 19014, 20121, 20122, 20123; PI: Cohn), totalling 277\,ks, into the analyses. 
The ACIS very faint mode (VFAINT) \revised{was used} to optimise background cleaning. \revised{See Table~\ref{t:chandra_obs}.}

\begin{table}
    \centering
    \resizebox{\columnwidth}{!}{%
    \begin{tabular}{ccccc}
    \hline
    Obs. ID & Time of observation & Exposure time (ks) & Instrument & Cycle \\
    \hline
    948     & 2000-05-15 04:36:02 & 29.47 & ACIS-S & 1\\
    6612    & 2006-02-10 22:48:48 & 37.97 & ACIS-S & 7\\
    19014   & 2017-07-02 03:27:25 & 98.81 & ACIS-S & 18\\
    19013   & 2017-07-24 09:33:12 & 43.20 & ACIS-S & 18\\
    20121   & 2017-07-25 17:04:15 & 18.26 & ACIS-S & 18\\
    20122   & 2017-07-29 09:00:43 & 67.22 & ACIS-S & 18\\
    20123   & 2017-07-30 23:53:18 & 49.46 & ACIS-S & 18\\
    \hline
    \end{tabular}%
    }
    \caption{{\it Chandra} observations used in this work.}
    \label{t:chandra_obs}
\end{table}

The {\it Chandra} data were reduced using the {\sc chandra interactive analysis of observations} software suite ({\sc ciao}; version 4.11 and CALDB 4.8.2). All data were first reprocessed by the {\sc ciao} {\tt chandra\_repro} script to align with the most up-to-date calibration, which generates new level-2 event files for further analyses. 

We chose the longest observation (Obs.\ ID 19014) as the reference frame, to which we calculated relative offsets of other observations based on the centroid locations of CX1, the brightest source. These offsets were then used to update the aspect solutions for each observation using the {\sc ciao} {\tt wcs\_update} tool. The shifted event files were then re-projected and combined using the {\sc ciao} {\tt merge\_obs} script, producing a merged event file, a combined exposure map, and X-ray images in a soft (0.5--2\,keV), a hard (2--7\,keV), and a broad (0.5--7\, keV) band. The X-ray images were re-binned to 
\revised{0\farcs25}
(i.e., 0.5 ACIS pixels) to reduce crowding and facilitate source detection. 

\begin{figure*}
    \centering
    \includegraphics[scale=0.40]{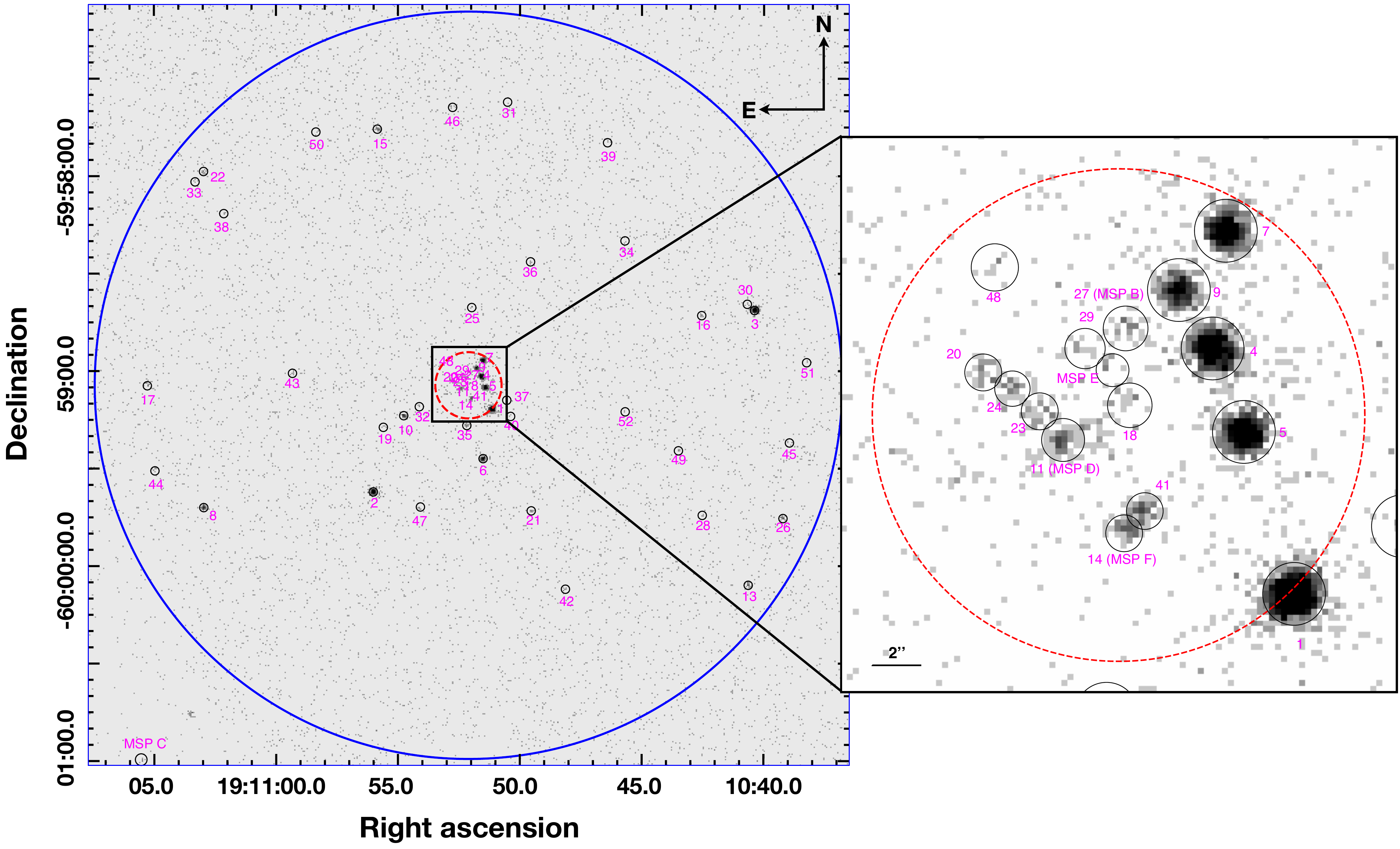}
    \caption{Combined 0.5--7\,keV X-ray image showing a 3\farcm9 $\times$ 3\farcm9 region centred on NGC 6752. The blue solid circle indicates the 1\farcm91 half-light radius, and the red dashed circle represents the 0\farcm17 core. All sources within the half-light radius ($r_\mathrm{h}$) are marked by their extraction regions (black circles) and are annotated by their CX IDs, including 39 sources from \citet{Forestell14} and 12 new sources (CX41-52) from this work. We also included the position of MSP C, which lies at a distance of $1.3r_\mathrm{h}$ from the cluster centre. The right panel is a $23\arcsec\times 23\arcsec$ square showing a zoomed-in view of sources in the the core, where sources are indicated with their extraction regions.}
    \label{f:x_ray_img}
\end{figure*}

\subsection{Source detection}
We then ran the {\sc ciao} {\tt wavdetect} tool \citep{Freeman02}\footnote{\url{http://cxc.harvard.edu/ciao/ahelp/wavdetect.html}} on the combined X-ray images to find and localise possible sources. The {\tt wavdetect} algorithm applies and correlates a ``Mexican Hat'' function with image pixels and identifies potential sources at different scale sizes based on the positive correlation values, while calculating source positions, detection significance, and other relevant information. To find sources at all possible size scales, we applied scale parameters of 1.0, 1.4, 2.0, 2.8, and 4.0. We set the detection threshold in all our runs to be $1.1\times10^{-6}$, the reciprocal of the area of the region, to limit the numbers of false detections to one in each run. Due to crowding in the core, some sources might be blended. While under-binning the images somewhat de-blends these sources, we also ran {\tt wavdetect} on different energy-filtered images, generating source lists for the above-defined soft, hard, and broad bands. This approach can be useful to decompose blended sources comprised of, for example, a soft and a hard source. The {\tt wavdetect} source positions are then used as input to the {\sc acis-exctract} software \citep{Broos10}\footnote{\url{http://personal.psu.edu/psb6/TARA/AE.html}} to obtain refined source positions. Finally, \revised{we combined} the energy-specific source lists. 

With the additional 277\,ks of observations, we detected 12 new sources within the 1\farcm91 half-light radius \citep[2010 edition]{Harris96}, of which CX14 was found to be blended with a relatively harder source, CX41 (Fig.~\ref{f:x_ray_img}). Positional information for these new sources and the previously detected sources is summarised in Table \ref{t:counterparts}. We note that CX18, CX29, CX30, CX33, CX34, and CX40 were not detected by our {\tt wavdetect} runs, so we kept their original positions from previous work \citep{Forestell14}. In Fig.~\ref{f:x_ray_img}, we show an X-ray image of the cluster half-light radius region overplotted with the 51 sources in our extended catalogue. \revised{Our combined observations reach to an average detection limit of $L_X=3\times10^{29}$ erg/s (this corresponds to 5 counts in 0.5-7 keV using our best faint-source spectral fit in the next section), though confusion in the core could hide a couple sources above that.}

\subsection{X-ray spectral analyses}
\label{sec:x_ray_spec_analyses}

We first extracted X-ray spectra from 
\revised{each}
Cycle 18 observation using the {\sc ciao} {\tt specextract}\footnote{\url{http://cxc.harvard.edu/ciao/ahelp/specextract.html}} script, and combined the spectra and response files for each source using the {\sc heasoft/ftools} {\tt addspec} task\footnote{\url{https://heasarc.gsfc.nasa.gov/ftools/}}.
We binned spectra with more than 1000 counts to 20 counts per bin, and spectra with less than 1000 but more than 100 counts to 10 counts per bin. These spectra were further analysed using $\chi^2$ statistics in {\sc heasoft/xspec} (version 12.10.1)\footnote{\url{https://heasarc.gsfc.nasa.gov/xanadu/xspec/}}, where the reduced $\chi^2$ ($\chi_\nu^2$) is 
a measure of \revised{fit} quality. 
Spectra with less than 100 counts \revised{were binned  to} at least one count per bin, and used C-statistics \citep{Cash1979}. 
To measure the fitting quality of the C-statistic, we used the {\tt goodness} command in {\sc xspec} to generate 1000 realisations of simulated spectra for each model, and determine the fraction of realisations that have a lower fit statistic than that of the data. 
High values of this fraction (e.g. 95\%) should be rare, unless the model is not a reasonable description of the data.
We use channels between 0.5 and 10 keV for all spectral fits.

We used the Tuebingen-Boulder ISM absorption model (TBabs in {\sc xspec}) to account for interstellar extinction, and adopted the {\tt wilm} abundances \citep{Wilms2000} in {\sc xspec}. The hydrogen column density ($N_\mathrm{H}$) was set as a free parameter for sources with more than 100 counts, 
while for fainter sources (<100\,counts), we did not get proper constraints on the absorption due to low counting statistics, so we froze $N_\mathrm{H}$ to the cluster value ($\approx 3.48\times10^{20}\,\mathrm{cm^{-2}}$) calculated using the reddening $E(B-V)=0.04$ from \citet[2010 edition]{Harris96} and a conversion factor of $2.81 \pm 0.13 \times 10^{21}~\mathrm{cm^{-2}}$ from \citet{Bahramian15}. 
For the only foreground chromospherically active binary (AB), CX8, we fixed $N_\mathrm{H}$ to zero.
For all sources, we adopt the cluster distance, 4\, kpc, so luminosities calculated for background active galactic nuclei (AGNs) and galaxies (GLXs) should be regarded as lower limits, while the luminosity for CX8, the foreground AB, should be regarded as an upper limit. 
We report \revised{spectral parameter} constraints
at the 90\% confidence level. 

For sources with $>100$ counts, we tried spectral fits with either an empirical power-law model ({\tt pow}) or a thermal plasma model ({\tt mekal}), adopting {\tt mekal} fits for CVs, ABs, and AB/CV candidates and power-law fits for GLXs and AGNs (this is based on source classification in \S\ref{s:classification}). For unidentified sources ($>10$\,counts), we include results from both {\tt pow} and {\tt mekal} fits as complementary information for further identification. In cases of sources with less than 10 counts between 0.5 and 10 keV, including CX18-19, CX29-34, CX38, CX40, and CX50-52, we cannot get proper constraints on either the power-law indices ($\Gamma$) or the plasma temperatures ($kT$) due to a dearth of counts. We therefore combined all faint source spectra and fit the combined spectrum to a {\tt mekal} model, obtaining an averaged plasma temperature $kT_\mathrm{avg}=2.0^{+2.0}_{-0.7}\,\kev$. This $kT_\mathrm{avg}$ is then applied to each faint source as a fixed parameter so only the normalisation of the {\tt mekal} model was fitted. CX33 was not detected in the Cycle 18 observations. We therefore adopt the result from fitting its Cycle 7 spectrum, applying the same method: use a $kT_\mathrm{avg}=4.0^{+3.3}_{-1.3}\,\kev$ obtained from fitting a {\tt mekal} model to the combined faint source spectrum from Cycle 7. The Cycle 18 observations also cover all 5 known MSPs, which allows us to extract and analyse their spectra. As a point of comparison with results from \citet{Forestell14}, we fit the MSP spectra with both a blackbody model ({\tt bbodyrad}) and a neutron star (NS) hydrogen atmosphere model ({\tt nsatmos}; \citealt{Heinke06}), constraining surface temperatures and effective sizes of emission regions. In all cases, we used the {\tt cflux} multiplicative component in {\sc xspec} to calculate fluxes in a soft (0.5-2 keV), a hard (2-7 keV) and a broad (0.5-7 keV) band. 

\begin{figure}
    \centering
    \includegraphics[width=\columnwidth]{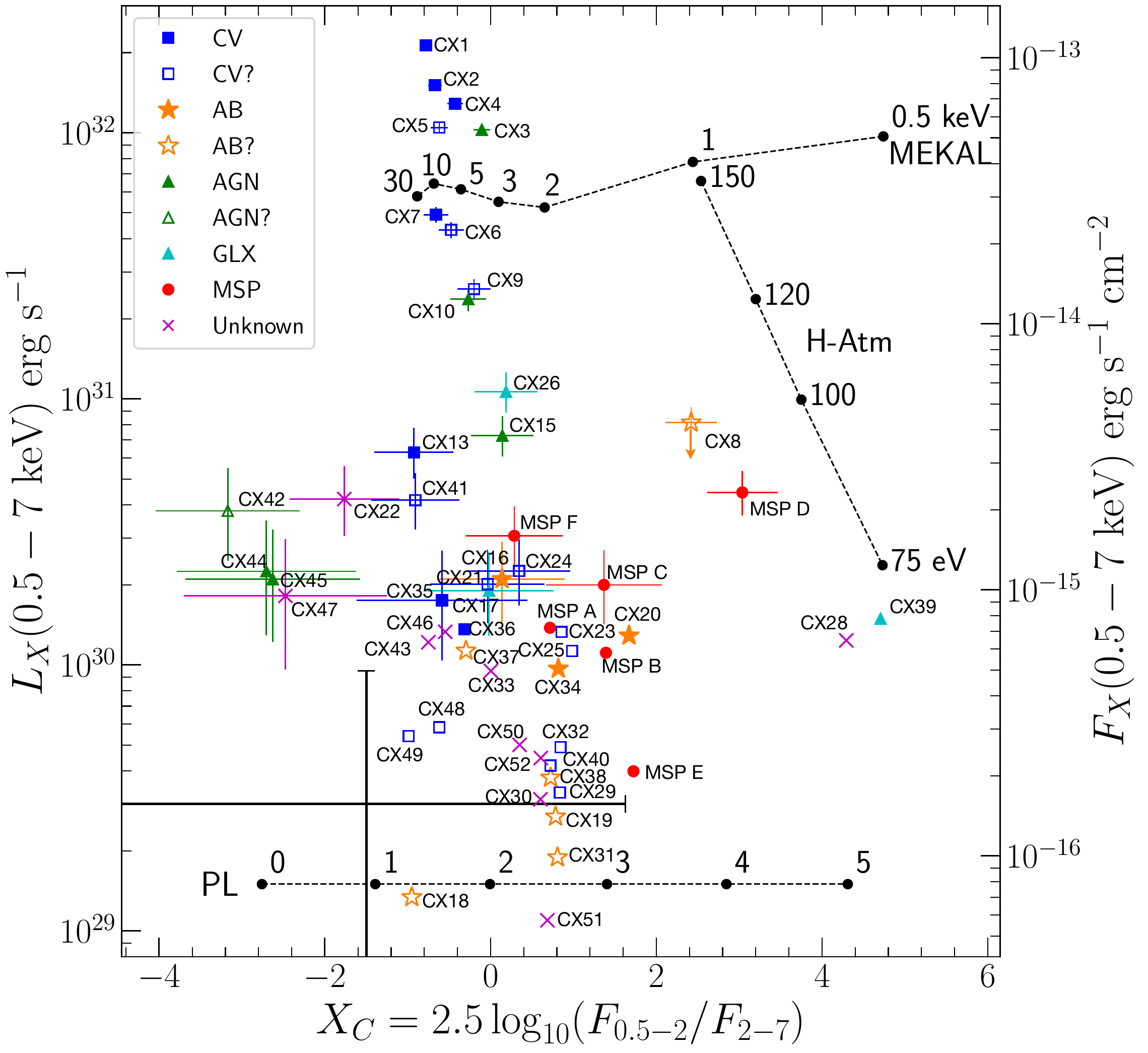}
    \caption{\mspf{X-ray colour-magnitude diagram plotting 0.5--7 $\kev$ X-ray luminosities vs. hardness ratios ($X_C$s), as defined in eq.(\ref{eq:x_ray_hardness}), for all 51 sources within the half-light radius and 2 MSPs outside (MSP A and MSP C). Source classes are distinguished by different markers (Table \ref{t:counterparts}). X-ray luminosities of AGNs and GLXs are lower limits (upward triangles), while the luminosity of CX8 (the foreground AB) is an upper limit (indicated with a downward arrow). For better readability, we only included error bars for sources brighter than $L_X=1.5\times 10^{30}~\ergs$, and plotted the average uncertainties of fainter sources with black bars. We also plotted tracks of {\tt nsatmos} models at different surface temperatures (assuming a $1.4\,\msun$, $12\,\km$ NS), the $X_C$s of {\tt mekal} models at different plasma temperatures, and power-law models with different photon indices ($\Gamma$s).}}
    \label{f:x_ray_cmd}
\end{figure}

We found acceptable fits for most sources, though some sources (CX5 and CX8) require more in depth spectral analyses (discussed in \S\ref{s:classification}). In Table \ref{t:spectral_analyses_51}, we summarise all best-fitting parameters, and in Fig. \ref{f:x_ray_cmd}, we show an X-ray colour-magnitude diagram (CMD) of these sources, defining an X-ray hardness ratio
\begin{equation}
    X_C \equiv 2.5\log(F_\mathrm{0.5-2}/F_\mathrm{2-7}),
    \label{eq:x_ray_hardness}
\end{equation}
where $F_\mathrm{0.5-2}$ and $F_\mathrm{2-7}$ are unabsorbed model fluxes (Table \ref{t:spectral_analyses_51}) in $0.5$-$2~\mathrm{keV}$ and $2$-$7~\mathrm{keV}$, respectively. 


\subsection{Inter-observational X-ray variability}
\label{s:inter-variability}
We checked for possible variability between different Cycles for only the bright sources, CX1-CX10, 
which have sufficient counts (>90) in Cycles 1 and 7 for spectral fitting (following the methods described in \S\ref{sec:x_ray_spec_analyses}). For each of these sources, we fit Cycle 1 and 7 spectra to the corresponding model presented in Table \ref{t:spectral_analyses_51}, obtaining a $0.5$--$7~\kev$ luminosity and an $X_C$ for each Cycle. We then 
searched for sources that 
\revised{significantly changed} luminosity and/or $X_C$. \revised{We expect that CVs and ABs may often (though not always) show strong X-ray variability (from accretion flickering and coronal flaring, respectively), while MSPs and quiescent LMXBs without accretion show little or no variability on these timescales \citep{Heinke05}.}

At  90\% confidence, we found clear variability in CX1 (CV), CX3 (AGN), CX4 (CV), CX7 (CV), and CX8 (AB). CX1 is brightest in Cycle 1,  
\revised{dimmed by}
a factor of 5 
in Cycle 7, then stayed at intermediate brightness in Cycle 18. CX3 is 
brighter in 
Cycle 7 
than in Cycles 1 and 18 (by factors of $1.8\pm 0.3$ and $1.8\pm 0.2$, respectively), with the $X_C$ being consistent. CX4 is brightest in Cycle 18 (by a factor of $1.6^{+0.3}_{-0.2}$ brighter 
\revised{vs.}
Cycle 7, 
\revised{and}
$1.3\pm 0.2$ brighter than Cycle 1), but the \revised{$X_C$s are consistent.} 
Similarly, CX7 shows consistent $X_C$ but is brightest in 
Cycle 7. 
Finally, 
CX8 is a factor of $1.9^{+0.7}_{-0.5}$ brighter in Cycle 1 than in Cycle 18, but 
the fit quality of the Cycle 18 data is not great (Table \ref{t:spectral_analyses_51}). 

We also compared our spectral fitting results for the 5 known MSPs with 
\revised{that of}
\citet{Forestell14}. 
MSP A \revised{appears variable}, where the fit to the Cycle 18 spectrum yields a somewhat higher $kT$ than for Cycle 7. MSP A, with a white dwarf companion \citep{Ferraro03a,Bassa03}, is not expected to show X-ray variability.
However, 
MSP A is far off-axis in both the Cycle 7 and Cycle 18 observations (about $6\arcmin$ off the aimpoint), which stretches the point spread function (PSF) and 
increases the relative background.
Although MSP A's Cycle 18 spectrum contains $\sim 20$\,counts, 
vs. $\sim 10$\, counts in Cycle 7, 
\craig{we cannot obtain clearer constraints.} 

\begin{figure*}
    \centering
    \includegraphics[scale=0.45]{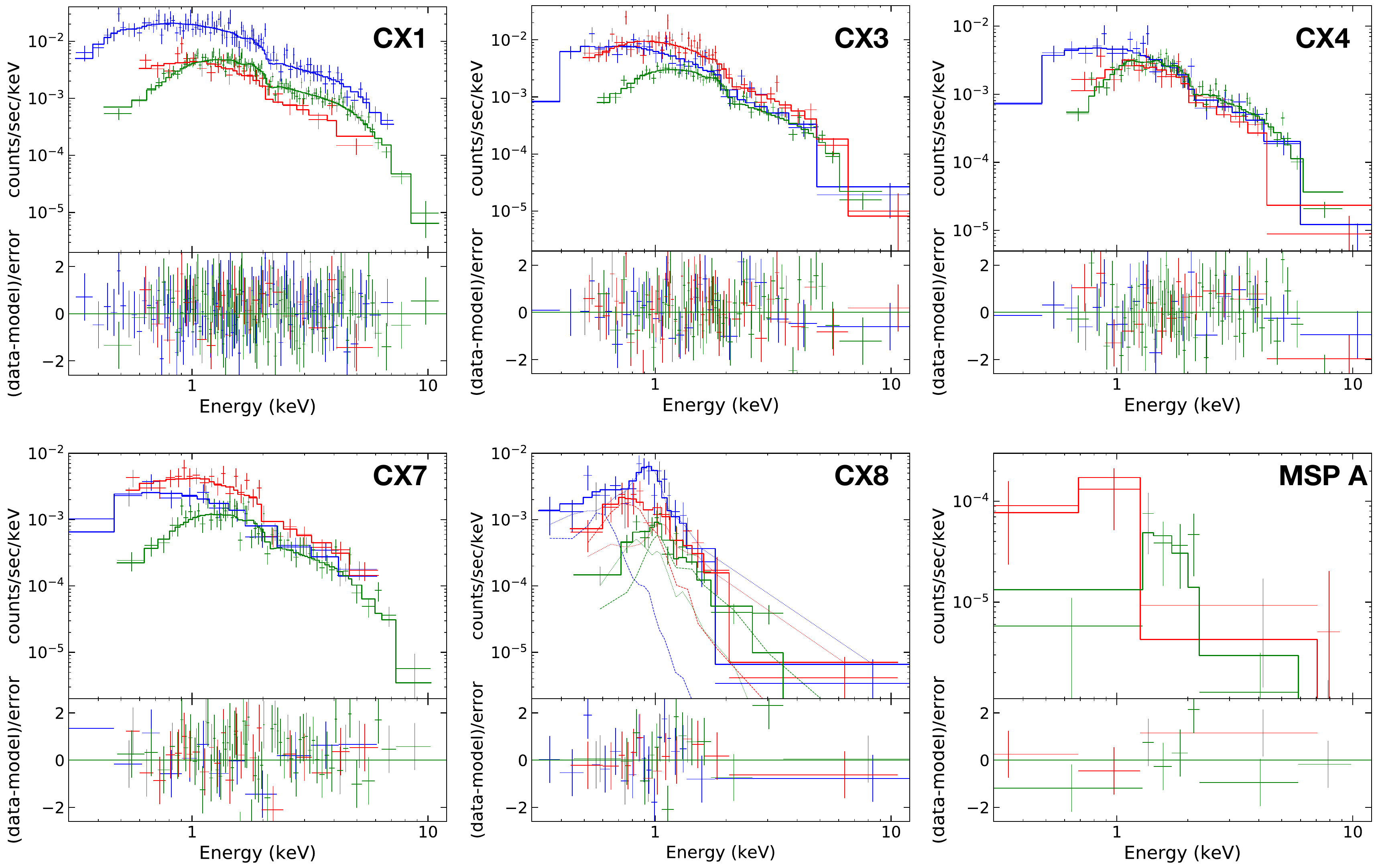}
    \caption{X-ray spectra from Cycle 1 (blue), Cycle 7 (red), and Cycle 18 (green), plotted with the corresponding best-fitting model (solid line with colour matching that of the spectrum) for sources with inter-observational spectral variability (CX1, CX3, CX4, CX7, CX8 and MSP A). The lower panel in each plot shows residuals defined as (data-model)/error. The best-fitting models from Table \ref{t:spectral_analyses_51} are used in each case.  For CX8, we plot the cool (dashed) and hot (dotted) {\tt mekal} components of the {\tt mekal+mekal} fit separately. \revised{Note that Cycle 18 spectra have specific count rates below the other two cycles at low ($\lesssim 1~\mathrm{keV}$) energies, due to loss of ACIS-S effective area over time.}}
    \label{f:x_ray_spec_inter_variab}
\end{figure*}

\begin{figure}
    \centering
    \includegraphics[scale=0.37]{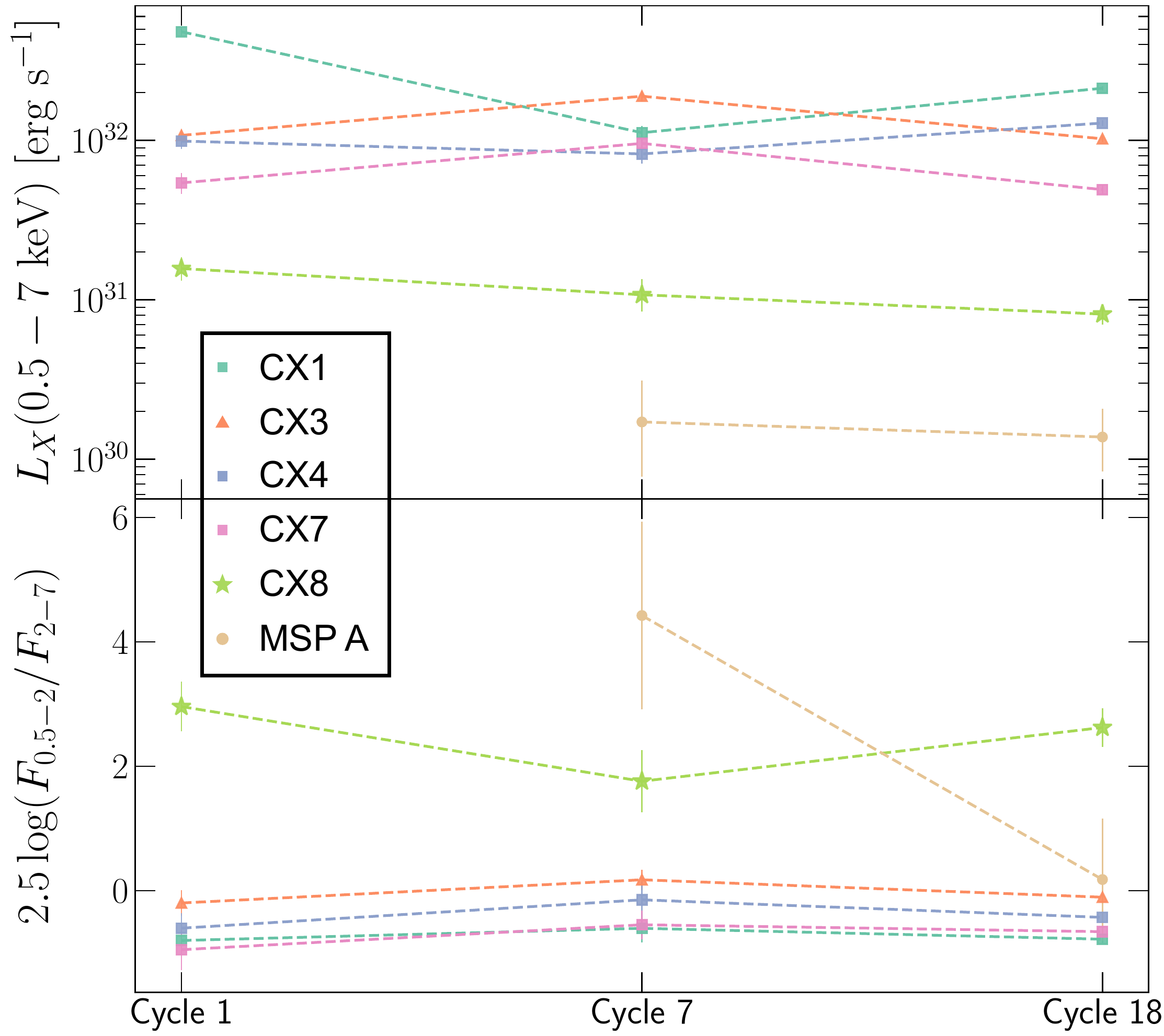}
    \caption{X-ray luminosities (top) and $X_C$s (eq.(\ref{eq:x_ray_hardness}); bottom) of sources with inter-observational variability plotted for different Cycles.}
    \label{f:x_ray_spectral_variability}
\end{figure}

In Fig. \ref{f:x_ray_spec_inter_variab}, we show source spectra from different Cycles plotted with the corresponding best-fit models. In Fig. \ref{f:x_ray_spectral_variability}, we show X-ray luminosities and $X_C$s in different Cycles for the above sources.

\begin{table*}
    \caption{Results of spectral analyses using the Cycle 18 observation.}
    \label{t:spectral_analyses_51}
    \centering
    \renewcommand{\arraystretch}{1.2}
    \begin{tabular}{lccccccc}
    \hline
       Source  & Model & $N_\mathrm{H}$ & Parameter 1$^a$ & Parameter 2$^b$ & $F_{0.5-2}$ & $F_{2-7}$ &  $\chi_\nu^2$ (dof) or Goodness $^c$\\
               &       & $(10^{22}~\mathrm{cm^{-2}})$ & & & \multicolumn{2}{c}{$(10^{-15}~\mathrm{erg~s^{-1}~cm^{-2}})$} & \\
    \hline
       CX1     & {\tt mekal} & $<0.03$ & $17.6^{+8.1}_{-4.7}$    & \nd & $36.2^{+1.2}_{-1.2}$ & $76.6^{+2.5}_{-2.5}$ & $0.90~(100)$ \\
       CX2     & {\tt mekal} & $<0.1$ & $9.1^{+3.2}_{-2.2}$     & \nd & $27.6^{+1.1}_{-1.0}$ & $51.2^{+2.0}_{-1.9}$ & $1.09~(77)$  \\
       CX3     & {\tt pow}   & $<0.1$ & \nd & $1.9^{+0.2}_{-0.1}$ & $25.5^{+1.2}_{-1.2}$ & $28.1^{+1.3}_{-1.3}$ & $0.86~(56)$  \\
       CX4     & {\tt mekal} & $0.2^{+0.1}_{-0.1}$ & $5.8^{+1.7}_{-1.1}$ & \nd & $27.1^{+1.2}_{-1.2}$ & $40.1^{+1.8}_{-1.8}$ & $1.17~(60)$ \\
       \hline
       \multirow{4}{*}{CX5}     & {\tt mekal} & $0.4^{+0.2}_{-0.1}$ & $7.2^{+3.7}_{-2.0}$ & \nd & $23.8^{+1.2}_{-1.2}$ & $38.3^{+1.8}_{-1.8}$ & $1.99~(53)$ \\
                                & {\tt pow}   & $0.6^{+0.2}_{-0.2}$ & \nd & $1.8^{+0.2}_{-0.2}$ & $32.2^{+1.6}_{-1.6}$ & $36.5^{+1.8}_{-1.8}$ & $1.91~(53)$ \\
                                & {\tt cutoffpl} & $<0.1$ & $1.5^{+0.4}_{-0.3}$ & $-0.5^{+0.4}_{-0.2}$ & $13.6^{+0.7}_{-0.7}$ & $36.0^{+1.7}_{-1.7}$ & $1.26~(52)$ \\
                                & {\tt gabs*mekal} & $0.3^{+0.1}_{-0.1}$ & $>21.5$ & $7.2^{+8.8}_{-0.7}$ & $19.7^{+0.9}_{-0.9}$ & $34.9^{+1.7}_{-1.7}$ & $1.21~(50)$\\
    \hline
       CX6    & {\tt mekal} & $<0.1$ & $6.3^{+2.1}_{-1.2}$ & \nd & $8.8^{+0.6}_{-0.6}$ & $13.7^{+1.0}_{-1.0}$ & $1.14~(53)$ \\
       CX7    & {\tt mekal} & $<0.1$ & $9.8^{+7.1}_{-2.9}$ & \nd & $9.1^{+0.6}_{-0.6}$ & $16.6^{+1.1}_{-1.1}$ & $1.27~(58)$ \\
    \hline
       \multirow{3}{*}{CX8}    & {\tt mekal} & $0^\dag$ & $0.9^\ast$ & \nd & $3.1^\ast$ & $0.3^\ast$ & $2.42~(13)$ \\
                               & {\tt pow}   & $0^\dag$ & \nd & $3.9^{+0.5}_{-0.5}$ & $6.5^{+1.0}_{-0.9}$ & $0.4^{+0.1}_{-0.1}$ & $1.88~(13)$ \\
                               & {\tt mekal+mekal} & $0^\dag$ & $1.2^{+0.5}_{-0.2}$ & $0.2^{+0.3}_{-0.1}$ & $4.4^{+0.6}_{-0.6}$ & $0.4^{+0.1}_{-0.1}$ & $1.86~(11)$\\
    \hline
      CX9    & {\tt mekal} & $<0.1$ & $5.7^{+3.5}_{-1.5}$ & \nd & $6.1^{+0.6}_{-0.5}$ & $7.4^{+0.7}_{-0.7}$ & $1.33~(32)$ \\
      CX10   & {\tt pow}   & $<0.4$ & \nd & $1.7^{+0.4}_{-0.3}$ & $5.4^{+0.5}_{-0.5}$ & $7.0^{+0.7}_{-0.7}$ & $1.31~(28)$ \\
      CX11   & {\tt nsatmos} & $0.03^\dag$ & $0.13^{+0.04}_{-0.03}$ & $<1.7$ & $2.2^{+0.4}_{-0.4}$ & $0.13^{+0.03}_{-0.03}$ & $21.5\%$ \\
      CX13   & {\tt mekal}   & $0.03^\dag$ & $>11.9$ & \nd & $1.0^{+0.2}_{-0.2}$ & $2.3^{+0.5}_{-0.5}$ & $40.3\%$ \\
      \hline
      \multirow{2}{*}{CX14} & {\tt pow}   & $0.03^\dag$ & \nd & $2.2^{+0.6}_{-0.5}$ & $0.9^{+0.3}_{-0.2}$ & $0.7^{+0.2}_{-0.2}$ & $16.8\%$ \\
                            & {\tt mekal} & $0.03^\dag$ & $4.1^{+8.4}_{-1.7}$ & \nd & $0.7^{+0.2}_{-0.2}$ & $0.9^{+0.3}_{-0.2}$ & $67.9\%$ \\
      \hline
      CX15   & {\tt pow} & $<0.4$ & \nd & $2.0^{+0.6}_{-0.4}$ & $2.0^{+0.4}_{-0.3}$ & $1.8^{+0.3}_{-0.3}$ & $9.5\%$\\
      CX16   & {\tt mekal} & $0.03^\dag$ & $2.9^{+10.3}_{-1.2}$ & \nd & $0.6^{+0.2}_{-0.2}$ & $0.5^{+0.2}_{-0.2}$ & $50.3\%$ \\
      CX17   & {\tt pow}  & $<0.4$ & \nd & $1.9^{+0.9}_{-0.7}$ & $0.5^{+0.2}_{-0.2}$ & $0.5^{+0.2}_{-0.2}$ & $16.9\%$ \\
      CX18   & {\tt mekal}  & $0.03^\dag$ & $2.0^\dag$ & \nd & $0.06^{+0.07}_{-0.04}$ & $0.1^{+0.2}_{-0.1}$ & $27.1\%$ \\
      CX19   & {\tt mekal}  & $0.03^\dag$ & $2.0^\dag$ & \nd & $0.1^{+0.1}_{-0.1}$ & $0.05^{+0.05}_{-0.03}$ & $81.2\%$ \\
      CX20   & {\tt mekal} & $0.03^\dag$ & $1.3^{+1.0}_{-0.5}$ & \nd & $0.6^{+0.2}_{-0.2}$ & $0.12^{+0.05}_{-0.04}$ & $55.6\%$ \\
      CX21   & {\tt mekal} & $0.03^\dag$ & $3.5^{+7.9}_{-1.5}$ & \nd & $0.5^{+0.2}_{-0.1}$ & $0.5^{+0.2}_{-0.2}$ & $82.1\%$ \\
      \hline
      \multirow{2}{*}{CX22}& {\tt pow} & $0.03^\dag$  & \nd & $0.7^{+0.5}_{-0.6}$ & $0.4^{+0.1}_{-0.1}$ & $1.8^{+0.6}_{-0.5}$ & $69.4\%$ \\
                           & {\tt mekal} & $0.03^\dag$ & $>11.9$ & \nd & $0.5^{+0.2}_{-0.1}$ & $1.3^{+0.4}_{-0.4}$ & $98.3\%$ \\ 
      \hline
      CX23   & {\tt mekal} & $0.03^\dag$ & $1.9^{+1.2}_{-0.5}$ & \nd & $0.5^{+0.2}_{-0.2}$ & $0.2^{+0.1}_{-0.1}$ & $22.2\%$ \\
      CX24   & {\tt mekal} & $0.03^\dag$ & $2.5^{+2.9}_{-1.0}$ & \nd & $0.7^{+0.2}_{-0.2}$ & $0.5^{+0.2}_{-0.1}$ & $69.6\%$ \\
      CX25   & {\tt mekal} & $0.03^\dag$ & $1.7^{+2.7}_{-0.6}$ & \nd & $0.4^{+0.2}_{-0.1}$ & $0.2^{+0.1}_{-0.1}$ & $68.4\%$ \\
      CX26   & {\tt pow} & $0.6^{+0.6}_{-0.6}$ & \nd & $2.1^{+0.7}_{-0.7}$ & $3.1^{+0.6}_{-0.5}$ & $2.5^{+0.5}_{-0.4}$ & $6.5\%$ \\ 
      CX27   & {\tt nsatmos} & $0.03^\dag$ & $0.2^{+0.2}_{-0.1}$ & $<0.5$ & $0.5^{+0.2}_{-0.2}$ & $0.13^{+0.05}_{-0.04}$ & $10.9\%$ \\
      \hline
      \multirow{2}{*}{CX28}  & {\tt pow} & $0.03^\dag$ & \nd & $4.9^{+2.5}_{-1.9}$ & $0.6^{+0.5}_{-0.3}$ & $0.01^{+0.01}_{-0.01}$ & $45.7\%$ \\
                             & {\tt mekal} & $0.03^\dag$ & $0.6^{+0.3}_{-0.4}$ & \nd & $0.3^{+0.2}_{-0.2}$ & $0.004^{+0.003}_{-0.002}$ & $66.1\%$ \\
      \hline
      CX29   & {\tt mekal} & $0.03^\dag$ & $2.0^\dag$ & \nd & $0.1^{+0.1}_{-0.1}$ & $0.06^{+0.06}_{-0.04}$ & $52.6\%$\\
      CX30   & {\tt mekal} & $0.03^\dag$ & $2.0^\dag$ & \nd & $0.1^{+0.2}_{-0.1}$ & $0.1^{+0.1}_{-0.1}$ & $8.3\%$ \\
      CX31   & {\tt mekal} & $0.03^\dag$ & $2.0^\dag$ & \nd & $0.1^{+0.2}_{-0.1}$ & $0.03^{+0.07}_{-0.03}$ & $15.9\%$ \\
      CX32   & {\tt mekal} & $0.03^\dag$ & $2.0^\dag$ & \nd & $0.2^{+0.2}_{-0.1}$ & $0.1^{+0.1}_{-0.1}$ & $17.2\%$ \\
      CX33   & {\tt mekal} & $0.03^\dag$ & $2.0^\dag$ & \nd & $0.3^{+0.4}_{-0.2}$ & $0.3^{+0.4}_{-0.2}$ & $49.8\%$ \\
      CX34   & {\tt mekal} & $0.03^\dag$ & $2.0^\dag$ & \nd & $0.4^{+0.5}_{-0.3}$ & $0.2^{+0.3}_{-0.1}$ & $37.0\%$ \\
      CX35   & {\tt mekal} & $0.03^\dag$ & $>2.3$ & \nd & $0.3^{+0.2}_{-0.1}$ & $0.6^{+0.3}_{-0.2}$ & $37.4\%$ \\
      CX36   & {\tt mekal} & $0.03^\dag$ & $>1.7$ & \nd & $0.3^{+0.2}_{-0.1}$ & $0.4^{+0.2}_{-0.2}$ & $38.0\%$ \\
      CX37   & {\tt mekal} & $0.03^\dag$ & $>1.8$ & \nd & $0.3^{+0.1}_{-0.1}$ & $0.3^{+0.2}_{-0.1}$ & $22.3\%$ \\
      CX38   & {\tt mekal}   & $0.03^\dag$ & $2.0^\dag$ & \nd & $0.1^{+0.2}_{-0.1}$ & $0.1^{+0.1}_{-0.1}$ & $43.5\%$ \\
      CX39   & {\tt pow}   & $0.03^\dag$ & \nd & $5.1^{+1.8}_{-1.5}$ & $0.8^{+0.5}_{-0.3}$ & $0.01^{+0.01}_{-0.00}$ & $18.1\%$ \\
      CX40   & {\tt mekal}   & $0.03^\dag$ & $2.0^\dag$ & \nd & $0.2^{+0.3}_{-0.1}$ & $0.1^{+0.1}_{-0.1}$ & $96.0\%$ \\
      CX41   & {\tt mekal} & $0.03^\dag$ & $>5.4$ & \nd & $0.7^{+0.2}_{-0.2}$ & $1.5^{+0.4}_{-0.3}$ & $33.3\%$ \\
      \hline
      \multirow{2}{*}{CX42} & {\tt pow}   & $0.03^\dag$ & \nd & $-0.4^{+0.9}_{-1.0}$ & $0.10^{+0.05}_{-0.04}$ & $1.9^{+0.8}_{-0.7}$ & $42.1\%$ \\
                          & {\tt mekal} & $0.03^\dag$ & $>18.2$ & \nd & $0.3^{+0.1}_{-0.1}$ & $0.8^{+0.4}_{-0.3}$ & $99.3\%$ \\
      \hline
    \end{tabular}%
\end{table*}

\begin{table*}
    \contcaption{Results of spectral analyses.}
    \centering
    \renewcommand{\arraystretch}{1.2}
    \begin{tabular}{cccccccc}
    \hline
       Source  & Model & $N_\mathrm{H}$ & Parameter 1 & Parameter 2 & $F_{0.5-2}$ & $F_{2-7}$ &  $\chi_\nu^2$ (dof) or Goodness$^c$\\
               &       & $(10^{22}~\mathrm{cm^{-2}})$ & & & \multicolumn{2}{c}{$(10^{-15}~\mathrm{erg~s^{-1}~cm^{-2}})$} & \\
    \hline
    \multirow{2}{*}{CX43} & {\tt pow}   & $0.03^\dag$ & \nd & $1.4^{+1.1}_{-0.9}$ & $0.2^{+0.1}_{-0.1}$ & $0.4^{+0.3}_{-0.2}$ & $29.4\%$ \\
                          & {\tt mekal} & $0.03^\dag$ & $>2.1$ & \nd & $0.2^{+0.1}_{-0.1}$ & $0.4^{+0.3}_{-0.2}$ & $21.2\%$ \\
    \hline
     CX44   & {\tt pow}  & $0.03^\dag$ & \nd & $-0.01^{+0.98}_{-1.10}$ & $0.09^{+0.05}_{-0.04}$ & $1.1^{+0.1}_{-0.5}$ & $4.9\%$ \\
     CX45   & {\tt pow}  & $0.03^\dag$ & \nd & $0.1^{+1.1}_{-1.3}$ & $0.09^{+0.05}_{-0.04}$ & $1.0^{+0.5}_{-0.4}$ & $30.4\%$ \\
    \hline
    \multirow{2}{*}{CX46} & {\tt pow}   & $0.03^\dag$ & \nd & $1.5^{+1.0}_{-0.9}$ & $0.3^{+0.1}_{-0.1}$ & $0.4^{+0.2}_{-0.2}$ & $32.8\%$ \\
                          & {\tt mekal} & $0.03^\dag$ & $>2.7$ & \nd & $0.2^{+0.1}_{-0.1}$ & $0.5^{+0.2}_{-0.2}$ & $29.1\%$ \\
    \hline
    \multirow{2}{*}{CX47} & {\tt pow}  & $0.03^\dag$  & \nd & $0.2^{+1.0}_{-1.2}$ & $0.09^{+0.06}_{-0.04}$ & $0.9^{+0.6}_{-0.4}$ & $1.8\%$ \\
                          & {\tt mekal} & $0.03^\dag$ & $>7.2$ & \nd & $0.2^{+0.1}_{-0.1}$ & $0.5^{+0.3}_{-0.2}$ & $99.5\%$ \\
    \hline
     CX48   & {\tt mekal} & $0.03^\dag$ & $>1.6$ & \nd & $0.1^{+0.1}_{-0.1}$ & $0.2^{+0.2}_{-0.2}$ & $63.9\%$ \\
     CX49   & {\tt mekal} & $0.03^\dag$ & $>0.7$ & \nd & $0.1^{+0.1}_{-0.1}$ & $0.2^{+0.3}_{-0.2}$ & $15.2\%$ \\
     CX50   & {\tt mekal}   & $0.03^\dag$ & $2.0^\dag$ & \nd & $0.1^{+0.2}_{-0.1}$ & $0.1^{+0.1}_{-0.1}$ & $82.3\%$ \\
     CX51   & {\tt mekal}   & $0.03^\dag$ & $2.0^\dag$ & \nd & $0.04^{+0.15}_{-0.04}$ & $0.02^{+0.08}_{-0.02}$ & $40.8\%$ \\
     CX52   & {\tt mekal}   & $0.03^\dag$ & $2.0^\dag$ & \nd & $0.1^{+0.3}_{-0.1}$ & $0.1^{+0.1}_{-0.1}$ & $2.0\%$\\
     MSP A  & {\tt nsatmos} & $0.03^\dag$ & $0.3^{+0.2}_{-0.1}$ & $<0.2$ & $0.5^{+0.2}_{-0.2}$ & $0.3^{+0.1}_{-0.1}$ & $55.2\%$ \\
     MSP C  & {\tt nsatmos} & $0.03^\dag$ & $0.2^{+0.1}_{-0.1}$ & $<0.6$ & $0.8^{+0.3}_{-0.2}$ & $0.2^{+0.1}_{-0.1}$ & $11.7\%$ \\
     MSP E  & {\tt nsatmos} & $0.03^\dag$ & $0.2^{+0.5}_{-0.1}$ & $<0.4$ & $0.2^{+0.2}_{-0.1}$ & $0.04^{+0.04}_{-0.02}$ & $4.6\%$ \\
     \hline
     \multicolumn{8}{l}{\makecell[tl]{$^a$Parameter 1 can be the temperature ($kT$), in $\kev$, for {\tt nsatmos} or {\tt mekal} model, or cutoff energy ($E_\mathrm{cut}$; in $\kev$) \\ of the {\tt cutoffpl} model used for CX5.}}\\
     \multicolumn{8}{l}{\makecell[tl]{$^b$Parameter 2 can be $\Gamma$ for {\tt pow} models, effective emission radii in $\km$ for {\tt nsatmos} models, line energy in $\kev$ \\ of the {\tt gabs} component used for CX5, or temperature in $\kev$ for the second {\tt mekal} component for CX8.}}\\
     \multicolumn{8}{l}{\makecell[tl]{$^c$We report the reduced $\chi^2$ ($\chi_\nu^2$) and the corresponding degrees of freedom (dof) for fits using $\chi^2$-statistics.\\ For fits using C-statistics, we report the goodness fractions described in \S\ref{sec:x_ray_spec_analyses}.}}\\
     \multicolumn{8}{l}{$^\dag$ superscripts indicate fixed parameters during the fits.}\\
     \multicolumn{8}{l}{$^\ast$ superscripts indicate parameters with no valid constraints due to $\chi_\nu^2>2$.}\\
     \multicolumn{8}{l}{Parameters reported with $>$ or $<$ signs exceed the pre-defined upper or lower limit in {\sc xspec}.}
    \end{tabular}%
    \label{tab:my_label}
\end{table*}

\subsection{Intra-observational X-ray variability}
\label{sec:intra_observational_variability}
We used the {\sc ciao} {\tt glvary} tool\footnote{\url{http://cxc.harvard.edu/ciao/ahelp/glvary.html}} to 
\revised{search for variation of sources within} 
observations. The {\tt glvary} tool utilises the Gregory-Loredo algorithm \citep{Gregory92} to look for significant differences between events in separate time bins, and 
assigns variability indices. 
Indices 
$\geq 6$ 
\revised{indicate}
high-confidence \craig{($P>90$\% conf.)} 
\revised{variability.}

\begin{figure*}
    \centering
    \includegraphics[scale=0.35]{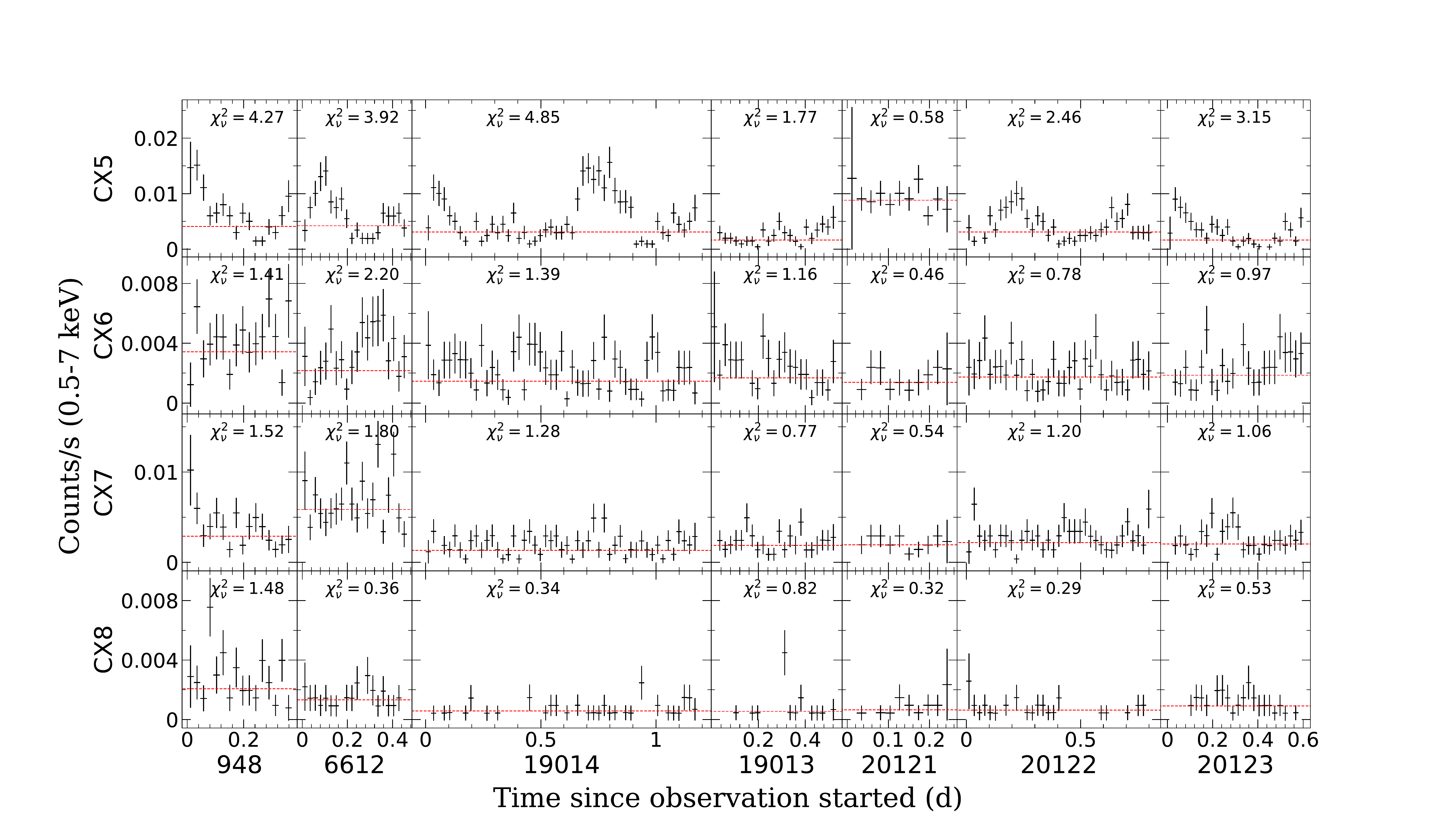}
    \caption{0.5-7 keV X-ray light curves of sources with variability indices greater or equal to 6 in one or more Obs. IDs (Table \ref{t:variability_indices}). Each time bin is $2000~\mathrm{s}$ long.}
    \label{f:all_lc}
\end{figure*}

We found strong variability in one or more observations for CX5 (CV), CX6 (CV), CX7 (CV), CX8 (AB), CX13 (CV), and CX21 (CV) 
(Table \ref{t:variability_indices}). 
CX5 shows the strongest variability with indices $\geq 6$ in all but one (20121) observation. 
Variability in the faint sources may not be detectable in the light curves, so we only extracted light curves (binned to 2000\,s) for the relatively bright sources using the {\sc ciao} {\tt dmextract} tool, which are shown in Fig. \ref{f:all_lc}. The red dashed lines 
\revised{are}
the best constant fits to the light curve, from which we calculate the reduced $\chi^2$ 
\revised{to measure variability.}

\begin{table}
    \caption{Variability indices of sources that show strong variability in one or more Obs. IDs.}
    \centering
    \resizebox{\columnwidth}{!}{%
    \begin{tabular}{lccccccc}
    \hline
        Obs. ID & 948 & 6612 & 19014 & 19013 & 20121 & 20122 & 20123 \\
        Source  & & & & & & & \\ 
    \hline
        CX5     &  8  &  8   &  10   &   6   &   0   &   8   &   8   \\
        CX6     &  0  &  6   &   0   &   0   &   0   &   0   &   0   \\
        CX7     &  1  &  6   &   0   &   0   &   0   &   0   &   0   \\
        CX8     &  0  &  0   &   0   &   7   &   0   &   6   &   2   \\
        CX13    &  0  &  6   &   0   &   1   &   0   &   1   &   0   \\
        CX21    &  1  &  1   &   6   &   1   &   0   &   0   &   0   \\
    \hline
    \end{tabular}%
    }
    \label{t:variability_indices}
\end{table}


\section{Radio observation and analyses}
\label{s:radio_observation}
We 
\revised{use} 
radio data from the MAVERIC (Milky Way ATCA and VLA Exploration of Radio sources In Clusters) survey (Project Code: C2877; see e.g., \citealt{Tremou18}). NGC 6752 was observed by the {\it Australia Telescope Compact Array} (ATCA) in three separate observing blocks spanning 2014-02-06 to 2014-02-09 (MJD 56694.88-56697.13) for a total on-source integration time of 20.1 hr. The observation was performed in the extended 6D configuration, with two $2~\mathrm{GHz}$-wide bands centred on $5.5$ and $9.0~\mathrm{GHz}$. 

We use the {\sc miriad} software \citep{Sault95} for preliminary calibration, 
\revised{via} 
standard procedures, and {\sc casa} \citep[version 4.2.0;][]{McMullin07} for imaging, 
\revised{making} 
radio images with noise levels of $3.8~\ujy/\mathrm{beam}$ and $4.3~\ujy/\mathrm{beam}$, with synthesised beam sizes of $2.8\arcsec\times 1.5\arcsec$ and $1.8\arcsec \times 1.0\arcsec$ at $5.5$ and $9.0~\mathrm{GHz}$, respectively.

We then generate $5\,\sigma$ source catalogues from the $5.5$ and $9.0~\mathrm{GHz}$ images, using the source detection software PyBDSF \citep[version 1.8.13;][]{Mohan15}. 
Details will be presented in 
Tudor et al. (in prep). For sources detected in both bands, we compute 
spectral indices, $\alpha$, defined by $S_\nu \propto \nu^\alpha$, where $S_\nu$ is the spectral flux density in units of $\mathrm{erg~s^{-1}~cm^{-2}~Hz^{-1}}$. Radio positional uncertainties 
from our 
detection workflow are typically underestimated, so we inflate the  
\revised{errors to $\geq$}1/10 the size of the beam.

\begin{figure*}
    \centering
    \includegraphics[scale=0.3]{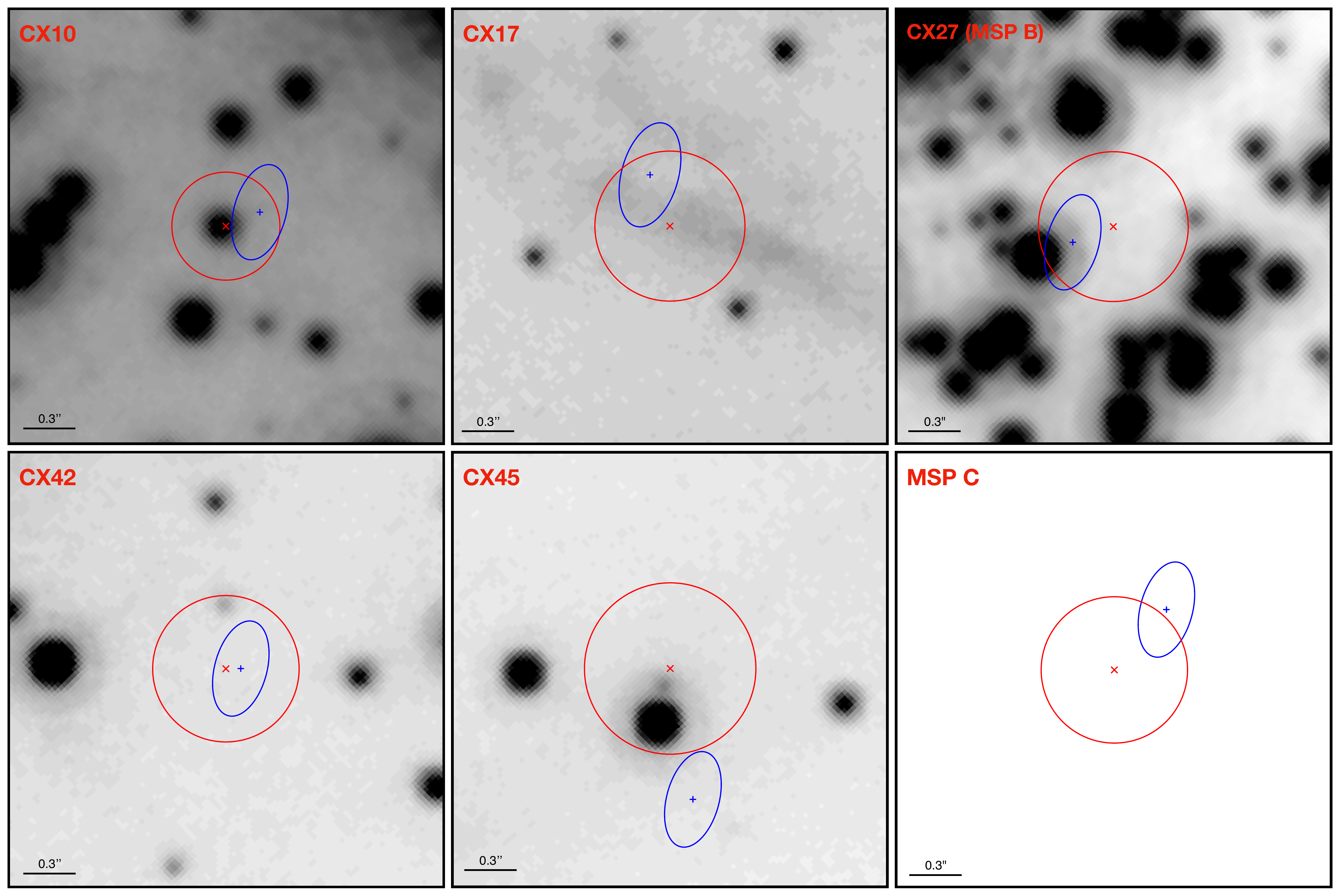}
    \caption{\coryrv{$2\farcs5 \times 2\farcs5$ H$\alpha$ finding charts for the radio-X-ray cross-matched sources including CX10 (AGN), CX17 (interacting galaxy), CX27 (MSP B), CX42 (AGN), CX45 (AGN), and MSP C; north is up and east is to the left. Note that MSP C is outside the ACS FOV. The nominal X-ray and radio positions are indicated with red x's and blue crosses. The red circles represent the 95\% X-ray confidence error regions according to \citet{Hong05}; the blue ellipses shows $1/10$ of the $5.5~\mathrm{GHz}$ radio beam as mentioned in \S\ref{s:radio_observation}.}} 
    \label{f:radio_xray_matched_finders}
\end{figure*}

\coryrv{Cross-matching the   $5\,\sigma$ source catalogue with our {\it Chandra} catalogue reveals 5 matches: CX17, CX27, CX42, CX45, and MSP C. There is also a $3\,\sigma$ catalogue source coincident with CX10 in the $5.5~\mathrm{GHz}$ image. This is a less confident match, so should be interpreted with caution. We summarise the radio fluxes and spectral indices of these sources in Table \ref{t:radio_matches}, and in Fig. \ref{f:radio_xray_matched_finders}, we present the corresponding H$\alpha$ finding charts 
radio and X-ray positions.}

\begin{table*}
    \centering
    \caption{Radio positions and fluxes of radio counterparts.}
    \begin{tabular}{lccccc}
    \hline
    Source$^a$ & \multicolumn{2}{c}{Radio Position}  & \multicolumn{2}{c}{$S_\nu~(\ujy)^b$}    & $\alpha$ \\
         & $\alpha_R$ & $\delta_R$ & $5.5~\mathrm{GHz}$ & $9~\mathrm{GHz}$ & $S_\nu \propto \nu^\alpha$ \\
    \hline
    CX10$^\ast$    & 19:10:54.73(4) & $-$59:59:13.8(3) & $15.3  \pm 4.1$   & $<13.2$          & \nd \\
    CX17           & 19:11:05.29(1) & $-$59:59:04.3(3) & $187.0 \pm 10.5$  & $131.0 \pm 19.1$ & $-0.7 \pm 0.3$ \\
    CX27 (MSP B)   & 19:10:52.10(1) & $-$59:59:01.0(3) & $20.5  \pm 3.6$   & $<13.0$          & \nd \\
    CX42           & 19:10:48.11(1) & $-$60:00:07.2(3) & $41.7  \pm 4.2$   & $33.4 \pm 4.5$   & $-0.5 \pm 0.3$ \\
    CX45           & 19:10:38.92(2) & $-$59:59:23.0(3) & $23.5  \pm 4.3$   & $23.4 \pm 5.0$   & $0.0 \pm 0.6$ \\
    MSP C          & 19:11:05.52(5) & $-$60:00:59.4(7) & $27.8  \pm 6.3$   & $<20.9$          & \nd \\
    \hline
    \multicolumn{6}{l}{$^a$An $^\ast$ indicates the radio counterpart is detected at the $3~\sigma$ level.}\\
    \multicolumn{6}{l}{$^b$The upper limits are at the $3~\sigma$ level.}
    \end{tabular}
    \label{t:radio_matches}
\end{table*}

\section{Optical/UV Observations}

\revised{We use} the \hst\ ACS/WFC photometric dataset described by L17 \revised{(Table~\ref{t:hst_obs})}, from imaging obtained in program GO-12254 (PI: Cool). This dataset provides F435W (\B), F625W (\R), and F656N (\ha) magnitudes for 68,439 stars within a mosaic that covers \revised{slightly more than} the half-light region of NGC 6752. 
\revised{We used the KS2 update of the photometric software}
developed for the ACS Globular Cluster Treasury project, 
described in \citet{Anderson08}.
We constructed CMDs and a colour-colour diagram from the GO-12254 photometry, employing the colours \br\ and \hr.  

As discussed in L17, the drizzle-combined ACS/WFC optical mosaics were rectified to the ICRS using approximately 600 astrometric standards from the USNO UCAC3 catalog. The RMS residual of the plate solution was 0\farcs09 in each coordinate. We determined a boresight correction for the Chandra source coordinates by computing the mean offsets between the \hst\ and \chandra\ coordinates for sources CX2, CX3, and CX4. 

To extend the analysis 
\revised{of}
L17, we employed the Hubble UV Globular Cluster Survey \citep[HUGS;][]{Piotto15,Nardiello18}. This provides imaging and photometry of NGC 6752 in F275W (\UV), F336W (\U), and F435W (\B). The imaging was obtained with the WFC3/UVIS (\UV\ and \U) and the ACS/WFC (\B). This latter ACS/WFC imaging is from program GO-12254, i.e.\ the same dataset as used for the optical analysis. However, one of the six visits was omitted in the HUGS analysis. The HUGS observations are also listed in Table~\ref{t:hst_obs}. The HUGS photometry was performed with the same KS2 software that was used for the optical analysis. We constructed CMDs from the HUGS photometry, employing the colours \uvu\ and \ub. Inclusion of the HUGS photometry provides several benefits in the search for counterparts to \chandra\ sources. First, the HUGS UV colours allow a further check on the blue objects detected in the (\br, \R) CMD of L17. In some cases, objects that are only marginally blue in \br\ have strong UV excesses. Second, the HUGS \UV\ and \U\ photometry allows the detection of very blue objects that cannot be detected at redder wavelengths. Third, the HUGS photometry allows the detection of faint blue objects that are situated in close proximity with much brighter red objects. In imaging at UV wavelengths, the contribution from these red objects is minimised.  

\revised{We note that some programs have used far-UV (FUV; $\lambda \lesssim 2000$\,\AA) observations in the search for counterparts to globular cluster X-ray sources. FUV imaging further reduces the contribution of red objects and emphasises hot objects, particularly CVs. These studies have investigated 47 Tuc \citep{Knigge02,Knigge03}, M70 \citep{Connelly06}, M15 \citep{Dieball07,Dieball10a,Haurberg10}, M80 \citep{Dieball10b,Thomson10}, NGC 6752 \citep{Thomson12}, and NGC 6397 \citep{Dieball17}. They have detected a number of CV candidates and dwarf novae, as well as other stellar exotica, such as BSSs, extreme horizontal branch (EHB) stars, ``blue hook'' stars, and helium-core WDs. In the CMDs that include FUV imaging, the CV candidates typically lie in the ``gap'' between the MS and the EHB. }

\begin{table}
    \caption{\hst\ observations used in this work.}
    \label{t:hst_obs}
    \centering
    \resizebox{\columnwidth}{!}{%
    \begin{tabular}{lclcr}
    \hline
    Program & Observation Date Range & Instrument & Filter & Exposure \\ 
    & & & & Time (s) \\
    \hline
    GO-12254          & 2011-05-19 to 2011-11-14 & ACS/WFC   & F435W & 4560 \\
    GO-12254          & 2011-05-19 to 2011-11-14 & ACS/WFC   & F625W & 4380 \\
    GO-12254          & 2011-05-19 to 2011-11-14 & ACS/WFC   & F658N & 18,528 \\
    GO-12311          & 2011-03-23 to 2011-04-03 & WFC3/UVIS & F275W & 4428 \\
    GO-12311          & 2010-05-05 to 2010-05-05 & WFC3/UVIS & F336W & 1000 \\
    GO-12254$\dagger$ & 2011-05-19 to 2011-11-14 & ACS/WFC   & F435W & 3800 \\
    \hline
    \multicolumn{5}{l}{$\dagger$The F435W frames used by the HUGS program include all of the GO-12254 visits,} \\ 
    \multicolumn{5}{l}{\quad except for 2011-09-07.} \\
    \end{tabular}%
    }
\end{table}

In order to find new optical identifications for the expanded X-ray source list, we examined the error circle of each X-ray source using the techniques described in L17. 
We determined 95\% confidence X-ray error circle radii using the prescription of \citet{Hong05}, 
given in Table~\ref{t:counterparts}. For strong, on-axis sources, the error circle radii approach a minimum of about 0\farcs3, and are larger for weaker and/or off-axis sources. In a few cases where the error circle contained no clearly convincing counterparts, we extended our search area somewhat beyond the formal 95\% error circle (see Table~\ref{t:counterparts}, column labelled ``Offset'').

\begin{table}
\revised{
    \caption{Chance Coincidence Analysis}
    \label{t:chance_coincidence}
    \centering
    \begin{tabular}{lrrrc}
    \hline
    Population & ${\no}^a$ & ${\np}^b$ & Excess$^c$ & Significance ($\sigma$)$^d$ \\[1pt]
    \hline
    MS      & 43 & 45.0 & $-2.0$ & 0.3 \\
    WD      &  4 &  1.0 &   3.0  & 1.3 \\
    Gap     & 15 &  5.8 &   9.2  & 2.5 \\
    RGB     &  4 &  2.5 &   1.5  & 0.4 \\
    SGB     &  1 &  3.9 & $-2.9$ & 2.0 \\
    \hline
    \multicolumn{5}{l}{$^a$Observed number of population members in all error circles} \\
    \multicolumn{5}{l}{$^b$Predicted number of population members in all error circles} \\
    \multicolumn{5}{l}{$^c$Excess (deficit if negative) of observed vs.\ predicted number} \\
    \multicolumn{5}{l}{\quad in all error circles} \\
    \multicolumn{5}{l}{$^d$Significance of excess or deficit expressed as Gaussian-equivalent} \\
    \multicolumn{5}{l}{\quad $\sigma$ level, based on \citet{Gehrels86} statistics} \\
    \end{tabular}%
}
\end{table}

\section{Classifications}
\label{s:classification}
\begin{figure}
\includegraphics[width=\columnwidth]{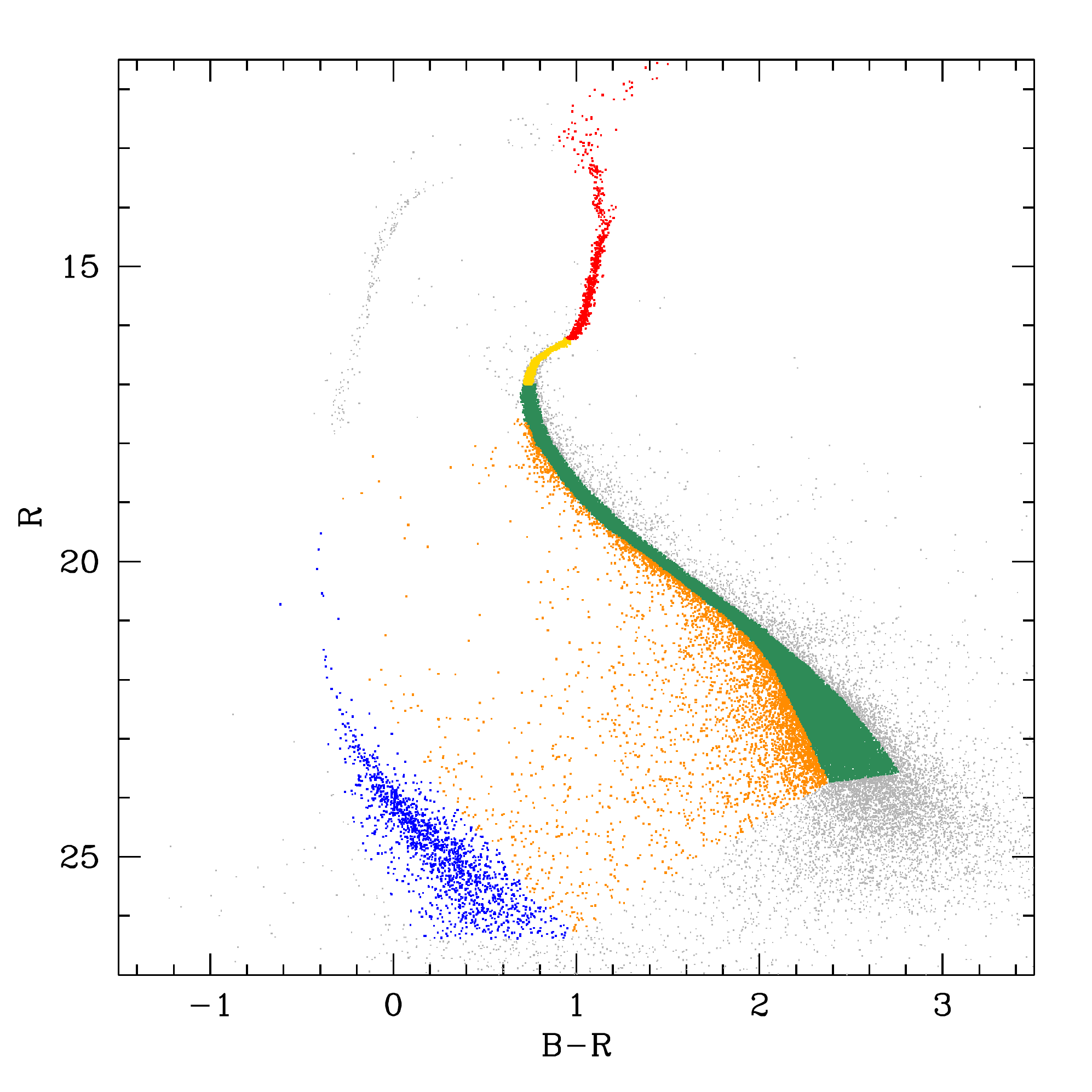}
\caption{Stellar population selection using Glue software. Color key: blue--WD; orange--``gap'' star; green--MS; gold--subgiant; red--red giant; grey--all other stars.}
\label{f:population_selection}
\end{figure}

\begin{figure}
\includegraphics[width=\columnwidth]{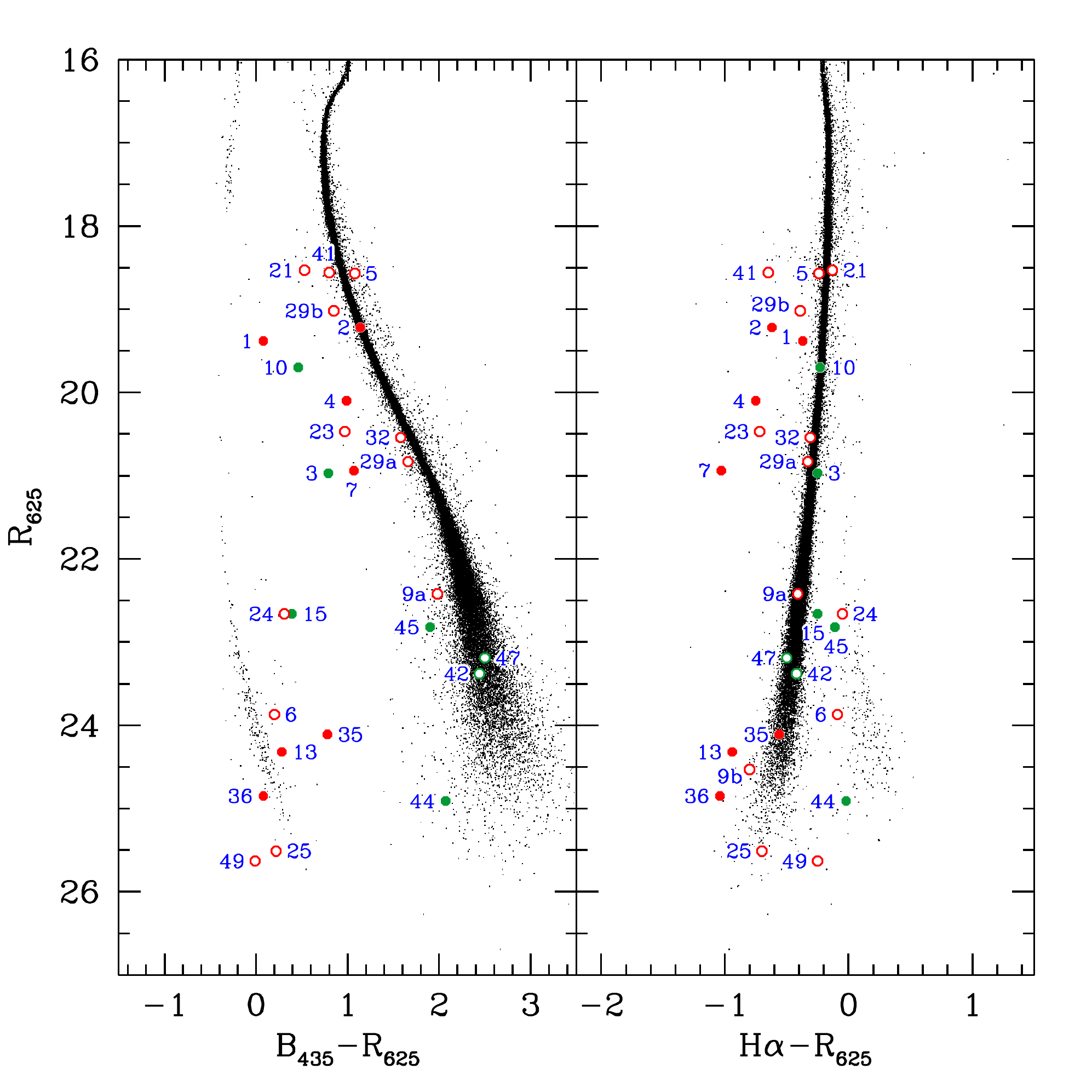}
\caption{Proper-motion cleaned colour-magnitude diagrams for stars within the half-light radius of NGC 6752 and CV (red symbols) or AGN (green symbols) identifications. The candidates have been selected based on their blue colour and/or \ha\ excess. Open symbols indicate less certain CV identifications, either due to a weak or absent \ha\ excess and/or uncertain photometry. Note that in the right panel, the bright CVs mostly lie to the \ha-excess side of the MS, while the faint CVs mostly lie to the \ha-excess side of the WD clump, which itself lies to the \ha-deficit side of the MS. All candidate counterparts are shown, independent of their proper-motion status. The counterparts to CX3, CX10, CX15, CX44, and CX45 have proper motions that are consistent with the extragalactic frame, leading to their identification as AGNs. 
\label{f:CMD_CV}}
\end{figure}

Figures \ref{f:finding_charts_1-6}--\ref{f:finding_charts_50-52} provide optical and UV finding charts of the error circle regions of all 51 sources. The \B, \R, and \ha\ images are from the GO-12254 dataset, and the \UV\ and \U\ images are from the HUGS database. For all objects that fell within the error circles of the \chandra\ sources, we determined their location in the colour-magnitude diagrams (CMDs) and the colour-colour diagram. Objects that fell on the main sequence (MS) were considered to be unlikely counterparts, given the frequency with which MS stars will land in X-ray error circles by chance. L17 found that the number of expected chance alignments of MS stars with error circles is within a factor of two of the observed number within error circles (see their \S4.3). \revised{We have revisited this issue with a similar approach to that of \citet{Zhao20b}. We first used the Glue software package \citep{Beaumont15,Robitaille17} to define regions in the (\R, \br) CMD\@, as shown in Fig.~\ref{f:population_selection}, including WDs, ``gap'' stars, MS stars, subgiants, and red giants. We then determined how many of each type of star fell within all of the error circles combined and compared this to the expected total number of chance coincidences based on the local density of such objects. The local densities for each group were computed by binning the counts for each group in logarithmically spaced radial bins. The results of this analysis are given in Table~\ref{t:chance_coincidence}, where the observed and predicted numbers of each population are given, along with the excess (or deficit) and the statistical significance of this excess (or deficit). The significance levels are computed following the precepts of \citet{Gehrels86}, who determined Poisson-statistics-based confidence limits for small numbers of events in astrophysical data. We start with equations (9) and (12) from \citet{Gehrels86}, which give the single-sided upper and lower limits, $\lambda_u$ and $\lambda_l$ respectively, for a sample size of $n$ and an equivalent Gaussian $\sigma$ level of $S$,
\begin{equation}
    \lambda_u \approx (n+1) \left[1 - \frac{1}{9(n+1)} + \frac{S}{3\sqrt{n+1}}\right]^3
    \label{eq:upper_confidence_limit}
\end{equation}
\noindent and
\begin{equation}
    \lambda_l \approx n \left[1 - \frac{1}{9n} - \frac{S}{3\sqrt{n}}\right]^3 .
    \label{eq:lower_confidence_limit}
\end{equation}

\noindent In the case that observed number of the members of a population that fall in the error circles, $\no$, exceeds the predicted number, $\np$, Eqn.~\ref{eq:upper_confidence_limit} can be solved for $S$ by setting $\lambda_u = \no$ and $n = \np$, with the result,
\begin{equation}
    S_\mathrm{excess} = 3{\no}^\frac{1}{3}(\np+1)^\frac{1}{6} - 3(\np+1)^\frac{1}{2} + \frac{(\np+1)^{-\frac{1}{2}}}{3} .
\end{equation}

\noindent Similarly, when $\no$ is less than $\np$, the significance level of this deficit can be solved for from Eqn.~\ref{eq:lower_confidence_limit}, with the result,
\begin{equation}
    S_\mathrm{deficit} = -3{\no}^\frac{1}{3}{\np}^\frac{1}{6} + 3{\np}^\frac{1}{2}-\frac{{\np}^{-\frac{1}{2}}}{3} .
\end{equation}

Table~\ref{t:chance_coincidence} indicates that the number of MS stars observed within the error circles closely agrees with the expected number---43 observed versus 45 expected---i.e.\ agreement to much better than a factor of two. This justifies our rejection of MS stars that fall within error circles as potential source counterparts. There is an excess number of WD-like objects within the error circles, but the difference is not statistically significant, given the small numbers. There is a significant excess of gap stars in error circles, at the 2.5-$\sigma$ level. As can be seen from Fig.~\ref{f:CMD_CV}, WD-like and gap objects comprise the likely and possible CVs. We note that if a WD-like object also shows other evidence for a CV identification, such as an \ha\ excess or variability, then the probability that it is a CV is significantly enhanced relative to the result of the analysis just presented. 
}

\begin{figure}
\includegraphics[width=\columnwidth]{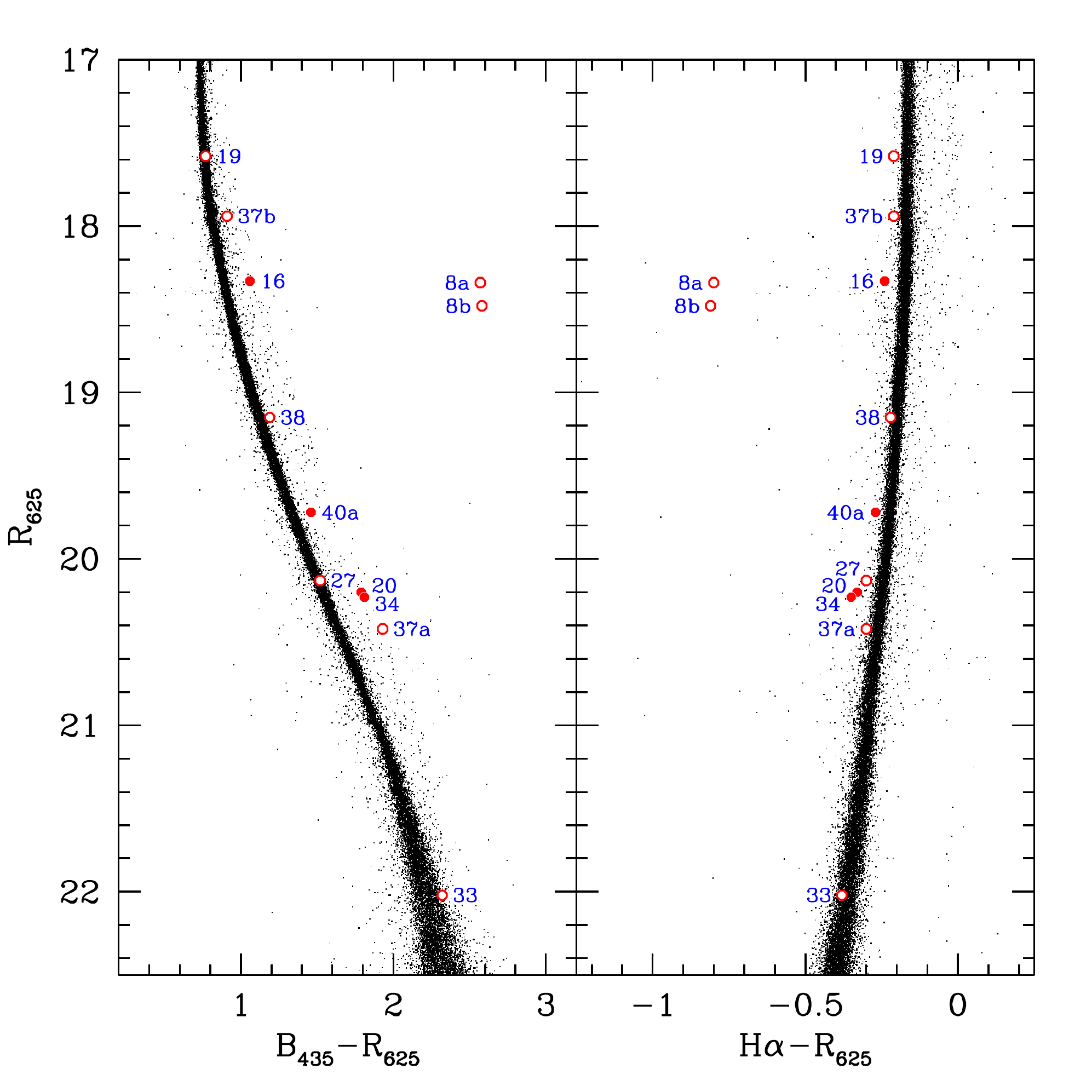}
\caption{Proper-motion cleaned colour-magnitude diagrams for stars within the half-mass radius of NGC 6752 and AB identifications.  The candidates have been selected based on their red colour and generally small \ha\ excess. All candidate counterparts are shown, independent of their proper motion status. The two counterparts to CX8 have an apparent proper motion that is inconsistent with the cluster mean, leading to their identification as foreground objects. 
\label{f:CMD_AB}}
\end{figure}

\begin{figure}
\includegraphics[width=\columnwidth]{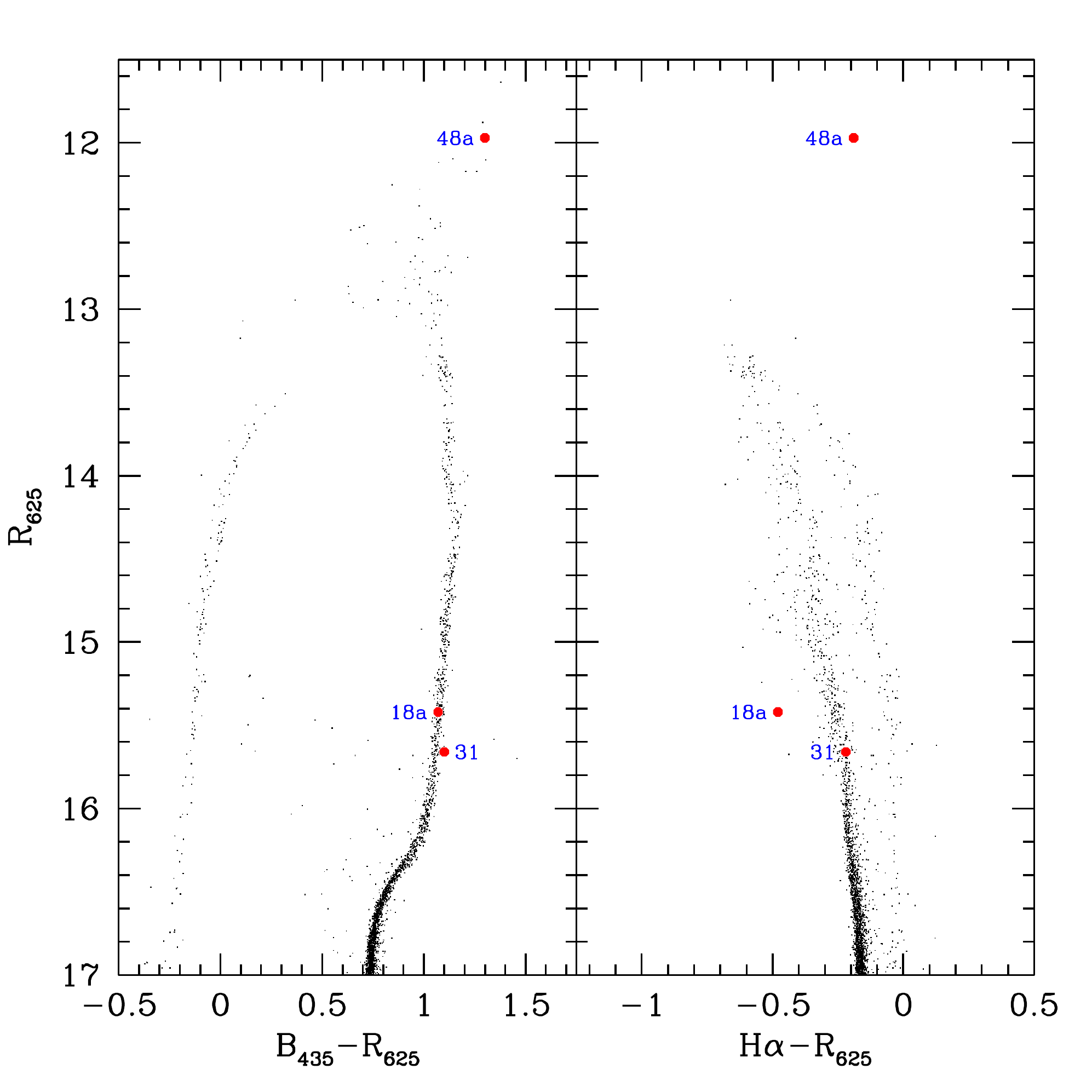}
\caption{Proper-motion cleaned colour-magnitude diagrams for stars within the half-light radius of NGC 6752 and red giant identifications. The counterpart to CX31 has an apparent proper motion that is inconsistent with the cluster mean, suggesting that it might be a foreground object. The counterpart to CX48 is saturated in \ha. 
\label{f:CMD_RG}}
\end{figure}

\begin{figure}
\includegraphics[width=\columnwidth]{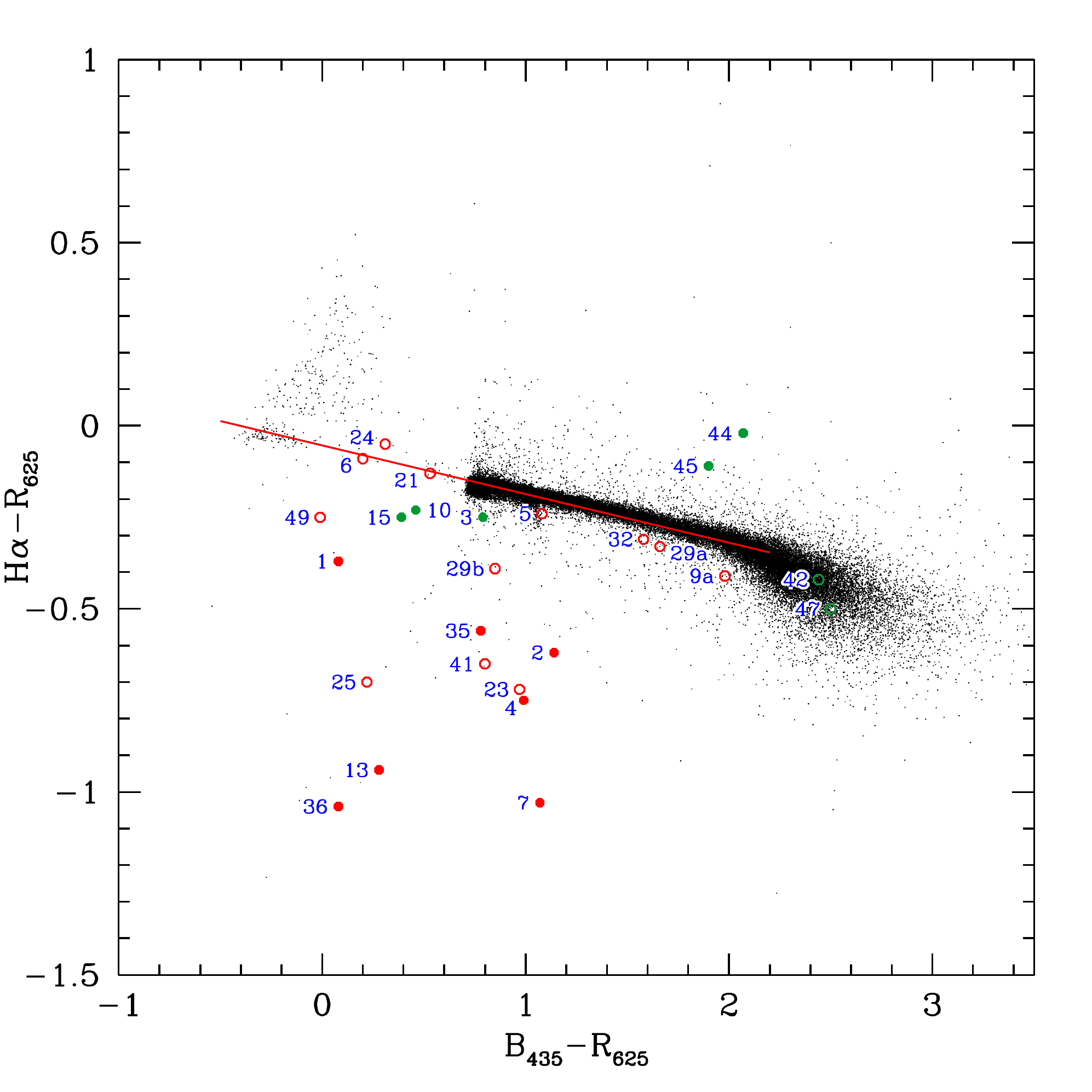}
\caption{Proper-motion cleaned colour-colour diagram for stars within the half-light radius of NGC 6752 and CV identifications.  The candidates are the same as in Fig.~\ref{f:CMD_CV}.  Open symbols indicate less certain CV identifications.  The red line is a linear regression of \hr\ on \br\ over the range $-0.5 \le \br \le 2.2$. All stars brighter than $\R = 15.5$ have been excluded, since saturation effects set in at the bright end.  The blue end of the colour-colour relation is populated by stars on the extreme blue horizontal branch. Note that all of the candidates except the counterpart to CX24 lie below (i.e.\ to the \ha-excess side of) the colour-colour relation.  
\label{f:CCD_CV}}
\end{figure}

\begin{figure}
\includegraphics[width=\columnwidth]{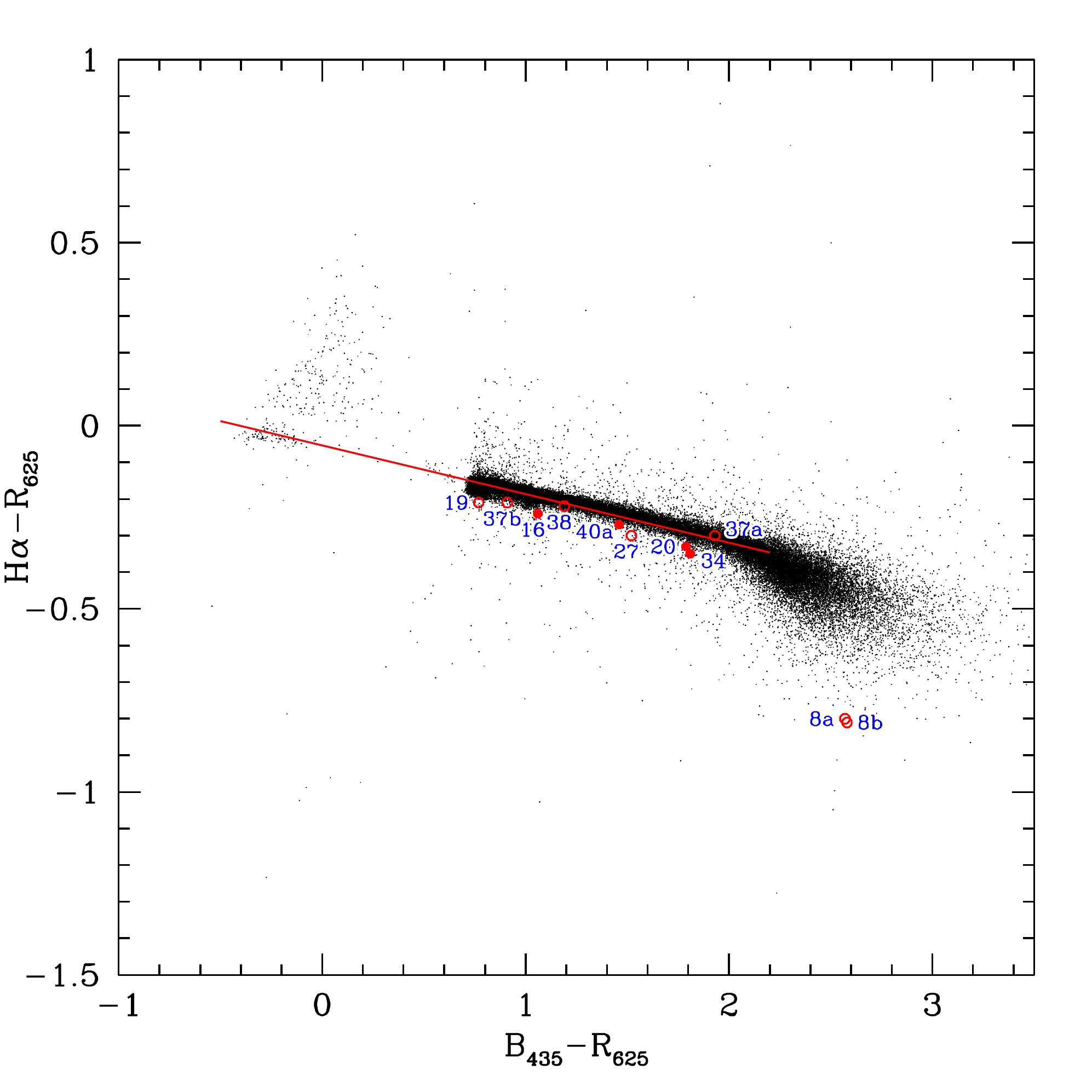}
\caption{Proper-motion cleaned colour-colour diagram for stars within the half-light radius of NGC 6752 and AB identifications.  The candidates are the same as in Fig.~\ref{f:CMD_AB}. Note that all of the candidates lie below (i.e.\ to the \ha-excess side of) the colour-colour relation.
\label{f:CCD_AB}}
\end{figure}

\begin{figure}
\includegraphics[width=\columnwidth]{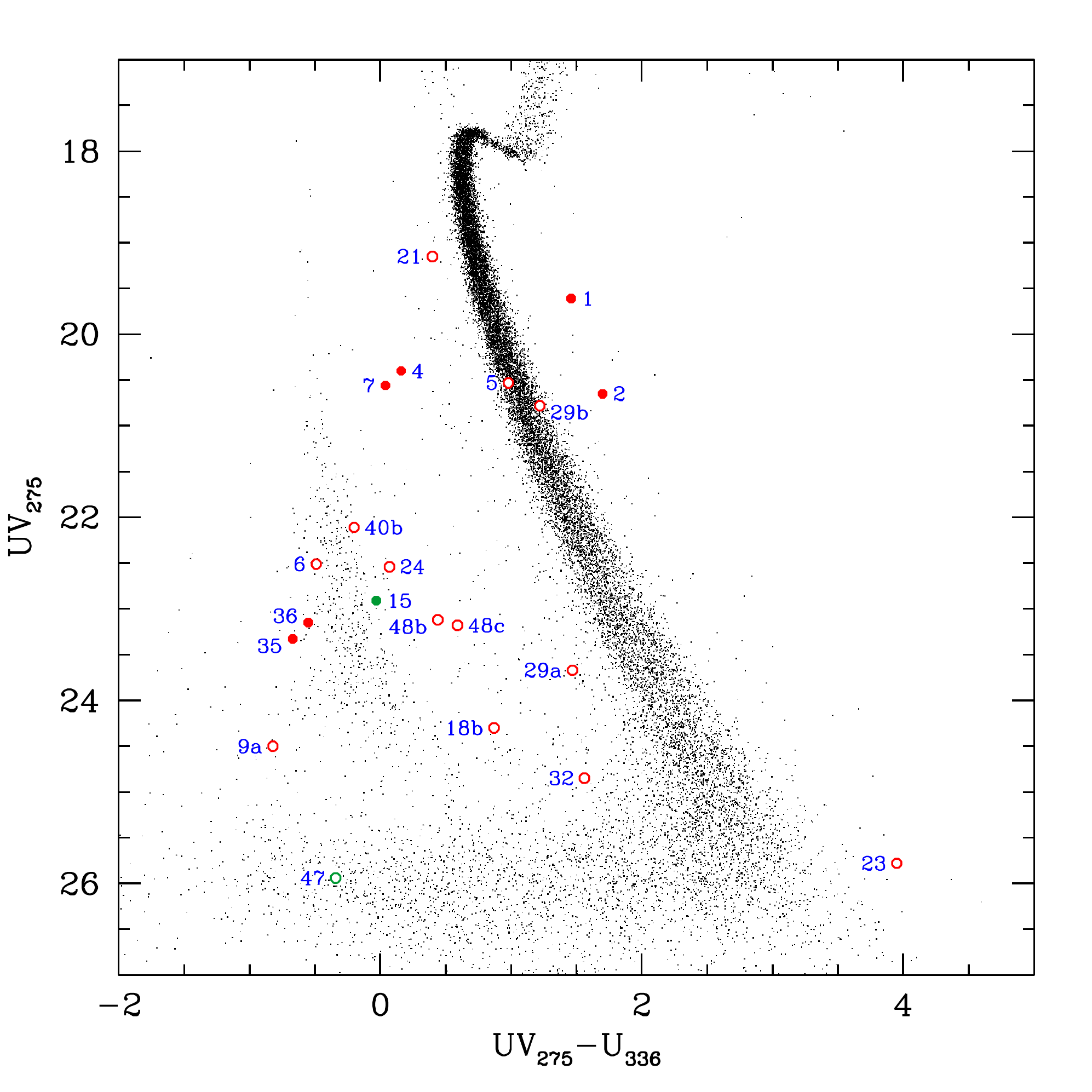}
\caption{(\uvu, \UV) colour-magnitude diagram for stars within the half-light radius of NGC 6752 and CV identifications.
\label{f:CMD_HUGS_UV-U_CV}}
\end{figure}

\begin{figure}
\includegraphics[width=\columnwidth]{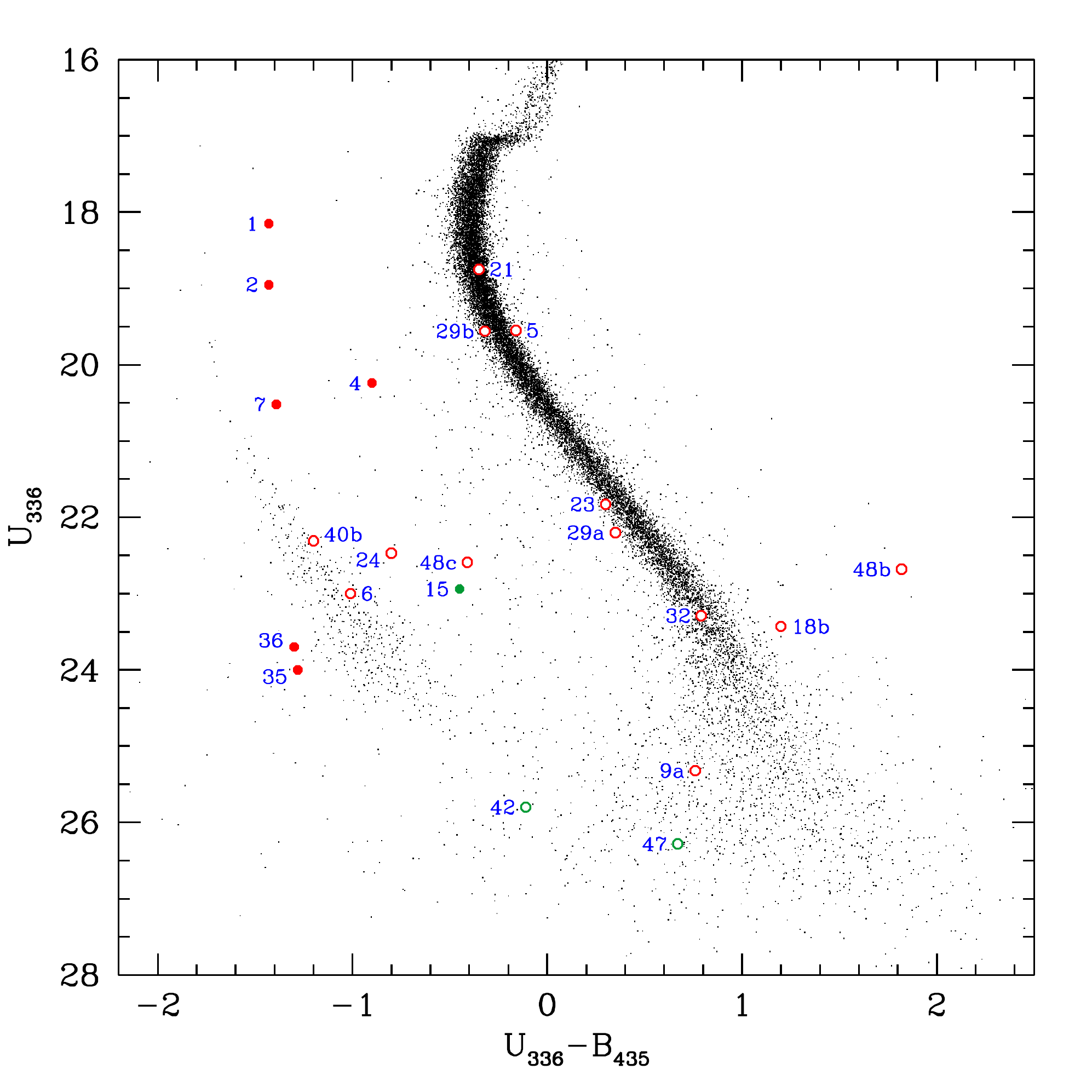}
\caption{(\ub, \U) colour-magnitude diagram for stars within the half-light radius of NGC 6752 and CV identifications.
\label{f:CMD_HUGS_U-B_CV}}
\end{figure}

\begin{figure}
\includegraphics[width=\columnwidth]{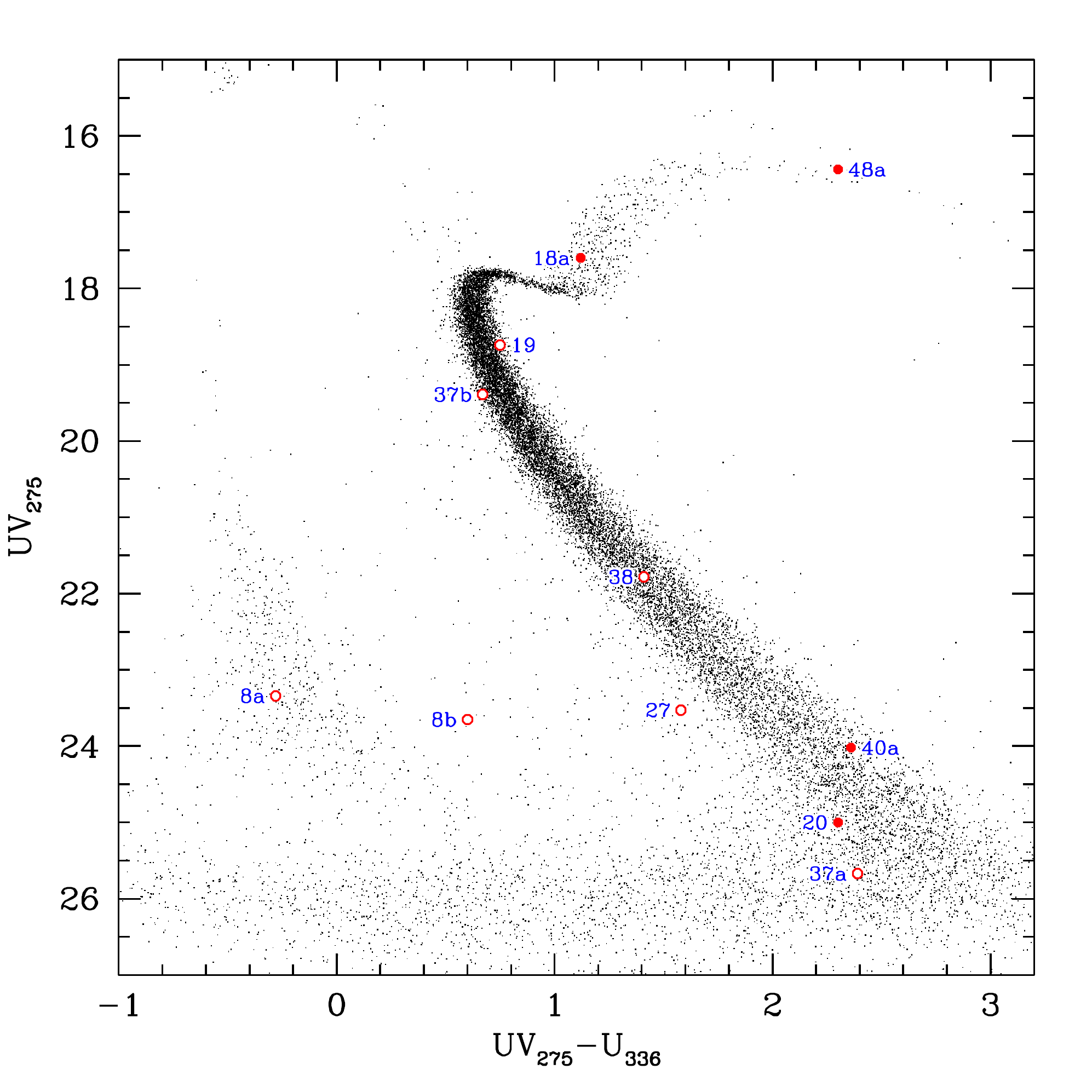}
\caption{(\uvu, \UV) colour-magnitude diagram for stars within the half-light radius of NGC 6752 and AB or red giant identifications.
\label{f:CMD_HUGS_UV-U_AB}}
\end{figure}

\begin{figure}
\includegraphics[width=\columnwidth]{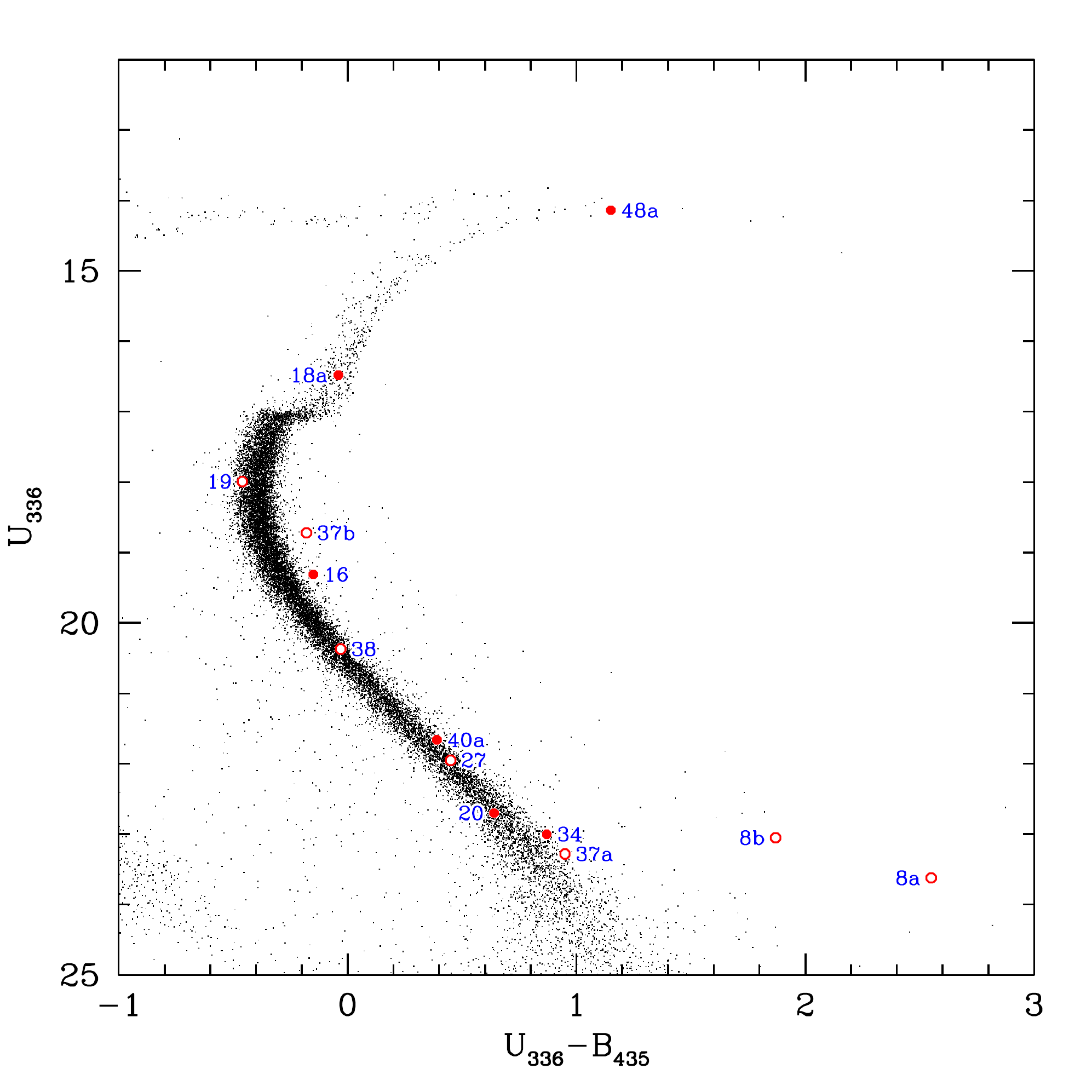}
\caption{(\ub, \U) colour-magnitude diagram for stars within the half-light radius of NGC 6752 and AB or red giant identifications.
\label{f:CMD_HUGS_U-B_AB}}
\end{figure}

We did note cases of red giants 
that fell within the error circles. While single giants are expected to have low X-ray to optical flux ratios, a giant with an unseen either MS or compact companion could be a plausible counterpart. Blue objects with significant \ha\ excesses were considered to be likely CVs, while red objects with modest \ha\ excesses were considered to be likely ABs. More ambiguous cases were classified as possible CVs and possible ABs. Fig.~\ref{f:CMD_CV} shows the optical CMDs for the CVs, Fig.~\ref{f:CMD_AB} shows the optical CMDs for the ABs, and Fig.~\ref{f:CMD_RG} shows the optical CMDs for the giants. As in L17, the \ha\ status for the CVs and ABs is clarified by examination of the colour-colour diagram, Figs.~\ref{f:CCD_CV} and \ref{f:CCD_AB}. Figs.~\ref{f:CMD_HUGS_UV-U_CV} and \ref{f:CMD_HUGS_U-B_CV} show the HUGS CMDs for the CVs, and Figs.~\ref{f:CMD_HUGS_UV-U_AB} and \ref{f:CMD_HUGS_U-B_AB} show the HUGS CMDs for the ABs and giants. The optical CMDs and the colour-colour diagram were proper-motion cleaned by rejecting stars that have total proper motions that exceed the central two-dimensional velocity dispersion by a factor of 3.5. The proper motions were computed as described in L17, using ACS/WFC imaging from two epochs spaced by 5.3 years. 

\begin{table*}
\revised{\caption{Optical Counterpart Summary}
\label{t:counterparts}
\resizebox{\textwidth}{!}{%
\begin{tabular}{lccccccp{108pt}}
\hline
Source$^a$ &
RA, Dec (J2000)$^b$ &
$\rerr~('')^c$ &
$r~(')^d$ &
Offset$^e$&
Type$^f$ &
PM$^g$ &
Notes \\
\hline
%
%
CX1    &  19:10:51.136  ~$-$59:59:11.91  &  0.29  &  0.18  &  0.02       &  CV        & c, c     &  \\
CX2    &  19:10:56.005  ~$-$59:59:37.34  &  0.29  &  0.73  &  0.02       &  CV        & c, c     &  \\
CX3    &  19:10:40.375  ~$-$59:58:41.45  &  0.30  &  1.52  &  0.10       &  AGN       & f, f     &  \\
CX4    &  19:10:51.586  ~$-$59:59:01.75  &  0.29  &  0.08  &  0.03       &  CV        & c, c     &  \\
CX5    &  19:10:51.413  ~$-$59:59:05.21  &  0.29  &  0.09  &  0.02       &  CV?       & c, c     &  slightly red, small \ha\ excess \\
CX6    &  19:10:51.502  ~$-$59:59:27.09  &  0.30  &  0.39  &  0.03       &  CV?       & n, n     &  very blue, small \ha\ excess \\
CX7    &  19:10:51.513  ~$-$59:58:56.88  &  0.30  &  0.15  &  0.15       &  CV        & c, c     &  \\
CX8    &  19:11:02.964  ~$-$59:59:42.15  &  0.32  &  1.49  &  0.29, 0.39 &  AB?       & n, n     &  resolved binary of nearly identical red, \ha-excess objects \\
CX9    &  19:10:51.772  ~$-$59:58:59.33  &  0.31  &  0.10  &  0.06       &  CV?       & c, c     &  alternative CV?? counterpart (offset: 0.37, PM: n) not detected by HUGS \\ 
CX10   &  19:10:54.760  ~$-$59:59:13.88  &  0.31  &  0.37  &  0.09       &  AGN       & f, f     &  \\
CX11   &  19:10:52.411  ~$-$59:59:05.54  &  0.33  &  0.04  &  \nd        &  MSP       & \nd      &  MSP D \\
CX13   &  19:10:40.624  ~$-$60:00:06.02  &  0.37  &  1.77  &  0.02       &  CV        & c, c     &  \\
CX14   &  19:10:52.074  ~$-$59:59:09.41  &  0.35  &  0.08  &  \nd        & \mspf{MSP} & \nd      &  MSP F \\
CX15   &  19:10:55.840  ~$-$59:57:45.70  &  0.34  &  1.39  &  0.04       &  AGN       & f, n     &  \\
CX16   &  19:10:42.542  ~$-$59:58:43.06  &  0.41  &  1.25  &  0.50       &  AB        & c, c     &  \\
CX17   &  19:11:05.273  ~$-$59:59:04.63  &  0.43  &  1.64  &  \nd        &  GLX       & \nd      &  asymmetric, extended object \\
CX18   &  19:10:52.043  ~$-$59:59:04.11  &  0.41  &  0.01  &  0.70       &  RG        & c, c     &  RG with apparent \ha\ excess; alternative CV?? counterpart (offset: 0.32, PM: n) only detected by HUGS has apparent \UV\ excess \\
CX19   &  19:10:55.595  ~$-$59:59:17.45  &  0.43  &  0.49  &  0.58       &  AB?       & c, c     &  normal \br, small \ha\ excess \\
CX20   &  19:10:52.852  ~$-$59:59:02.75  &  0.37  &  0.10  &  1.24       &  AB        & c, c     &  \\
CX21   &  19:10:49.528  ~$-$59:59:43.14  &  0.37  &  0.72  &  0.31       &  CV?       & c, c     &  blue, very slight \ha\ excess in colour-colour diagram \\
CX22   &  19:11:02.966  ~$-$59:57:58.71  &  0.37  &  1.74  &  \nd        &  \nd       & \nd      &  only MS stars present \\
CX23   &  19:10:52.540  ~$-$59:59:04.36  &  0.37  &  0.05  &  0.17       &  CV?       & c, c     &  uncertain photometry \\ 
CX24   &  19:10:52.691  ~$-$59:59:03.42  &  0.37  &  0.07  &  1.57       &  CV?       & n, n     &  weakly detected in \R\ and \ha \\ 
CX25   &  19:10:51.969  ~$-$59:58:40.60  &  0.38  &  0.40  &  1.62       &  CV?       & n, \nd   &  weakly detected in \R\ and \ha; not detected by HUGS \\ 
CX26   &  19:10:39.201  ~$-$59:59:45.46  &  0.37  &  1.75  &  0.02       &  GLX       & f, n     &  extended elliptical image \\
CX27   &  19:10:52.066  ~$-$59:59:00.92  &  0.43  &  0.06  &  \nd        &  MSP       & \nd      &  MSP B; alternative AB? counterpart (offset: 1.13, PM: c, c) has normal \br, small \ha\ excess \\
CX28   &  19:10:42.513  ~$-$59:59:44.52  &  0.50  &  1.38  &  \nd        &  \nd       & \nd      &  only MS stars present \\
CX29   &  19:10:52.290  ~$-$59:59:01.76  &  0.45  &  0.05  &  0.21       &  CV?       & f, n     &  alternative CV? counterpart (offset: 1.27, PM: c, c) \\
CX30   &  19:10:40.670  ~$-$59:58:39.56  &  0.55  &  1.49  &  \nd        &  \nd       & \nd      &  undetected faint object \\
CX31   &  19:10:50.503  ~$-$59:57:37.36  &  0.48  &  1.47  &  0.31       &  RG        & f, n     &  \\
CX32   &  19:10:54.121  ~$-$59:59:11.10  &  0.44  &  0.27  &  0.33       &  CV?       & n, n     &  slightly blue, slight \ha\ excess, small \uvu\ excess \\
CX33   &  19:11:03.311  ~$-$59:58:01.88  &  0.76  &  1.74  &  0.79       &  AB?       & c, \nd   &  not detected by HUGS \\
CX34   &  19:10:45.690  ~$-$59:58:20.06  &  0.53  &  1.09  &  0.52       &  AB        & c, c     &  \\
CX35   &  19:10:52.171  ~$-$59:59:16.88  &  0.39  &  0.21  &  0.65       &  CV        & n, n     &  \\
CX36   &  19:10:49.559  ~$-$59:58:26.50  &  0.41  &  0.71  &  0.74       &  CV        & n, n     &  \\
CX37   &  19:10:50.535  ~$-$59:59:09.11  &  0.43  &  0.21  &  1.13       &  AB?       & c, c     &  alternative AB? counterpart (offset: 0.49, PM: c, c) \\
CX38   &  19:11:02.136  ~$-$59:58:11.67  &  0.50  &  1.53  &  0.31       &  AB?       & c, c     &  slightly red, slight \ha\ excess \\
CX39   &  19:10:46.404  ~$-$59:57:49.84  &  0.52  &  1.43  &  0.91       &  GLX       & f, n     &  extended elliptical image \\
CX40   &  19:10:50.374  ~$-$59:59:14.10  &  0.64  &  0.27  &  1.29       &  AB        & c, c     &  alternative CV? counterpart (offset: 0.94, PM: n) only detected by HUGS \\
\hline
\end{tabular}
}
}
\end{table*}
\begin{table*}
\revised{
\contcaption{Optical Counterpart Summary}
\resizebox{\textwidth}{!}{%
\begin{tabular}{lccccccp{108pt}}\hline
Source$^a$ &
RA, Dec (J2000)$^b$ &
$\rerr~('')^c$ &
$r~(')^d$ &
Offset$^e$&
Type$^f$ &
PM$^g$ &
Notes \\
\hline
%
%
CX41   &  19:10:51.960  ~$-$59:59:08.49  &  0.34  &  0.07  &  0.85       &  CV?       & c, \nd   &  uncertain photometry; not detected by HUGS \\
CX42   &  19:10:48.124  ~$-$60:00:07.25  &  0.42  &  1.16  &  \nd        &  AGN?      & \nd      &  radio detection \\ 
CX43   &  19:10:59.323  ~$-$59:59:00.82  &  0.41  &  0.90  &  \nd        &  \nd       & \nd      &  only one MS star present \\
CX44   &  19:11:04.964  ~$-$59:59:30.82  &  0.46  &  1.66  &  0.43       &  AGN       & f, \nd   &  not detected by HUGS \\
CX45   &  19:10:38.940  ~$-$59:59:22.22  &  0.49  &  1.68  &  0.23       &  AGN       & f, \nd   &  not detected by HUGS \\
CX46   &  19:10:52.755  ~$-$59:57:39.00  &  0.42  &  1.43  &  \nd        &  \nd       & \nd      &  only one MS star present \\
CX47   &  19:10:54.075  ~$-$59:59:42.37  &  0.43  &  0.68  &  0.69       &  \nd       & n, n     &  faint star with hint of \UV\ excess \\
CX48   &  19:10:52.788  ~$-$59:58:58.39  &  0.48  &  0.13  &  0.28       &  RG        & c, c     &  uncertain photometry of bright RG; two alternative CV? counterparts (offset: 0.30, PM: n) and (offset:1.54, PM: n) only detected by HUGS \\                         
CX49   &  19:10:43.493  ~$-$59:59:24.64  &  0.54  &  1.13  &  0.42       &  CV?       & n, \nd   &  uncertain photometry in \ha; not detected by HUGS \\
CX50   &  19:10:58.360  ~$-$59:57:46.58  &  0.50  &  1.51  &  \nd  &  \nd  & \nd &  only one MS star present \\
CX51   &  19:10:38.235  ~$-$59:58:57.50  &  0.96  &  1.74  &  \nd  &  \nd  & \nd &  empty search area \\
CX52   &  19:10:45.677  ~$-$59:59:12.67  &  0.64  &  0.82  &  \nd  &  \nd  & \nd &  only one MS star present \\
\hline
\multicolumn{8}{l}{\makecell[tl]{$^a$Extension of \citet{Forestell14} numbering system. Sources CX41--CX52 are newly reported in this study.}}\\
\multicolumn{8}{l}{\makecell[tl]{$^b$\chandra\ source positions have been boresight corrected to align with the drizzled image coordinate system, \\ \hspace*{3pt} which is tied to the ICRS via UCAC3 astrometric standards.}}\\
\multicolumn{8}{l}{\makecell[tl]{$^c$95\% confidence X-ray error circle radius, \rerr, based on the prescription of \citet{Hong05}.}}\\
\multicolumn{8}{l}{\makecell[tl]{$^d$Projected distance from cluster centre in arcmin.}}\\
\multicolumn{8}{l}{\makecell[tl]{$^e$Offset of primary counterpart from X-ray position in units of \rerr.}}\\
\multicolumn{8}{l}{\makecell[tl]{$^f$Type of primary counterpart (types of alternative counterparts are given in Notes column): CV = cataclysmic variable; \\ \hspace*{3pt} AB = active binary; RG = red giant; GLX = galaxy; AGN = active galactic nucleus; ? indicates less certain classification.}}\\
\multicolumn{8}{l}{\makecell[tl]{$^g$Proper-motion membership determinations from the optical study and the HUGS study: c = consistent with  cluster; \\ \hspace*{2pt} f = consistent with field; n = no proper-motion measurement.}}\\
\end{tabular}
}
}\end{table*}

\begin{table}
\caption{\textbf{Optical Counterpart Photometry}\label{t:photometry}}
\begin{tabular}{lcccccc}
\hline
Source & 
\B & 
\R & 
\ha &
\UV & 
\U  &
\B  \\
\hline
%
%
CX1    &  19.46  &  19.38  &  19.01  &  19.61  &  18.15  &  19.58  \\
CX2    &  20.36  &  19.22  &  18.60  &  20.65  &  18.95  &  20.38  \\
CX3    &  21.76  &  20.97  &  20.72  &  \nd    &  \nd    &  22.01  \\
CX4    &  21.09  &  20.10  &  19.35  &  20.40  &  20.24  &  21.14  \\
CX5    &  19.65  &  18.57  &  18.33  &  20.53  &  19.55  &  19.71  \\
CX6    &  24.07  &  23.87  &  23.78  &  22.51  &  23.00  &  24.01  \\
CX7    &  22.01  &  20.94  &  19.91  &  20.56  &  20.52  &  21.91  \\
CX8a   &  20.91  &  18.34  &  17.54  &  23.34  &  23.62  &  21.07  \\
CX8b   &  21.06  &  18.48  &  17.67  &  23.65  &  23.05  &  21.18  \\
CX9a   &  24.40  &  22.42  &  22.01  &  24.50  &  25.32  &  24.56  \\ 
CX9b   &  \nd    &  24.53  &  23.73  &  \nd    &  \nd    &  \nd    \\
CX10   &  20.16  &  19.70  &  19.47  &  20.02  &  \nd    &  20.37  \\
CX11   &  \nd    &  \nd    &  \nd    &  \nd    &  \nd    &  \nd    \\
CX13   &  24.60  &  24.32  &  23.38  &  \nd    &  \nd    &  \nd    \\
\mspf{CX14}&  \nd    &  \nd    &  \nd    &  \nd    &   \nd   &  \nd    \\
CX15   &  23.05  &  22.66  &  22.41  &  22.91  &  22.94  &  23.29  \\
CX16   &  19.39  &  18.33  &  18.09  &  \nd    &  19.31  &  19.46  \\
CX17   &  \nd    &  \nd    &  \nd    &  \nd    &  \nd    &  \nd    \\
CX18a  &  16.49  &  15.42  &  14.94  &  17.60  &  16.48  &  16.52  \\
CX18b  &  \nd    &  \nd    &  \nd    &  24.30  &  22.43  &  22.23  \\
CX19   &  18.35  &  17.58  &  17.37  &  18.74  &  17.99  &  18.45  \\
CX20   &  21.99  &  20.20  &  19.87  &  25.00  &  22.70  &  22.06  \\
CX21   &  19.06  &  18.53  &  18.40  &  19.15  &  18.75  &  19.10  \\
CX22   &  \nd    &  \nd    &  \nd    &  \nd    &  \nd    &  \nd    \\
CX23   &  21.44  &  20.47  &  19.75  &  25.78  &  21.83  &  21.53  \\
CX24   &  22.97  &  22.66  &  22.61  &  22.54  &  22.47  &  23.27  \\
CX25   &  25.73  &  25.51  &  24.81  &  \nd    &  \nd    &  \nd    \\
CX26   &  25.16  &  23.00  &  22.55  &  \nd    &  \nd    &  25.44  \\
CX27   &  21.65  &  20.13  &  19.83  &  23.53  &  21.95  &  21.50  \\
CX28   &  \nd    &  \nd    &  \nd    &  \nd    &  \nd    &  \nd    \\
CX29a  &  22.49  &  20.83  &  20.50  &  23.67  &  22.20  &  21.85  \\
CX29b  &  19.87  &  19.02  &  18.63  &  20.78  &  19.56  &  19.88  \\
CX30   &  \nd    &  \nd    &  \nd    &  \nd    &  \nd    &  \nd    \\
CX31   &  16.76  &  15.66  &  15.44  &  19.03  &  \nd    &  17.18  \\
CX32   &  22.10  &  20.54  &  20.23  &  24.85  &  23.29  &  22.50  \\
CX33   &  \nd    &  \nd    &  \nd    &  \nd    &  \nd    &  \nd    \\
CX34   &  22.04  &  20.23  &  19.88  &  \nd    &  23.00  &  22.13  \\
CX35   &  24.89  &  24.11  &  23.35  &  23.33  &  24.00  &  25.28  \\
CX36   &  24.93  &  24.85  &  23.81  &  23.15  &  23.70  &  25.00  \\
CX37a  &  22.35  &  20.42  &  20.12  &  25.67  &  23.28  &  22.33  \\
CX37b  &  18.85  &  17.94  &  17.73  &  19.39  &  18.72  &  18.90  \\
CX38   &  20.34  &  19.15  &  18.93  &  21.78  &  20.37  &  20.40  \\
CX39   &  24.86  &  22.26  &  21.89  &  \nd    &  \nd    &  25.36  \\
CX40a  &  21.18  &  19.72  &  19.45  &  24.02  &  21.66  &  21.27  \\
CX40b  &  \nd    &  \nd    &  \nd    &  22.11  &  22.31  &  23.51  \\
CX41   &  19.36  &  18.56  &  17.91  &  \nd    &  \nd    &  \nd    \\
CX42   &  25.82  &  23.38  &  22.96  &  \nd    &  25.80  &  25.91  \\
CX43   &  \nd    &  \nd    &  \nd    &  \nd    &  \nd    &  \nd    \\
CX44   &  26.98  &  24.91  &  24.89  &  \nd    &  \nd    &  \nd    \\
CX45   &  24.72  &  22.82  &  22.71  &  \nd    &  \nd    &  \nd    \\
CX46   &  \nd    &  \nd    &  \nd    &  \nd    &  \nd    &  \nd    \\
CX47   &  25.69  &  23.19  &  22.69  &  25.94  &  26.28  &  25.61  \\
CX48a  &  13.27  &  11.97  &  11.78  &  16.44  &  14.13  &  12.99  \\
CX48b  &  \nd    &  \nd    &  \nd    &  23.12  &  22.68  &  20.86  \\
CX48c  &  \nd    &  \nd    &  \nd    &  23.18  &  22.59  &  23.00  \\
CX49   &  25.62  &  25.63  &  25.38  &  \nd    &  \nd    &  \nd    \\
CX50   &  \nd    &  \nd    &  \nd    &  \nd    &  \nd    &  \nd    \\
CX51   &  \nd    &  \nd    &  \nd    &  \nd    &  \nd    &  \nd    \\
CX52   &  \nd    &  \nd    &  \nd    &  \nd    &  \nd    &  \nd    \\
\hline
\end{tabular}
\end{table}

\subsection{Classifications and refinements among sources 1-40}
\label{sec_classification_1_40}
In the following paragraphs, we discuss the classifications of the sources that result from the combination of the optical photometry with the HUGS photometry. 
In some cases, we refer to the X-ray and/or radio results to aid in the classification process. The results of this analysis are summarised in Table~\ref{t:counterparts} and the photometry of the proposed counterparts is listed in Table~\ref{t:photometry}. \revised{In cases where alternative counterparts are present, the properties of the primary counterpart are listed in Table~\ref{t:counterparts} in the columns labelled ``Offset,'' ``Type,'' and ``PM.'' These three properties are given for the alternative counterparts in the column labelled ``Notes.''}

\subsubsection{CX1-CX4}

It can be seen in Fig.~\ref{f:CMD_CV} that the CX1 counterpart is quite blue in the (\br, \R) CMD, while the CX2 counterpart is on the MS\@. L17 classified both of these as likely CVs, based on their \ha\ excesses and high X-ray to optical flux ratios. Interestingly, both of these counterparts lie 
to the \emph{red} side of the MS in the (\uvu, \UV) CMD (Fig.~\ref{f:CMD_HUGS_UV-U_CV}), but well to the blue of the MS in the (\ub, \U) CMD (Fig.~\ref{f:CMD_HUGS_U-B_CV}). \craig{The 
\revised{large discrepancies in these}
colours \revised{are likely} 
due to variability,} as the \UV\ and \U\ imaging were done approximately one year apart. As noted in L17, the CX1 and CX2 counterparts both show substantial variability on a timescale of months, with a peak-to-trough amplitude of over one magnitude for CX1\@. \revised{Such variations in the NUV are common among CVs \citep[e.g.][]{Dieball17,RiveraSandoval18}, while we are not aware of plausible alternative explanations for such unusual near-UV spectra.} \revised{We note that \citet{Thomson12} found that CX1 is a likely dwarf nova.}
We also note that \citet{Gottgens19} have used MUSE integral-field spectrograph observations to confirm the CV identification of the optical counterpart to CX2, by the detection of strong \ha, H$\beta$, and \ion{He}{I} emission lines.  

The CX3 counterpart, which L17 classified as an active galactic nucleus (AGN), on the basis of its blue colour and proper motion consistency with the external galaxy frame, is outside of the \UV\ and \U\ fields in the HUGS imaging, and thus the HUGS UV photometry does not provide any additional information. Its HUGS \B\ magnitude is 0.25 mag fainter than its GO-12254 \B\ magnitude. Comparison of the HUGS \B\ magnitudes to ours indicates that the former are $0.09 \pm 0.20$ mag (std.\ dev.) fainter. The median magnitude offset is 0.08 mag and the semi-interquartile range is 0.07 mag. A magnitude difference as large as 0.25 mag might indicate some photometric variability, rather than simply measurement uncertainty, since the HUGS dataset does not include all of the GO-12254 data (see footnote to Table \ref{t:hst_obs}). \revised{The lack of a radio detection of a relatively X-ray bright AGN is mildly unusual, but the radio/X-ray flux ratio is not outside the range observed for AGN by \citet{Maccarone12}.}

L17 classified the CX4 counterpart as a likely CV, based on its blue \br\ colour and its \ha\ excess. This classification is supported by the UV excesses of this object in both HUGS colour indices, \uvu\ and \ub\ (Figs.~\ref{f:CMD_HUGS_UV-U_CV} and \ref{f:CMD_HUGS_U-B_CV}). \revised{As noted by L17, \citet{Kaluzny09} showed that the $B$-band light curve of this source resembles those of ordinary dwarf novae, which implies that this system is likely a dwarf nova.}

\subsubsection{CX5}
L17 noted that the CX5 counterpart 
falls slightly to the red side of the MS in the (\br, \R) CMD, rather than falling to the blue side, as expected for a CV\@. Nevertheless, L17 classified it as a likely CV, based on a small \ha\ excess seen in the optical CMD (Fig.~\ref{f:CMD_CV}) and in the optical colour-colour diagram (Fig.~\ref{f:CCD_CV}), and a moderately high X-ray to optical flux ratio (see their Fig.~10). The HUGS CMDs (Figs.~\ref{f:CMD_HUGS_UV-U_CV} and \ref{f:CMD_HUGS_U-B_CV}) indicate that the CX5 counterpart falls on the blue edge of the MS in the (\uvu, \UV) CMD and on the red edge of the MS in the (\ub, \U) CMD\@.

\begin{figure}
    \centering
    \includegraphics[scale=0.5]{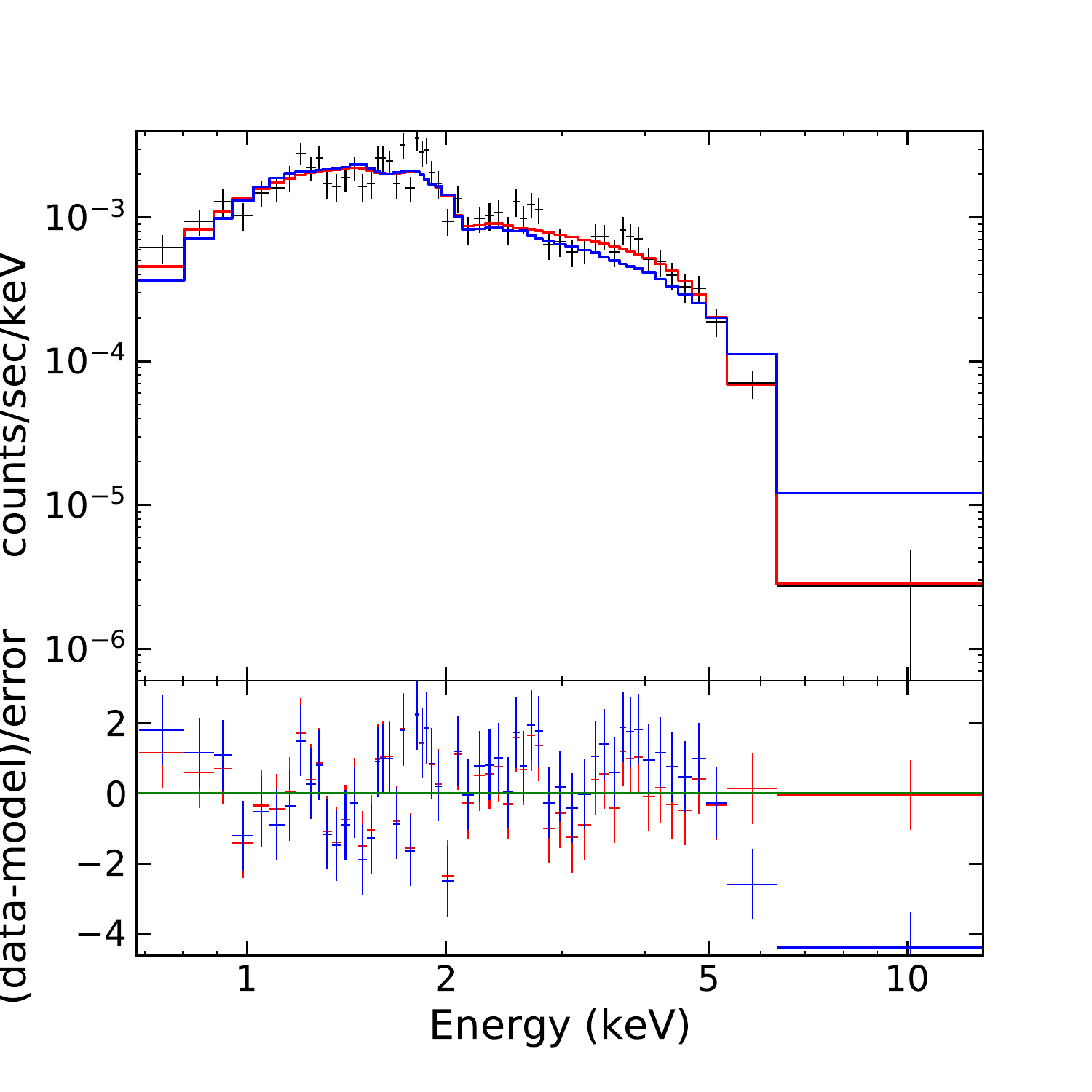}
    \caption{Cycle 18 {\it Chandra} spectrum of CX5, fit with an absorbed {\tt mekal} model (blue), and with the same {\tt mekal} model but modified by a Gaussian absorption line ({\tt gabs}) at $7.2^{+8.8}_{-0.7}~\mathrm{keV}$ (red), as described in \S\ref{s:classification}. The lower panel shows the corresponding residuals colour-coded according to the models in the top panel. Without the {\tt gabs} component, the {\tt mekal} model over-predicts spectra values above $\approx 5~\mathrm{keV}$.}
    \label{f:cx5_spec_cycle17}
\end{figure}

\cory{The Cycle 18 X-ray spectrum of CX5 
is poorly fit by 
a single {\tt mekal} model  ($\chi^2/\mathrm{dof}=105.42/53=1.99$). 
The X-ray spectrum declines 
above $\sim 5\,\kev$, suggesting fitting the continuum with a cut-off power-law model ({\tt cutoffpl} in {\sc xspec}). With one less degree of freedom, this gives a slightly better fit with $\chi^2/\mathrm{dof}=65.36/52=1.26$, with a cutoff energy at $1.5^{+0.4}_{-0.3}~\mathrm{keV}$. The high-energy cutoff can also be fit by convolving the {\tt mekal} model with 
a Gaussian absorption line component ({\tt gabs}), which yields a better fit ($\chi^2/\mathrm{dof}=60.69/50=1.21$), with a higher plasma temperature ($kT>21.5\,\kev$) and an absorption line at $7.2^{+8.8}_{-0.7}\,\kev$. 
In Fig. \ref{f:cx5_spec_cycle17}, we show the Cycle 18 spectrum of CX5 fitted with the {\tt gabs}-modified {\tt mekal} model. \coryrv{CX5 shows marked X-ray variability  (\S\ref{sec:intra_observational_variability}) on time scales of order $ 0.2~\mathrm{day}$ (Fig. \ref{f:all_lc}). We used the {\tt pdm} function from the {\sc pwkit} Python package \citep{Williams17} to perform the phase dispersion minimisation algorithm (PDM; \citealt{Stellingwerf78}) on the Cycle 18 light curves, but found no periodic features.} 
}

\craig{CX5's collection of strange features---a complete lack of blue colours even in ultraviolet CMDs, a small H$\alpha$ excess, an unusual X-ray spectrum, and rapid X-ray variability---are difficult to explain as a CV.  
The high X-ray/optical flux ratio argues for emission from a compact object, while the lack of an UV excess (and only weak H$\alpha$ emission) argues against a substantial accretion disc. \revised{A magnetic CV, without an accretion disc, might seem plausible, but in such systems the hot white dwarf produces a bright UV flux; see the known polar X10/W27 in 47 Tuc, which shows a 1-magnitude blue excess in \br, a 0.5 magnitude excess in \hr, and a 4-magnitude excess in a $U_{300}$ vs.\ $B_{390}$ CMD \citep{RiveraSandoval18}.}
The high X-ray luminosity
\revised{and low mass transfer rate (indicated by the lack of a UV excess)}
argues for a very deep gravitational potential well, and thus a neutron star or black hole accretor. Alternatively, ``redback'' millisecond pulsars (with nondegenerate, mass-losing companions of $>$0.1 \msun) can also match all the data; several candidate redbacks have recently been identified in other globular clusters without the detection of radio pulsations  \citep{Zhao20a,Urquhart20}.  Our ATCA radio observations allow a (3$\sigma$) 5 GHz radio luminosity limit of $1.1\times10^{27}$ erg s$^{-1}$, which along with the 1-10 keV $L_X$ of $8.7\times10^{31}$ erg s$^{-1}$, are not deep enough to rule out a black hole \citep[e.g.][]{Rodriguez20},  a transitional millisecond pulsar \citep{Bogdanov18}, or a redback millisecond pulsar \citep[such as NGC 6397 A,][]{Zhao20a}. 
A spectroscopic radial velocity campaign, 
\revised{which would require}
 adaptive optics (due to crowding), \revised{e.g. MUSE in narrow-field mode}, could shed light on this bizarre system. }


\subsubsection{CX6-CX7}
The CX6 counterpart shows a strong UV excess in the HUGS CMDs, putting it on the white dwarf (WD) cooling sequence. L17 had previously classified it as a candidate CV, based on its blue colours and \craig{relative \ha\ excess compared to other WDs, which exhibit broad \ha\ absorption lines} (Fig.~\ref{f:CCD_CV}). \craig{The lack of \ha\ emission or absorption could also signal an AM CVn system, \revised{which might be verified through narrow-band He line photometry, or spectroscopy}.} 

The CX7 counterpart shows a strong UV excess in the HUGS CMDs, which, along with its previously noted blue \br\ colour and large \ha\ excess, make it a likely CV\@.
\revised{\citet{Bailyn96} suggested this to be a CV (and found a likely 3.7 hour period), and \citet{Thomson12} observed a 6-magnitude near-UV outburst, confirming it to be a dwarf nova.}

\subsubsection{CX8}
The striking pair of CX8 counterparts (see Fig.~\ref{f:finding_charts_7-13}), which L17 interpreted as a resolved pair of active binaries (ABs) \craig{at a distance of 500-800 pc,} shows unusual behaviour in the HUGS photometry. While the two objects fall far to the blue of the MS in the (\uvu, \UV) CMD, they nevertheless fall far to the red of the MS in the (\ub, \U) CMD\@. 
The CX8 counterparts show some variability, but only at the level of a few hundredths of a magnitude over timescales of months (L17). 
\craig{The UV emission from M dwarfs such as these is entirely chromospheric in nature, for which such colours are not unusual \citep{Stelzer13,Heinke20}.}

\cory{
Neither single {\tt mekal} ($\chi^2/\mathrm{dof}=31.45/13=2.42$) nor  power-law model  ($\chi^2/\mathrm{dof}=35.05/13=2.70$) fits to CX8 give good fits. The {\tt mekal} fit shows an apparent excess $\lesssim 1\,\kev$, requiring a second component. We thus added another {\tt mekal} component and found that with two fewer degrees of freedom, the fit  improves to $\chi^2/\mathrm{dof}=19.27/11=1.75$, suggesting low-temperature ($0.2^{+0.3}_{-0.1}~\kev$) and high-temperature ($1.2^{+0.5}_{-0.2}~\kev$) components (Table \ref{t:spectral_analyses_51}). The two-component {\tt mekal} model also fits the Cycle 1 and Cycle 7 spectra (\S\ref{s:inter-variability}, Fig. \ref{f:x_ray_spectral_variability}),} \craig{as found by L17, and is consistent, in temperatures and luminosity, with typical X-ray spectra of nearby ABs \citep{Dempsey97}.} 


\subsubsection{CX9-CX13}
We find two alternative counterparts to CX9, both within the error circle. L17's suggested counterpart, 9a, is a faint star that shows a small \br\ excess and a small \hr\ excess in the colour-colour diagram. It shows a very strong UV excess in \uvu\ and a small excess in \ub\@. Counterpart 9b is a faint extension of the image of counterpart 9a, with measured magnitudes in \R\ and \ha\ that yield a small \hr\ excess. (See Fig.~\ref{f:finding_charts_7-13}.) KS2 did not report a \B\ magnitude for counterpart 9b and we did not attempt to determine one using PSF-fitting photometry, due to the complexity of the image. The HUGS photometry database does not detect the 9b counterpart. Given the strong UV excess of counterpart 9a, we suggest that this is the more likely identification of CX9, \craig{and that it is} 
a possible \craig{CV.}

The CX10 counterpart, which L17 classified as an AGN, on the basis of its blue \br\ colour and its proper motion consistency with the external galaxy frame (see Fig.~5 of L17), is in the \UV\ field of the HUGS imaging but outside of the \U\ field. It shows a strong \uvb\ excess. Its HUGS \B\ magnitude is 0.21 mag fainter than the GO-12254 \B\ magnitude. As for the case of CX3, this difference may either indicate some photometric variability or else photometric uncertainty. \coryrv{If the $3\,\sigma$ radio detection here is genuine (\S\ref{s:radio_observation}), the AGN nature could be further corroborated.} 

As L17 noted, only MS stars lie in the error circle of CX11\@. As noted by \citet{Forestell14}, source CX11 is coincident with the isolated MSP D, and thus is not  expected to have an optical/UV counterpart. There is a faint, blue star with a \hr\ deficit located 0\farcs71 from the centre of the error circle, 
 or 2.1 times the error circle radius.
Based on this large offset and its location in the CMDs and the colour-colour diagram, we suggest that this object is a He-core WD that is coincidentally near the error circle. 

As noted by L17, source CX12 from \citet{Pooley02} has been replaced by sources CX20, CX23, and CX24. 

The counterpart to CX13 that L17 proposed lies near the WD cooling sequence in the (\br, \R) CMD and shows a large \hr\ excess, relative to both the MS and the WD sequence, making it a strong candidate as a CV\@. This object is outside of the HUGS \UV\ and \U\ frames. 


\subsubsection{CX14}
\mspf{A recent {\it MeerKAT} survey of globular clusters discovered a new isolated MSP, PSR J1910$-$5959F (MSP F), in the core \citep{Ridolfi21}, and the timing solution identify MSP F with CX14. Because MSP F is isolated, the MS stars found in the error circle of CX14 (L17) and a blue star seen in the HUGS images are not the actual counterparts. We tried both power-law and blackbody-like fits to CX14's X-ray spectrum, and noted that the power-law model fits better to the X-ray spectrum, which gives a photon index $\Gamma=2.2^{+0.6}_{-0.5}$. The hardness is consistent (at 90\% confidence) with 3 other isolated MSPs (B, C, E) in NGC 6752, despite the relatively large uncertainties.} 

\subsubsection{CX15-CX16}
As in the cases of CX3 and CX10, L17 classified CX15 as an AGN, on the basis of its blue colour and proper motion consistency with the extragalactic frame. Unlike CX3 and CX10, CX15 is present in all three HUGS fields of view. It has a large UV excess in both HUGS CMDs (Figs.~\ref{f:CMD_HUGS_UV-U_CV} and \ref{f:CMD_HUGS_U-B_CV}). 

L17 classified the CX16 counterpart as an AB, based on its red \br\ colour and its \ha\ excess. It is outside of the HUGS \UV\ field of view. Its \ub\ color puts it a small distance redward of the MS in the (\ub, \U) CMD, consistent with the optical results.  

\subsubsection{CX17}
As L17 noted, we do not find a starlike object in the error circle for CX17\@. Instead, there is an an amorphous object, suggesting an interacting galaxy as seen in Fig.~\ref{f:finding_charts_14-19}. As \citet{Pooley02} have noted, this object is coincident with a radio source, detected by an ATCA observation \citep{Frater92}. 
\cory{Our ATCA observation also reveals a radio source that lies $0\farcs3$ from the nominal X-ray position of CX17 (Fig. \ref{f:radio_xray_matched_finders}) with a moderately steep spectral index ($\alpha = -0.7\pm 0.3$; Table \ref{t:radio_matches}) typical of AGNs.}

\subsubsection{CX18-CX21}
L17 noted that the error circle of CX18 contains only MS stars. With a shift in the centre of the error circle, it now contains a red giant with an apparent \ha\ excess. While the sequences in the (\hr, \R) CMD are less well defined at $\R \sim 15.5$ than at somewhat fainter magnitudes, the apparent \ha\ excess of this giant does stand out (see Fig.~\ref{f:CMD_RG}). A possible interpretation is that CX18 is an RS CVn binary with an active chromosphere. Examination of the HUGS CMDs indicates that this object lies on the red giant sequence in both diagrams, (see Figs.~\ref{f:CMD_HUGS_UV-U_AB} and \ref{f:CMD_HUGS_U-B_AB}). In addition, the HUGS photometry detects an additional faint object with an apparent \UV\ excess, that is not detected in the optical photometry. We include this object in the HUGS CMDs for CVs, although its classification as a possible CV is quite tenuous, given its rather weak detection. 

As L17 noted, the CX19 counterpart shows a small \hr\ excess, but falls on the MS in the (\br, \R) CMD\@. It falls on the red edge of the MS in the (\uvu, \UV) CMD (Fig.~\ref{f:CMD_HUGS_UV-U_AB}) and on the blue edge of the MS in in the (\ub, \U) CMD (Fig.~\ref{f:CMD_HUGS_U-B_AB}). Thus, the HUGS CMDs provide mixed evidence and we continue to classify the CX19 counterpart as a possible AB rather than as a likely AB\@. We note that \citet{Gottgens19} have used MUSE integral-field spectrograph observations to detect the optical counterpart to CX19 as an object with filled \ha\ absorption, which is consistent with our photometric finding of a small \hr\ excess. Thus, their study supports its classification as a possible AB\@. 

L17 classified the counterpart to CX20 as a likely AB, based on its red \br\ colour and its \hr\ excess. In the HUGS CMDs (Figs.~\ref{f:CMD_HUGS_UV-U_AB} and \ref{f:CMD_HUGS_U-B_AB}), it lies on the fairly broad lower part of the main sequence. Thus, it is not possible to discern a possible small deviation from the MS in these cases, using the HUGS photometry. 

L17 classified the counterpart to CX21 as a possible CV, based on its blue \br\ colour and its very slight \hr\ excess in the colour-colour diagram (see Fig.~\ref{f:CCD_CV}). While it falls on the MS in the (\ub, \U) CMD, it falls to the blue side of the MS in the (\uvu, \UV) CMD\@. \craig{Its proper motion also marks it as a cluster member.} Thus, the HUGS data support the classification of this object as a possible CV\@.

\subsubsection{CX22}
The optical photometry for the CX22 error circle indicates the presence of one MS star. The circle is outside the \UV\ and \U\ fields of view. The HUGS database indicates the detection in \B\ of a second, faint star in the error circle. However, in the absence of other optical or UV photometry for this second star, there is no information on its colour, and thus no reason to believe it is the counterpart.

\cory{The X-ray spectrum of CX22 can be reproduced by a hard ($\Gamma=0.7^{+0.5}_{-0.6}$) power-law model when $N_\mathrm{H}$ is fixed to the cluster value; a {\tt mekal} fit to CX22 with the same fixed $N_\mathrm{H}$ gives a worse fit (goodness fraction of $98.3\%$); however, if we apply a free $N_\mathrm{H}$, the fit will be more acceptable (goodness$=25.4\%$), yet it suggests an $N_\mathrm{H}=4.8^{+3.6}_{-3.2}\times10^{22}~\mathrm{cm^{-2}}$, which is above the cluster value. Either the power-law or the free-$N_\mathrm{H}$ {\tt mekal} fit points to a hard X-ray spectrum, which is unlikely from an AB. It is possible that CX22 is an AGN,} but in the absence of more compelling evidence we leave this source unclassified.

\subsubsection{CX23-CX27\label{sec_cx23-cx27}}
While the optical photometry for the CX23 counterpart indicates clear excesses in \br\ and \hr, L17 classified this star only as a possible CV due to photometric uncertainty owing to its location near several much brighter stars (see Fig.~\ref{f:finding_charts_20-25}), which necessitated aperture photometry in place of KS2 photometry. The HUGS photometry places the CX23 counterpart to the \emph{red} of the MS in the (\uvu, \UV) CMD, while it is slightly to the blue of the MS in the (\ub, \U) CMD\@. Evidently, photometry of this object is also challenging at UV wavelengths, and we leave it classified as a possible CV.

Our candidate counterpart to CX24 is a very blue object in \br\ that lies at 1.6\,\rerr\ from the centre of the error circle. It shows a small \ha\ deficit in the colour-colour diagram (Fig.\ref{f:CCD_CV}). 
L17 classified it as a possible CV given its location between the WD sequence and the MS in the (\br, \R) CMD\@. The HUGS CMDs show it consistently slightly redward of the CO WD sequence (Figs.~\ref{f:CMD_HUGS_UV-U_CV} and \ref{f:CMD_HUGS_U-B_CV}). 
\craig{We suggest four possible interpretations: this may be a normal He-core WD coincidentally located near the CX24 error circle in the crowded core (see Fig.~\ref{f:finding_charts_20-25}); it could be a CV composed of a hot WD and a faint, undetected M star, with little accretion; it could be an AM CVn double-degenerate, explaining the lack of \ha\ emission; or it could be another millisecond pulsar (with radio pulsations not yet detected) with a He-core WD companion.}

Our candidate counterpart to CX25 lies at 1.6\,\rerr\ from the centre of the error circle. As in the case of CX24, it is weakly detected in \R\ and \ha. Nonetheless, it \craig{lies on the WD sequence in the (\br, \R) CMD} and registers as having a large \hr\ excess 
for its blue colour (see Fig.~\ref{f:CCD_CV}). L17 thus classified it as a possible CV\@. This object does not appear in the HUGS database. Examination of the stacked \UV\ image suggests that the object is detected there and is thus quite blue in UV colours, consistent with its location on the WD sequence in the (\br, \R) CMD\@. 

The CX26 counterpart has an extended, galaxian appearance in the optical imaging, and L17 identified it as a galaxy based on its morphology and proper motion, which is discordant from the cluster motion. It lies outside of the \UV\ and \U\ fields of view. 

CX27 is positionally coincident with the isolated MSP NGC~6752~B and has a location in the X-ray CMD (Fig.~\ref{f:x_ray_cmd}) that is consistent with this association. L17 identified a possible counterpart to CX27 that lies on the MS in the (\br, \R) CMD and has a small \hr\ excess. L17 classified it as a possible AB, based on the latter. This star falls just outside of the error circle at 1.1\,\rerr. In the HUGS photometry, the candidate AB falls on the MS in the (\ub, \U) CMD and modestly blueward of the MS in the (\uvu, \UV) CMD\@. \coryrv{The ATCA radio counterpart is detected at $5.5~\mathrm{GHz}$ but not at  $9.9~\mathrm{GHz}$, consistent with a steep radio spectrum as observed in other MSPs \citep{Bates13}. The candidate optical counterpart is outside the radio error ellipse (Fig. \ref{f:radio_xray_matched_finders}), confirming it is not associated with the MSP. 
CX27 could be a mixture of X-rays from this possible AB and MSP NGC~6752~B, but given the superior agreement of the X-ray position with the radio position, most or all the X-rays are likely due to MSP B.}

\subsubsection{CX28}
The error circle of CX28 is empty; there are four MS stars in a region extending out to 2.5\,\rerr. Thus, there is no hint of a plausible optical/UV ID for CX28. 

\begin{figure}
    \centering
    \includegraphics[scale=0.5]{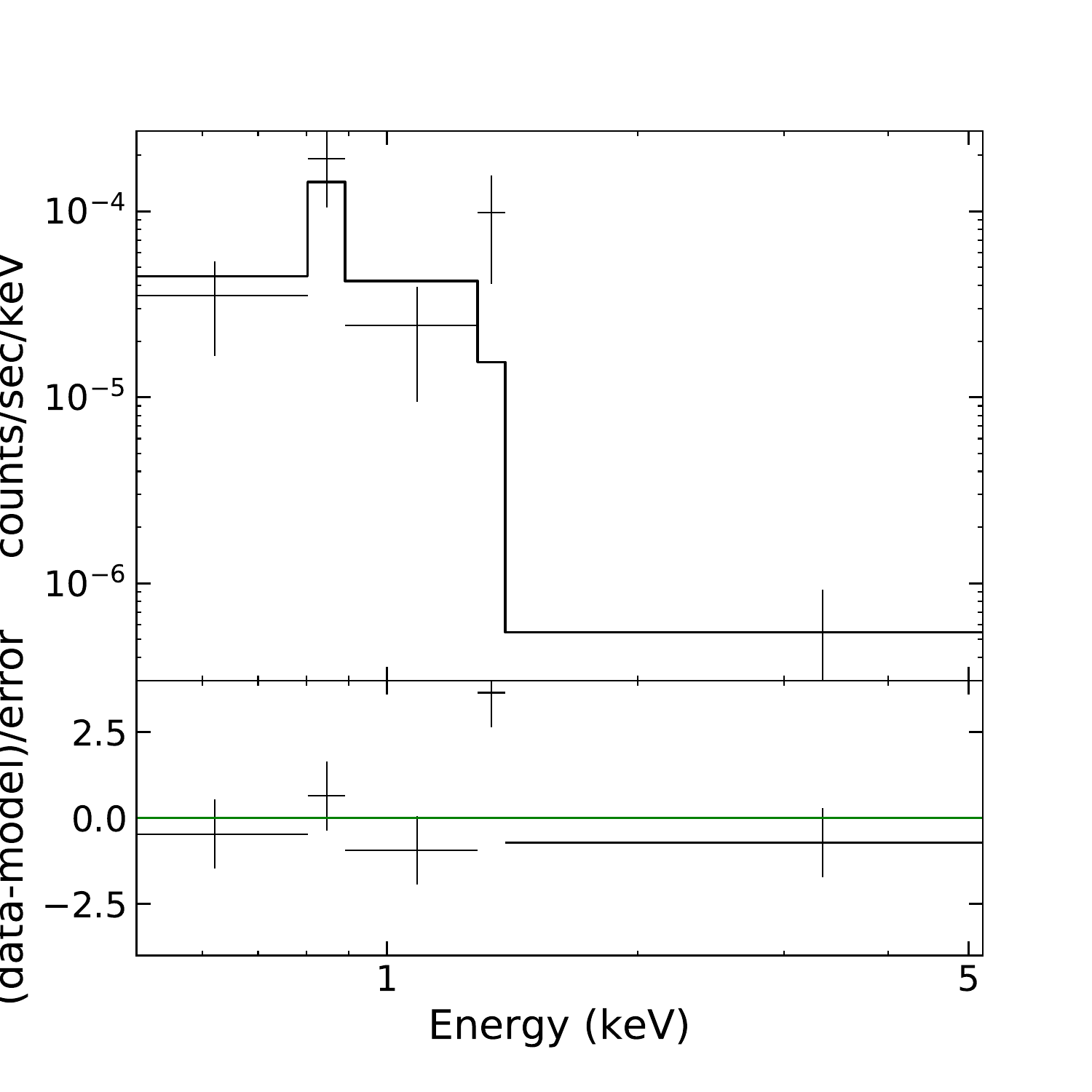}
    \caption{Combined ACIS-S spectrum (from all 3 cycles) of CX28. The black solid line represents the best-fitting {\tt mekal} model. The spectrum is rebinned only for plotting purpose.}
    \label{fig:cx28_aciss_spec}
\end{figure}

\cory{CX28 is a soft source (Fig. \ref{f:x_ray_cmd}), which can be reproduced by either a power-law ($\Gamma = 4.9^{+2.5}_{-1.9}$) or a {\tt mekal} model ($kT = 0.6^{+0.3}_{-0.4}~\kev$). The Cycle 18 X-ray spectrum shows some evidence of an emission feature at $\sim 0.8~\kev$ (Fig. \ref{fig:cx28_aciss_spec}), reminiscent of the Fe L-shell line observed in some faint CVs or ABs \citep{Heinke05}. We also find some very mild evidence of an emission feature at $\sim 1.3~\kev$, of almost the same strength as the one at $0.8~\kev$. The line energy overlaps the K$\alpha$ emission of Mg but cannot be reproduced by the model at the best-fitting plasma temperature and solar abundances. This line is a bit more prominent if we combine ACIS-S spectra from all 3 Cycles (Fig. \ref{fig:cx28_aciss_spec}); a {\tt mekal} fit to this combined spectrum is still acceptable (goodness$=46.5\%$) with a similar $kT = 0.5^{+0.4}_{-0.2}~\kev$.} 

\cory{Considering the position of CX28 being far off the cluster centre ($=1.37\arcmin=0.72r_\mathrm{h}$) and the empty error circle, 
CX28 is a 
candidate for a 
distant AGN, in which case the likely line at $1.3~\kev$ could then be a red-shifted feature. Given its very soft X-ray spectrum and the emission feature at $0.8~\kev$, it is also possible that CX28 is a very faint BY Draconis system (ABs composed of two main sequence stars) with optical counterparts fainter than the limiting magnitude of the given observations. We thus consider CX28 to be unclassified.}


\subsubsection{CX29-CX40}
L17 identified CX29 with a slight extension of the image of a much brighter neighbour. The photometry of this extension indicated that it was a possible CV\@. There is an alternative counterpart that registers as a CV in the optical photometry, which is more cleanly detected. It is located at 1.3\,\rerr, i.e.\ somewhat outside of the new error circle. While this new possible counterpart, 29b, also has a brighter neighbour, it is much better separated from its neighbour than the original counterpart, 29a (see Fig.~\ref{f:finding_charts_26-31}). L17 noted that 29a had a discordant proper motion, relative to the cluster distribution, possibly due to measurement errors. In the HUGS database, the membership probability of 29a, based on proper motion, is reported as not available. The proper motion of 29b is consistent with that of the cluster. In the HUGS photometry, 29a has a small blue excess in both CMDs (Figs.~\ref{f:CMD_HUGS_UV-U_CV} and \ref{f:CMD_HUGS_U-B_CV}). In contrast, 29b lies on the MS in these HUGS CMDs. Thus, it remains unclear which of these two alternative counterparts, if either, is the more likely identification of CX29. 

CX30 lies outside of the \UV\ and \U\ fields of view. The only object within the error circle for CX30 is a faint star that is in the wings of a far brighter MS turnoff star that lies just outside of the error circle. KS2 did not detect the faint star in either the optical photometry or the HUGS reductions. We did not attempt to extract magnitudes for this barely visible faint star with PSF fitting.

L17 noted that there is an apparent red giant in the search area for CX31\@. This object, which remains within the error circle, is slightly red relative to the giant branch in \br, but does not show an \ha\ excess. As noted in Table~\ref{t:counterparts}, it has a proper motion that is not consistent with the cluster mean, indicating that it is a foreground object that is projected on the cluster. CX31 lies outside of the \U\ field of view. While the HUGS photometry resolves the apparent single optical object image into two separate objects, 
this is evidently an artefact of observing a large-proper-motion object at multiple epochs (see Fig.~\ref{f:finding_charts_26-31}). We have plotted the photometry for only one of these apparent objects. 
\craig{Gaia gives a parallax distance estimate for this star of $1.6^{+1.3}_{-0.6}$ kpc \citep{Bailer-Jones18}. }

L17's counterpart to CX32 is slightly blue in \br\ and has a small \hr\ excess. In the HUGS CMDs, it shows a larger blue excess in \uvu. In contrast it falls on the MS in the (\ub, \U) CMD\@. The \UV\ excess tends to support the identification of CX32 as a CV\@.

L17 reported an empty search area for CX33\@. With the slightly larger error circle used in the present study, one star is present within the error circle in the optical imaging. While the star appears in the HUGS database, it is outside of the \UV\ and \U\ fields. The star is on the red edge of the MS in the (\br, \R) CMD and the \ha\ excess side of the MS in the (\hr, \R) CMD. Thus, there is a slight indication that this star may be an AB.  

L17 classified the counterpart to CX34 as a likely AB, based on its optical photometry. It is just outside of the HUGS \UV\ field of view, but inside the \U\ field of view. In the (\ub, \U) CMD, it falls slightly redward of the MS, consistent with its AB identification. 

The counterpart to CX35 that L17 identified was not detected by KS2 in the optical imaging. We visually selected it as a faint shadow and performed aperture photometry. We classified it as a possible CV, based on its apparent extremely blue \br\ colour and large \hr\ excess relative to the WD sequence, but added the cautionary note that the photometry was uncertain. This counterpart is well detected in both the \UV\ and \U\ images and has extremely blue colours in both of the HUGS CMDs (Figs.~\ref{f:CMD_HUGS_UV-U_CV} and \ref{f:CMD_HUGS_U-B_CV}). Thus, we now consider it to be a likely CV\@. 

L17 identified a faint object with an extremely blue \br\ colour and a large \hr\ excess as the counterpart to CX36. We thus classified it as a likely CV\@. It is strongly detected in both the \UV\ and \U\ images and has extremely blue colours in all of the HUGS CMDs\@. Thus, the HUGS photometry corroborates our optical classification on this object. 

L17's identification of CX37 was with a possible AB that fell significantly to the red of the MS in the (\br, \R) CMD and thus fell to the \ha-\emph{deficit} side of the mean colour-colour relation, notwithstanding its small apparent \ha\ excess in the (\hr, \R) CMD\@. We now add, as an alternative counterpart, another apparent AB that lies closer to the centre of the new error circle and that falls to the \ha-excess side of the mean colour-colour relation. The 37a and 37b counterparts are visible in both the \UV\ and \U\ frames (see Fig.~\ref{f:finding_charts_32-37}). Both objects are slightly blue in \uvu. In \ub, 37a is slightly red, while 37b is significantly red. 
The deviation of both counterparts from the MS, in the latter HUGS CMD, generally supports their classification as possible ABs.

L17 classified the counterpart to CX38 as a possible AB, based on its slightly red \br\ optical colour and slight \ha\ excess. This object falls on the MS in both of the HUGS CMDs (Figs.~\ref{f:CMD_HUGS_UV-U_AB} and \ref{f:CMD_HUGS_U-B_AB}). Thus, the HUGS photometry does not provide any additional information on the nature of this object. Given the absence of any other objects, in or near the error circle, it appears that this star is the likely counterpart of CX38.

The CX39 counterpart has an extended, galaxian appearance in the optical imaging (see Fig.~\ref{f:finding_charts_38-43}), and L17 identified it as a galaxy based on its morphology and proper motion. It is outside of the field of view in the \UV\ and \U\ imaging.

L17's identification of CX40 was with a likely AB, with a  red \br\ colour and an \ha\ excess, that falls at 1.3\,\rerr\ from the X-ray position. The HUGS photometry for this object places it to the red side of the MS in both CMDs (Figs.~\ref{f:CMD_HUGS_UV-U_CV} and  \ref{f:CMD_HUGS_U-B_CV}). The HUGS \UV\ and \U\ imaging reveals an extremely blue object that falls on the error circle, somewhat closer to the X-ray position. (See Fig.~\ref{f:finding_charts_38-43}.) This new object, 40b, falls on or near the WD sequence in both of the HUGS CMDs. Since 40b is not visible in the optical, its \ha\ status is unknown. It is unclear which of the two counterparts is the more likely identification. We list 40a as the primary counterpart, given its likely AB status, based on the CMD and colour-colour diagram, versus the uncertain CV status of 40b. 

\subsubsection{MSP C}
\coryrv{MSP C is outside the ACS and WFC FOVs, so we cannot identify an optical counterpart. Our radio observation reveals a faint ($S_\nu \approx 28~\mathrm{\mu Jy}$) radio source in the $5.5~\mathrm{GHz}$ image consistent with the X-ray error circle (Fig. \ref{f:radio_xray_matched_finders}); whereas no detection was found in $9~\mathrm{GHz}$. Similar to CX27, this source is consistent with a steep radio spectrum, so is likely to be the actual radio counterpart.}

\subsection{Classifications among the new sources 41-52}
\label{sec_classification_41_52}
In the optical imaging, the probable counterpart to CX41 appears as a slight extension to the image of a  neighbouring star that is 2 mag brighter. Its position in the optical CMDs is consistent with a bright CV, falling to the blue side of the MS in \br\ and well to the \ha-excess side of the \hr\ sequence. The CV interpretation is further supported by the colour-colour diagram location of this object, in which it registers as having a clear \hr\ excess. While this object is visible in the \UV\ and \U\ frames, indicative of a UV excess that further supports the CV interpretation, it was not detected by KS2 in those frames and does not appear in the HUGS database. We list it as a possible CV due to uncertain photometry associated with crowding. 

There is only one object within the error circle of CX42 and it lies on the MS in both optical CMDs. Although this object appears in the HUGS database and has a listed \U\ magnitude that gives it a pronounced blue colour in \ub, visual inspection of the stacked \U\ image (see Fig.~\ref{f:finding_charts_38-43}) indicates that its apparent detection there may be an artefact of the sky noise fluctuations, so it is unclear whether this object is actually UV-bright. 

\cory{The X-ray spectrum of CX42 can be fit by a hard power-law ($\Gamma = -0.4^{+0.9}_{-1.0}$; Table \ref{t:spectral_analyses_51}) when $N_\mathrm{H}$ is fixed to the cluster value. This is typical of sources with high $N_\mathrm{H}$, where 
the fixed $N_\mathrm{H}$ is too low to account for the lack of soft X-rays. A high $N_\mathrm{H}$ could be from a torus around an AGN, or from an accretion disc in a CV (when viewed edge-on).

We find 
a radio source $0\farcs1$ from the X-ray position of CX42 (Fig.~\ref{f:radio_xray_matched_finders}), with a spectral index of $-0.5\pm 0.3$ (Table \ref{t:radio_matches}). The possibly UV-bright object, however, lies marginally outside the radio error ellipse (Fig.~\ref{f:radio_xray_matched_finders}).
This optical object has a proper motion consistent with cluster membership, and thus is not an AGN\@. We still include it in the CMDs for CVs, but consider the Chandra source CX42 to be a likely AGN, given its radio detection which would be unlikely for a CV at the distance of NGC~6752. }\craig{The potential optical counterpart, in that case, is not associated with CX42.}

Only one object falls in the error circle of CX43, which lies on the MS in both optical CMDs. It is detected in the \UV\ image, but is outside the \U\ field of view. It falls on the red edge of the MS in the (\uvb, \UV) CMD, making it a likely MS star with no obvious association with the X-ray source. 

There are two objects within the CX44 error circle. The brighter object lies on the MS in both optical CMDs. It is in the HUGS database and falls on the blue edge of the lowermost part of the MS in both HUGS CMDs. We interpret this as indicating that this object is most likely a MS star. The fainter object is somewhat blue in \br\ but shows an \hr\ deficit in both the CMD and the colour-colour diagram. We measure a proper motion for this object that is discordant from the cluster distribution, and is similar to the external galaxy frame. It appears that this object is an AGN\@. It is not detected by the HUGS photometry. 

There are two objects within the CX45 error circle. The brighter object lies on the MS in both optical CMDs. The fainter object is somewhat blue in \br\ but shows an \hr\ deficit in both the CMD and the colour-colour diagram. We measure a proper motion for this object that is discordant from the cluster distribution, and agrees well with the external galaxy frame. CX45 lies outside of the field of view of the \UV\ and \U\ frames, so the HUGS database does not provide additional information. Like the fainter object in the CX44 error circle, this fainter object also appears to be an AGN\@. 

\cory{Indeed, our ATCA observation found a flat-spectrum ($\alpha=0.0\pm 0.6$) radio source  $0\farcs8$ south of the nominal X-ray position of CX45 (Fig. \ref{f:radio_xray_matched_finders}). The radio error ellipse marginally matches the X-ray error circle, but is not consistent with the optical counterpart. The ATCA source could then be a radio lobe produced by the AGN. Moreover, the X-ray spectrum of CX45 can be fit by a hard power-law ($\Gamma = 0.1^{+1.1}_{-1.3}$; Table \ref{t:spectral_analyses_51}). Similar to CX22 and CX42, this could indicate enhanced absorption from an AGN torus. We thus classify CX45 as an AGN.}

There is one star at the edge of the CX46 error circle, lying on the MS in both optical CMDs. It falls outside the \U\ field of view, but is visible in the \UV\ frame. It also lies on the MS in the (\uvb, \UV) CMD and thus appears to be a MS star with no apparent association with the X-ray source. 

There are two bright objects and one faint object in the CX47 error circle. The bright objects lie on the MS in the optical and HUGS CMDs. The faint object lies on the MS in the optical CMD and the HUGS (\ub, \U) CMD\@. However, in the (\uvu, \UV) CMD, it lies to the blue of the WD sequence. Visual inspection of the stacked \UV\ image indicates that this apparent pronounced \UV\ excess may likely be an artefact of sky noise fluctuation (see Fig.~\ref{f:finding_charts_44-49}). Indeed, the faint object falls in the broad distribution of questionably measured faint objects located across the bottom of the HUGS CMDs. Thus, while we plot the faint star in the CV CMDs (Figs.~\ref{f:CMD_CV}, \ref{f:CMD_HUGS_UV-U_CV}, and \ref{f:CMD_HUGS_U-B_CV}), we do not consider it to have a high likelihood of being a CV\@. \cory{The X-ray spectrum of CX47 can also be fit by a hard power-law ($\Gamma=0.2^{+1.0}_{-1.2}$). Following the same logic as for CX22, CX47 is also likely an AGN; this is, however, not as robust a conclusion as that for CX42 (hard X-ray spectrum and radio detection), so we keep this source unclassified.}

The error circle for CX48 contains a very bright red giant.
Since the 
optical photometry 
is saturated, it is not possible to determine precisely how far this star might fall  from the red giant branch in \br\ or \hr. 
This star falls near the tip of the giant branch in both HUGS CMDs (Figs.~\ref{f:CMD_HUGS_UV-U_AB} and \ref{f:CMD_HUGS_U-B_AB}), and  
\craig{Gaia gives a proper motion and distance consistent with cluster membership \citep{Gaia18,Bailer-Jones18}.}
Examination of the HUGS photometry reveals two additional potential counterparts, one inside the error circle and one outside of it (see Fig.~\ref{f:finding_charts_44-49}). The inside object, 48b, is very blue in \uvu, while it is very red in \ub. 
Visual inspection of the HUGS frames suggests that 48b is only definitely detected in \UV. The outside object, 48c, at 1.5\,\rerr\ is quite blue in both of the HUGS CMDs, falling between the MS and the WD sequence. Given the lack of optical photometry for 48b and 48c, it is difficult to assess the likelihood of these  possible counterparts. We consider it more likely than not that one of the faint, UV-bright objects is the optical counterpart, but this is clearly uncertain.  Such bright RGB stars are rare enough that the fact that one lands in an X-ray error circle is suggestive of a possible association.  

There are three objects inside the CX49 error circle, two MS stars and a faint WD like object with a substantial apparent \ha\ excess in the colour-colour diagram. This latter object, which lies closest to the centre of the error circle, appears to be the counterpart. However, due to uncertainty in its \ha\ magnitude, we classify it as a possible rather than likely CV\@. It is outside of the \U\ field of view, but is well detected visually in the \UV\ frame (see Fig~\ref{f:finding_charts_44-49}). Nonetheless, it does not appear in the HUGS database, so no additional information is provided. 

We find no compelling counterparts for CX50-52. The CX50 error circle contains one MS star that lies outside of the \UV\ field of view, but is detected in the \U\ frame and has a \ub\ colour consistent with the MS\@.

The large error circle of CX51 does not contain any objects in the optical imaging and does not appear in the \UV\ and \U\ fields of view. The objects in the immediate vicinity of the error circle fall on the MS\@. 

The CX52 error circle contains one object that falls on the MS in both optical CMDs. It is detected in \UV\ and \U, and falls on the blue edge of the MS in the (\uvu, \UV) CMD and on the red edge of the MS in (\ub, \U) CMD\@. 
We conclude that this object is an MS star.

\section{Spatial Distribution and Object Masses}

The relative spatial distributions for different classes of objects in the cluster can be used to estimate object masses following a Jeans equations type of approach, as reviewed by \citet{Cohn10}. It is useful to outline the theory here. For a spherical stellar system with an isotropic velocity dispersion tensor, the first-order Jeans equation can be written as,
\begin{equation} 
    \frac{d (n \sigma^2)}{dr} = - n \frac{d\Phi}{dr},
\end{equation}
where $n$ is the spatial density of a particular mass component, $\sigma$ is the corresponding one-dimensional velocity dispersion of that component, and $\Phi$ is the gravitational potential \citep[][see their eqn.~4.221]{Binney08}. For a multi-component system, it follows for components $i$ and $j$ that,
\begin{equation}
    \frac{d(n_j \sigma_j^2)}{d(n_i \sigma_i^2)} = \frac{n_j}{n_i}.
\end{equation}
We now make the approximations, based on the Fokker-Planck simulations of post-core-collapse clusters by \citet{Murphy90}, that the velocity dispersion profile of each component is relatively flat, so that we may neglect its radial gradient, and that the mass components above the main-sequence turnoff (MSTO) mass are in approximate thermal equilibrium, so that $m_i \sigma_i^2 = m_j \sigma_j^2$. \revised{This is consistent with the short half-mass relaxation time of NGC~6752, $t_\mathrm{rh} < 0.7\,\mathrm{Gyr}$ \citep{Belloni19}, which provides sufficient time for mass segregation to be established}. It follows from these approximations that,
\begin{equation}
\label{eqn:Diff_eqn}
    \frac{d \ln n_j}{d \ln n_i} = \frac{m_j}{m_i}.
\end{equation}
Integrating eqn.~\ref{eqn:Diff_eqn} we find that,
\begin{equation}
    n_j \propto {n_i}^q 
\end{equation}
\noindent where $q = m_j/m_i$ is the mass ratio. As in our previous studies \citep{Cohn10,Lugger17}, we consider density profiles of the ``generalised King model'' form, which we have also called a ``cored power law.'' This takes the form,
\begin{equation}
    n(r) = n_0 \left[1 + \left(\frac{r}{r_0}\right)^2\right]^{\beta/2}.
\end{equation}
The corresponding surface density profile is given by,
\begin{equation}
\label{eqn:Cored_PL} 
S(r) = S_0 \left[1 + \left({\frac{r}{r_0}}\right)^2 \right]^{\alpha/2},
\end{equation}
where $\alpha = \beta -1$. The core radius $r_c$ is related to the scale parameter $r_0$ by,
\begin{equation}
r_c = \left(2^{-2/\alpha} -1 \right)^{1/2} r_0\,.
\end{equation}

By fitting Eqn.~\ref{eqn:Cored_PL} to the profiles of various components, their characteristic masses may be estimated relative to the MSTO mass.  We perform the fits using a maximum-likelihood approach to fit the cumulative radial surface density distribution. The fitting software was originally developed by \citet{Slavin03}. We chose the half-light radius $r_\mathrm{h} = 115\arcsec$ as the limiting radius of the maximum-likelihood fits. The fitting results are given in Table~\ref{t:Cored_PL_fits}. 

\begin{table*}
\caption{Cored-Power-Law Model Fit Results} 
\label{t:Cored_PL_fits}
\begin{tabular}{lrcrcccr}
\hline
Sample & 
$N^a$ &
$q$ & 
$r_c~(\arcsec)$~~~ &
$\alpha$ &
$m~(\msun)$ &
$\sigma^b$ &
K-S prob$^c$ \\
\hline
%
%
MSTO       &10016 &   1.0            & $12.2 \pm 1.2$ & $-1.27 \pm 0.03$ & $0.80 \pm 0.05$ & \nodata & \nodata \\
\chandra\ sources
           &   51 &  $1.25 \pm 0.10$ & $ 9.2 \pm 0.9$ & $-1.83 \pm 0.24$ & $1.00 \pm 0.08$ & 2.5     & 0.34\%  \\
bright CV  &   10 &  $2.10 \pm 0.33$ & $ 5.8 \pm 0.7$ & $-3.77 \pm 0.76$ & $1.68 \pm 0.26$ & 3.3     & 0.0091\% \\
faint CV   &    8 &  $1.44 \pm 0.23$ & $ 7.9 \pm 1.3$ & $-2.28 \pm 0.53$ & $1.15 \pm 0.18$ & 1.9     & 8.5\%    \\
AB         &    9 &  $1.09 \pm 0.25$ & $10.8 \pm 4.4$ & $-1.48 \pm 0.57$ & $0.87 \pm 0.20$ & 0.3     & 79\%    \\
BSS (blue) &   25 &  $1.37 \pm 0.12$ & $ 8.4 \pm 0.8$ & $-2.10 \pm 0.28$ & $1.10 \pm 0.10$ & 2.7     & 2.7\% \\
BSS (red)  &   17 &  $1.53 \pm 0.15$ & $ 7.5 \pm 0.7$ & $-2.48 \pm 0.33$ & $1.22 \pm 0.12$ & 3.2     & 0.0047\% \\
\hline
\multicolumn{8}{l}{\makecell[tl]{$^a$Size of sample within 115\arcsec\ of cluster centre}}\\
\multicolumn{8}{l}{\makecell[tl]{$^b$Significance of mass excess above MSTO mass in sigmas}}\\
\multicolumn{8}{l}{\makecell[tl]{$^c$K-S probability of consistency with MSTO group}}\\
\end{tabular}
\end{table*}

\begin{figure}
\includegraphics[width=\columnwidth]{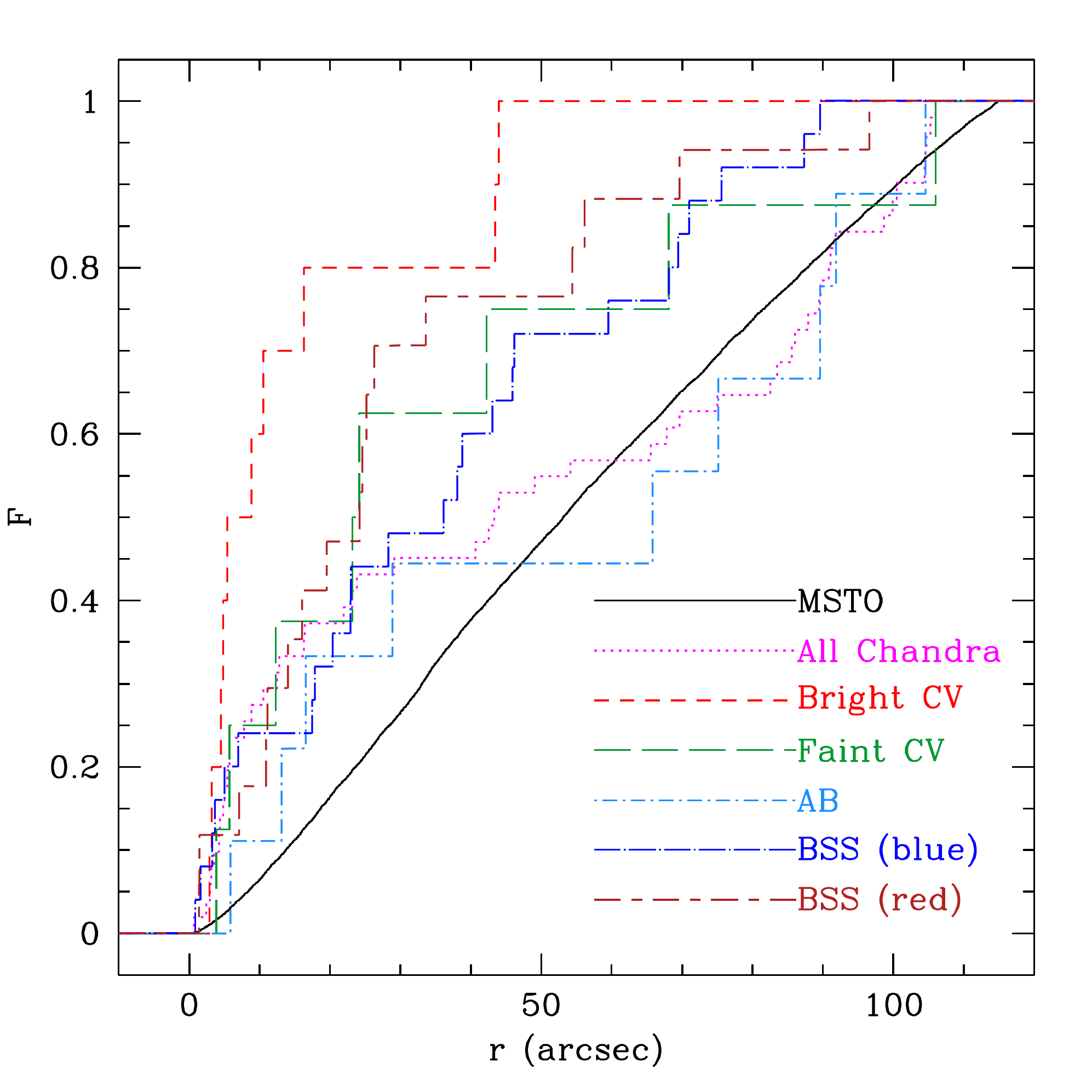}
\caption{Cumulative radial distributions for selected stellar groups.  Note that the \chandra\ sources, the bright CVs, and the BSSs show significant central concentration ($p \lesssim 1\%$) relative to the MSTO group.  Fitting information and K-S sample comparisons for these stellar groups are given in Table~\ref{t:Cored_PL_fits}.  
\label{f:cumulative_radial_distributions}}
\end{figure}

\begin{figure}
\includegraphics[width=\columnwidth]{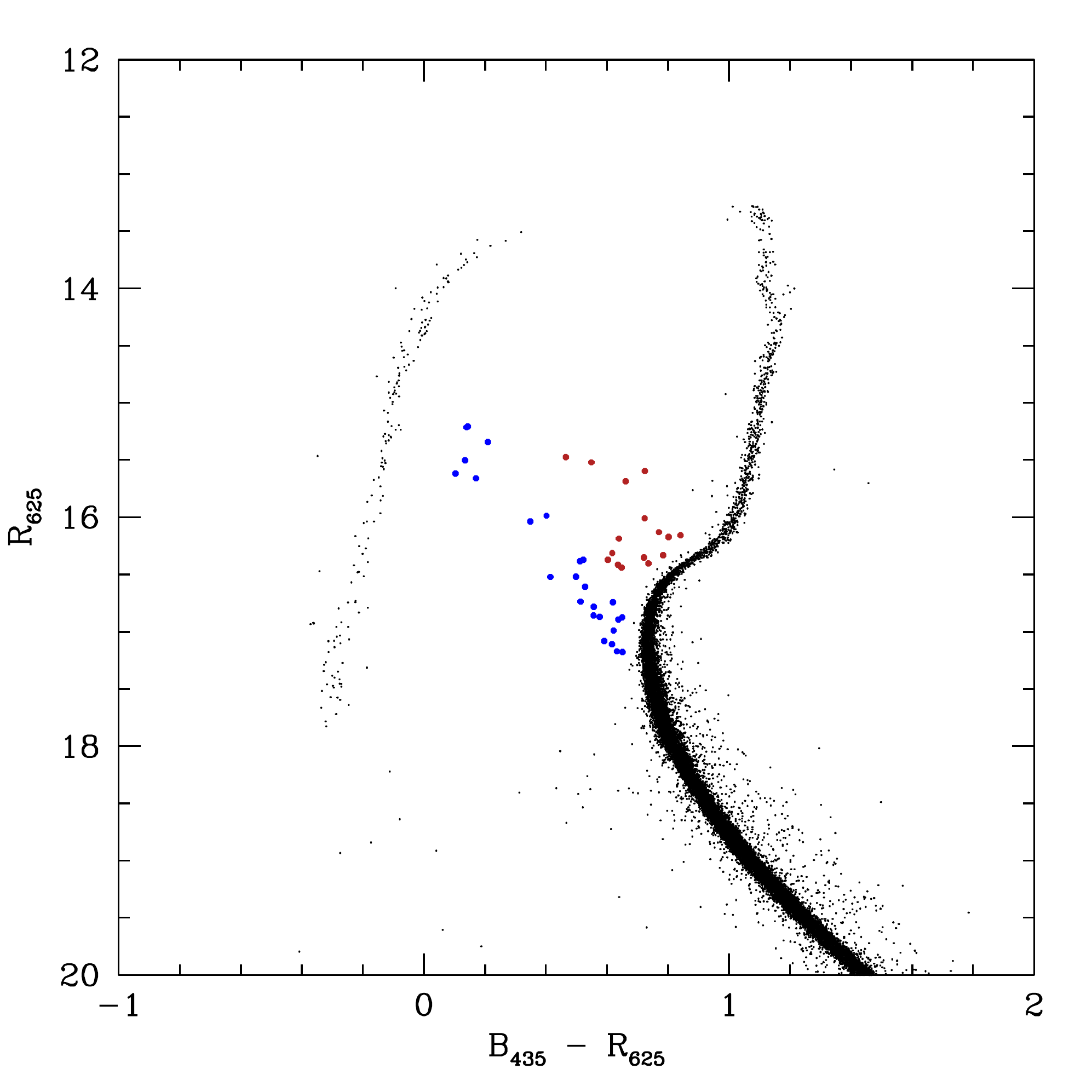}
\caption{\revised{BSS selection from the (\R, \br) CMD\@. The blue and red BSS sequences (color coded) were selected by visual appearance, using the Glue software package \citep{Beaumont15,Robitaille17}. We closely emulated the approach of \citet{Ferraro09}, who discovered the double BSS sequence in the core-collapsed cluster M30.}
\label{f:BSS_selection}}
\end{figure}

We first examined the radial profiles of a number of different stellar groups by comparing the cumulative radial distributions shown in Fig.~\ref{f:cumulative_radial_distributions}.  As in L17, we define a MSTO sample by selecting stars with magnitudes in the range $16 \le \R < 18.5$, which extends to 2 mag below the MSTO\@. We took the MSTO mass to be $0.80 \pm 0.05\,\msun$, based on the study of \citet{Gruyters14} who find a MSTO mass of 0.79\,\msun. In addition to the MSTO group, we consider all of the 51 \chandra\ sources within $r_\mathrm{h}$, the \revised{10} brightest CVs, the \revised{8} faintest CVs, the ABs, and  \revised{two blue straggler star (BSS) groups---a blue sequence and a red sequence---selected from the (\R, \br) CMD as illustrated in Fig.~\ref{f:BSS_selection}. The selection of the BSSs is described below.} As can be seen in Fig.~\ref{f:cumulative_radial_distributions}, the \chandra\ sources, the bright CVs, and \revised{both BSS groups} show strong central concentration relative to the MSTO group. In order to quantify the significance of the differences in the distributions, we performed Kolmogorov-Smirnov (K-S) comparisons of each sample with the MSTO group.  The results are given in Table~\ref{t:Cored_PL_fits}, where the probability, $p$, of the two samples being drawn from the same parent distribution is listed in the last column.  The distributions of the \chandra\ sources, the bright CVs, and \revised{the red BSSs} differ very significantly ($p \ll 1\%$) from that of the MSTO group. \revised{The distribution of the blue BSSs differs significantly $(p<5\%$) from that of the MSTO group.} The faint CV and AB distributions do not differ from the MSTO group at a significant level \revised{($p=8.5\%$ and $p=79\%$, respectively), although the $p$ value for the faint CVs approaches the 5\% cutoff for significance. A direct comparison of the bright and faint CVs indicates that these two groups do not differ at a significant level ($p=24\%$), although Fig.~\ref{f:cumulative_radial_distributions} suggests that the bright CVs are somewhat more centrally concentrated than are the faint CVs. The blue and red BSS sequences similarly do not differ at a significant level ($p=33\%$), although the red sequence appears more centrally concentrated than is the blue sequence (see Fig.~\ref{f:cumulative_radial_distributions}).} 

\revised{We note that L17 excluded two CV candidates from the faint CV group, CX9a and CX24. These objects are intermediate in \R\ magnitude between the bright group and the faintest six CV candidates, as can be seen Fig.~\ref{f:CMD_CV}. The exclusion of these two objects produced a more significant difference between the radial distributions of the bright and faint CV groups than that found in the present study. We now choose to include these two objects with the other faint CV candidates for consistency with the results of \citet{Pala20}. Based on a large-scale \emph{Gaia} survey of CVs within 150~pc, they find that those that lie below the period gap have $9 \lesssim M_R \lesssim 12$, while those above the period gap have $4 \lesssim M_R \lesssim 8$. Since the CX9a and CX24 candidate counterparts have $M_R \sim 9.5$, this places them within the magnitude range below the period gap.}

\revised{We define the blue and red BSS sequences following the work of \citet{Ferraro09} for M30, \citet{Dalessandro13} for NGC~362, and \citet{Beccari19} for M15\@. As can be seen in Fig.~\ref{f:BSS_selection}, the blue sequence is a narrow extension of the MS beginning at the turnoff point. The red sequence is somewhat broader and begins above the subgiant branch. \citet{Ferraro09} interpret the blue sequence BSSs as the products of direct stellar collisions and the red sequence BSSs as the result of close binary evolution with mass transfer. They find that the red sequence is more centrally concentrated than is the blue sequence, as we find here (see Fig.~\ref{f:cumulative_radial_distributions}), although, as we note above, the difference between the two radial distributions is not statistically significant in the present case of NGC~6752. \citet{Xin15} have investigated the formation of the red BSS sequence by modelling the evolution of an initial 0.9\,\msun\ MS star + 0.5\,\msun\ WD close binary system and find that they are able to well reproduce the red BSS sequence observed in M30.}

For each of the stellar groups shown in Fig.~\ref{f:cumulative_radial_distributions}, we carried out maximum likelihood fits of Eqn.~\ref{eqn:Cored_PL} to the surface-density distribution. In the thermal equilibrium approximation introduced above, the space density slope of a component is given by $\beta = q \beta_{\rm to}$ where $\beta_{\rm to}$ is the slope value for the MSTO group and $q = m/m_{\rm to}$. It follows that the surface density slope of the component is given by,
\begin{equation}
    \alpha = q\, (\alpha_{\rm to} - 1) + 1.
\end{equation}

Table~\ref{t:Cored_PL_fits} gives the results of maximum-likelihood fits of Eqn.~\ref{eqn:Cored_PL} to the turnoff-mass stars, \chandra\ sources, CVs, ABs, and BSSs.  By definition, the $q$ value for the MSTO sample is unity. For all of the other groups, the $q$ value is determined by maximising the likelihood with respect to $q$, for fixed $r_0$ and $\alpha_\mathrm{to}$ determined by the MSTO sample fit. As can be seen from the table, the $q$ values for all other groups exceed unity, indicating that the characteristic masses exceed the turnoff mass (assumed to be 0.80~\msun).  For the \chandra\ sources, bright CVs, and BSSs, the excesses are significant at the \revised{$2.5-3.3\,\sigma$} level.  We note that the results of this analysis are supported by the K-S comparison results given in the last column of the table.

The bright candidate CV sample size only increased by one object from that in L17 \revised{(CX41) while the faint candidate CV sample increased by three objects, one newly detected candidate CV (CX49) plus two intermediate brightness candidate CV counterparts that were previously excluded (CX9a and CX24).} Our inferred characteristic masses for these groups did not change significantly, staying within the range of uncertainty of the previous estimates. The 1.7\,\msun\ average mass of the bright CVs suggest that they typically contain relatively massive WDs. The \R\ magnitudes of these systems imply MS component masses on the order of $0.6 \pm 0.2\,\msun$ at most \revised{\citep[based on evolutionary models for metal-poor low-mass stars by][]{Baraffe97}}, leaving $\gtrsim\!1.0\,\msun$ to be accounted for by the WD. This is substantially higher than that of WDs currently being produced in the cluster, the masses of which are on the order of $0.53 \pm 0.03\,\msun$ \citep{Moehler04}. \revised{The faint CVs have combined MS+WD masses of $\sim\!1.2\,\msun$. Since the luminosity of the faintest CV candidates is dominated by that of WD, as indicated by the CMD locations of these objects, the inferred masses of the companion stars is exceedingly low, $\lesssim\!0.1\msun$. Thus, the implied WD masses in the faint CV candidates is also $\gtrsim\!1.0\,\msun$.}

\revised{It is useful to compare our estimates for the CV component masses to those made for the CV population in the Milky Way disk. As reviewed by \citet{Zorotovic11}, the average WD mass in CVs in the Galactic disk has long been known to lie in the range 0.8--1.2\,\msun, which is substantially higher than the value of $\sim 0.6\,\msun$ for single WDs. \citet{Zorotovic11} and \citet{Schreiber16} give updated values of $0.83 \pm 0.23\,\msun$ for the mean WD mass in CVs and $0.67 \pm 0.21\,\msun$ for single WDs, where the uncertainties are the sample standard deviations rather than the smaller errors of the mean. \citet{McAllister19} similarly find a mean CV WD mass of $0.81 \pm 0.02\,\msun$ ($\sigma = 0.13\,\msun$), where the uncertainty is the error of the mean and $\sigma$ is the sample standard deviation. This mean CV WD mass is somewhat smaller than the value of $\gtrsim\!1.0\,\msun$ that we estimate for the NGC~6752 CVs. This may reflect differences in CV evolution between the cluster and field environments. In dense clusters, such as NGC~6752, stellar interactions may play a role in promoting mass transfer in CVs, leading to a somewhat faster WD mass increase.}

\revised{
\begin{figure}
\includegraphics[width=\columnwidth]{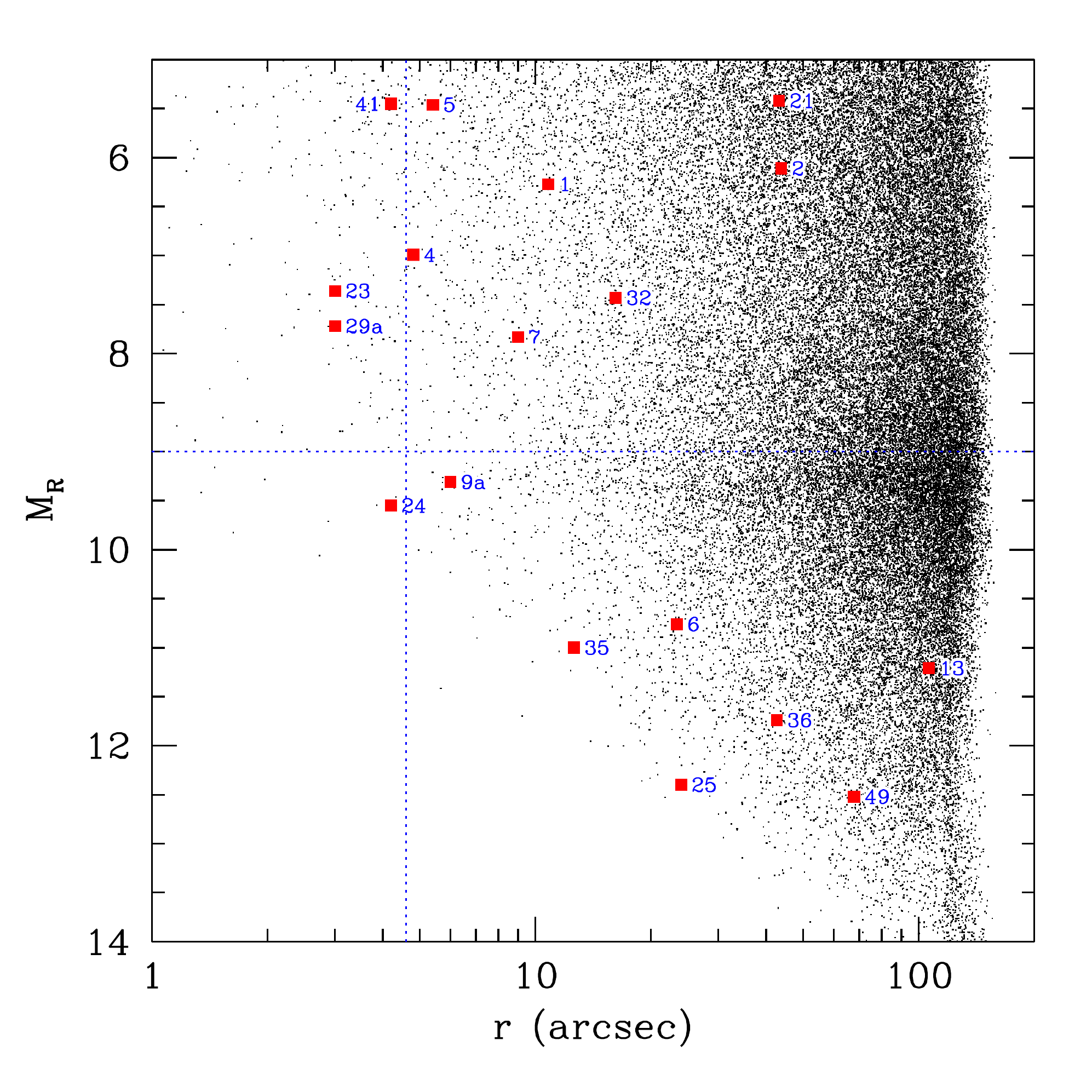}
\caption{\revised{Absolute \R\ magnitude vs.\ projected radial distance from the cluster centre for candidate CV counterparts (red symbols) and the entire stellar distribution (black symbols). The CX number is indicated for each source. The horizontal dashed line at $M_R = 9$ separates the bright and faint CV populations. The vertical dashed line represents the core radius, $r_c = 4\farcs6$, determined by L17. Note the general lack of stars below a diagonal boundary defined by CX24, CX35, and CX25.}
\label{f:M_R-radius_plot}}
\end{figure}

To further investigate the possible correlation between candidate CV optical brightness and radial distribution, we generated a plot of absolute magnitude $M_R$ (where $R$ denotes \R) vs.\ projected radial offset $r$ from the cluster centre (Fig.~\ref{f:M_R-radius_plot}). Visual inspection of this plot supports the impression from the cumulative distribution plot (Fig.~\ref{f:cumulative_radial_distributions}) that optically bright CV candidates cluster at small $r$ while the distribution of faint CV candidates extends to much larger $r$. However, the plot also suggests that incompleteness may play a role in the apparent correlation, inasmuch as the small-$r$ region of the plot appears to be progressively underpopulated at faint $M_R$. Some of this apparent depletion may be the result of mass segregation, as $M_R$ is correlated with mass along the MS, but some 
may also be due to incomplete detection of faint stars at small $r$. In this case, some optically faint objects that are the actual counterparts of \chandra\ sources may have been undetected in favour of brighter apparent counterparts. We note that all of the unclassified sources listed in Table~\ref{t:counterparts} lie at large offsets from the cluster centre ($r>41\arcsec = 9\,r_c$) and thus do not represent undetected faint sources at small $r$. A robust approach to investigating the degree of photometric incompleteness would be by use of artificial star experiments. However, this is beyond the scope of the current investigation and we defer it for further study. 
}

\section{Conclusions}

Our deep {\it Chandra} observation revealed 12 new sources within the half-light radius, extending the original catalogue to a total of 51 sources. We performed spectral analyses for all these sources using the Cycle 18 data, finding that most sources can be fitted either with a thermal plasma ({\tt mekal}) or a power-law model. 
\craig{CX5 is a particularly unusual object, which shows a high-energy cutoff to its X-ray spectrum, strong X-ray variability on timescales of $\sim$0.2 days, and an optical counterpart showing only weak \ha\ emission and colors of a main-sequence star, even in UV filters. The lack of UV evidence for an accretion disk, with an $L_X\sim10^{32}$ erg s$^{-1}$, suggests the accretor may be a neutron star or black hole; the lack of a radio detection does not exclude these possibilities.
CX8 is best-fit by a double thermal plasma model, typical of chromospheric emission from ABs, matching the photometric evidence that this is a (pair of) foreground binaries.} 
We 
found inter-observational X-ray variability in CX1, CX3, CX4, CX7 and CX8, and perhaps MSP A, though the latter may be spurious. 

We used both the optical \hst\ photometry reported in L17 and the HUGS photometry \citep{Piotto15,Nardiello18} to search for counterparts to the 51 \chandra\ sources within the half-light radius of NGC 6752. This led to several refinements of the identifications for the previous sources CX1--CX40 and identifications for some of the new sources CX41--CX52. The HUGS photometry proved useful for detecting UV-excess objects that are too faint to be detected in the optical, and revealed a few such objects that are potential CVs. However, it is not possible to assess the \ha\ status of objects, such as CX40b, that are not detected in \ha\ and \R. Not including the potential CVs from the HUGS photometry, the total numbers of identifications are \revised{18} CVs, 9 ABs, 3 RGs, 3 GLXs, and 5 AGNs. \revised{In addition, 3 of the sources are associated with MSPs.}  
Cross-matching our X-ray catalogue with our ATCA 5 GHz radio catalogue revealed three matches, CX17, CX42, and CX45, all of which we interpret as extragalactic sources. 
 
We note that the significant increase in \chandra\ exposure time in this study, relative to L17 (though with reduced effective area at low energies), did not result in the detection of a significant number of new CVs; \revised{one additional bright CV (CX41) and one additional faint CV (CX49) were detected.} Thus, the numbers of bright and faint CVs remained roughly comparable, similar to the situation in NGC 6397 \citep{Cohn10}. This is contrary to the prediction of a factor of two or more larger number of faint CVs than bright CVs by the simulations of \citet{Ivanova06} \revised{or \citet{Belloni19}}, \craig{and observed in the globular cluster 47 Tuc \citep{RiveraSandoval18}, although we have reached the necessary depths of $M_V=11.6$ ($R\sim24.5$) and $L_X=3\times10^{29}$ erg s$^{-1}$.} This suggests that as core-collapsed clusters, NGC 6397 and NGC 6752 both have experienced either a recent burst of bright CV formation or else have undergone faint binary destruction/ejection.  \craig{ \citet{Belloni19} argue that strong interactions forming new, bright CVs will eject them from the cluster cores, so that the present radial distributions are governed entirely by mass segregation. 

\revised{There are interesting arguments for and against the possibility of significant dynamical destruction of CVs after they are formed. This is distinct from the destruction of CV progenitors, which is well-established, \citep[e.g.][]{Belloni19,Haggard09}. On the one hand, \citet{Leigh16} argue that dynamical destruction of binaries in relatively massive globular clusters should be rather rare (less than 1\% of binaries in clusters \hp{with masses above $10^5$ \msun}). However, \hp{they} consider a very limited range of cluster densities. The argument in favor of dynamical destruction of existing CVs flows from consideration of MSPs in globular clusters, which show strong evidence for dynamical destruction during or after formation in the densest clusters \citep{Verbunt14}. This evidence comes in multiple forms\hp{:} the substantial eccentricities of all MSPs with orbital periods $>\,$1 day; the strong cluster density dependence of the single vs.\ binary fractions of MSPs\hp{---with singles predominating in the densest clusters---}implying that single MSPs have been separated from their companions \hp{there}; and the strong density dependence of apparently ``young,'' \hp{high-magnetic-field-strength pulsars,} which \citet{Verbunt14} argue convincingly have been disrupted during the process of mass transfer, leaving them only partly recycled \hp{and thus with magnetic fields that have not undergone ``burial.''}

The question is whether this evidence in favour of dynamical destruction of MSPs in the densest clusters (which we can associate with core-collapsed clusters) also translates to a significant probability of dynamical destruction of CVs in dense clusters. Against this possibility, one could point to the higher masses of NSs in MSP systems, which would make them more mass segregated into the highest-density cores than \hp{are the} lower-mass \hp{CV} binaries; and the longer (few days) orbital periods of a fraction of MSPs with He WD companions as allowing easier disruption, compared to CVs. In favor of the possibility, we note that many core-collapsed clusters have all single MSPs, or MSPs showing evidence of major interactions during or after mass transfer. \hp{The prototypical core-collapsed cluster M15 has 7 single pulsars and one binary pulsar, PSR 2127+11C, which presently has a degenerate companion---a NS or WD \citep[see,][]{Anderson90,Prince91,Anderson93}. PSR 2127+11C lies well outside of the central region of M15, providing evidence for an interaction that ejected it from there. \citet{Prince91} find that this MSP has a short characteristic age of $10^8$\,yr and argue that this indicates that it likely underwent at least two collisional interactions during this time. In their scenario, 
the present degenerate companion was likely captured during an exchange encounter, which interrupted the recycling phase of the MSP, given the short amount of time that the original mass-donating companion would have had to evolve to a degenerate state.
Similarly, the core-collapsed cluster NGC~6752 has 4 single pulsars and one binary pulsar that was ejected nearly out of the cluster \citep{Colpi02}. 
In the least dense clusters (presumably those where destruction is the least common), we see a range of orbital periods of the MSPs, but still most $P_\mathrm{orb}$ values are less than 0.5 day, which is also the case for most CVs.  
Thus, we infer that most MSPs in globular clusters were formed in short-period binaries. Yet in the core-collapsed clusters, we find that most MSPs are single, implying that these short-period binaries were disrupted. This therefore suggests that core-collapsed clusters are capable of disrupting CVs, which have similar orbital periods.}
We also have some suggestive evidence for dynamical disruption of CVs in core-collapsed clusters: \citet{Bahramian13} finds smaller numbers of X-ray sources (in a $L_X$ range dominated by CVs) in core-collapsed clusters, compared to \hp{non-core-collapsed} clusters with similar stellar encounter rates. Finally, there is evidence of a relative lack of faint CVs (compared to bright CVs) in core-collapsed globular clusters \citep[][this work]{Cohn10}, relative to non-core-collapsed clusters with similar depth of data. In our opinion, the question of whether dynamical destruction of CVs post-formation is a significant effect is not completely settled, but the evidence tilts toward suggesting that it plays a role.}

A caveat is that the extreme crowding in the Chandra image of NGC 6752's core is likely to combine real X-ray sources. Core sources CX18, CX27, CX29, and CX48 all have multiple possible optical counterparts, and the X-ray emission of each may arise from multiple sources. \revised{Several objects also have uncertain classification (notably CX5, and several CV candidates seen only in the UV).}
New high-resolution, higher-sensitivity X-ray instruments (e.g. Lynx, \citealt{Lynx18}; or AXIS, \citealt{AXIS18}) could dramatically improve our view.} \revised{Further evidence confirming candidates as CVs, such as spectroscopy, or identification of dwarf-nova outbursts (e.g. \citealt{Thomson12,Modiano20}) would also be of great value. Existing MUSE observations have provided insight into CX2 and CX19 (\citealt{Gottgens19}), and the existing spectra may be useful for up to 6 more of the brightest objects, while narrow-field mode and longer observations would be needed for others.}

\revised{NGC 6397 also has many more known ABs (42; \citealt{Cohn10}) than does NGC 6752 (9). Part of the difference is certainly due to the steep luminosity function of ABs; half the ABs identified in NGC 6397 have $L_X<3\times10^{29}\,\ergs$ \citep{Bogdanov10}, our $L_X$ limit in this study. The other factor of two in the different AB numbers may be due to the differing binary content; \citet{Milone12} show that the binary fraction in NGC 6752 is 1.0$\pm0.6$\%, while that in NGC 6397 is 2.4$\pm0.6$\%. As ABs are thought to be produced primordially (e.g.\ \citealt{Bassa04}), their numbers should roughly scale with the binary fraction, though ABs have shorter orbital periods than average binaries \citep[e.g.][]{Albrow01}.}

We examined the radial distributions of several different populations of objects in NGC 6752 in order to determine the masses of these objects relative to the MSTO mass of 0.8\,\msun, which we adopted from an isochrone fitting analysis by \citet{Gruyters14}. Our analysis is based on the assumption that the populations more massive than the MSTO mass are in thermal equilibrium. In this case, the higher the characteristic mass of a population, the higher its degree of central concentration. This analysis resulted in characteristic mass estimates that significantly exceed the MSTO mass for the entire set of \chandra\ sources ($1.0\pm0.1\,\msun$), the bright CVs ($1.7\pm0.3\,\msun$), \revised{the blue BSSs ($1.1\pm0.1\,\msun$), and the red BSSs ($1.2\pm0.1\,\msun$).} \revised{The  characteristic masses of the faint CVs ($1.2\pm0.2\,\msun$) and the ABs ($0.9\pm0.2\,\msun$) do not differ from the MSTO mass at a significant level, although the mass excess for the faint CVs approaches the $2\sigma$ level. The bright CVs appear to be somewhat more centrally concentrated and thus more massive than the faint CVs, 
as found by L17, although the effect is no longer statistically significant as a consequence of the inclusion of two intermediate-brightness CVs in the faint group. We note that photometric incompleteness may play a role in the observed depletion of optically faint CVs in the crowded central region of the cluster. Nonetheless, the apparent difference between the radial distributions of the bright and faint CVs likely indicates the action of mass segregation, as discussed by \citet{Belloni19}.}


\section*{Acknowledgements}

HC and PL acknowledge the support of NASA/SAO grant GO7-18020X to Indiana University and discussions with D.~Nardiello. CH acknowledges discussions with T.~J.~Maccarone, A.~Bahramian, and G.~R.~Sivakoff, and support from NSERC Discovery Grant RGPIN-2016-04602, and a Discovery Accelerator Supplement. VT acknowledges a CSIRS scholarship from Curtin University. JCAM-J is the recipient of an Australian Research Council Future Fellowship (FT140101082). J.S. acknowledges support from the Packard Foundation and NSF grants AST-1308124 and AST-1514763. The International Centre for Radio Astronomy Research is a joint venture between Curtin University and the University of Western Australia, funded by the state government of Western Australia and the joint venture partners. 


\section*{Data Availability}

The \chandra\ data used in this paper are available in the \chandra\ Data Archive (https://cxc.harvard.edu/cda/) by searching the Observation ID listed in Table~\ref{t:chandra_obs} in the Search and Retrieval interface, ChaSeR (https://cda.harvard.edu/chaser/). The \hst\ data used in this work can be retrieved from the Mikulski Archive for Space Telescope (MAST) Portal (https://mast.stsci.edu/portal/Mashup/Clients/Mast/Portal.html) by searching the proposal IDs listed in Table~\ref{t:hst_obs}. The ATCA observations are available at https://atoa.atnf.csiro.au (project C2877).




\bibliographystyle{mnras}
\bibliography{NGC6752}


\appendix

\section{Optical/UV Finding Charts}

(This material is intended for online distribution only.)


\newpage
\begin{figure*}
\includegraphics[width=\textwidth]{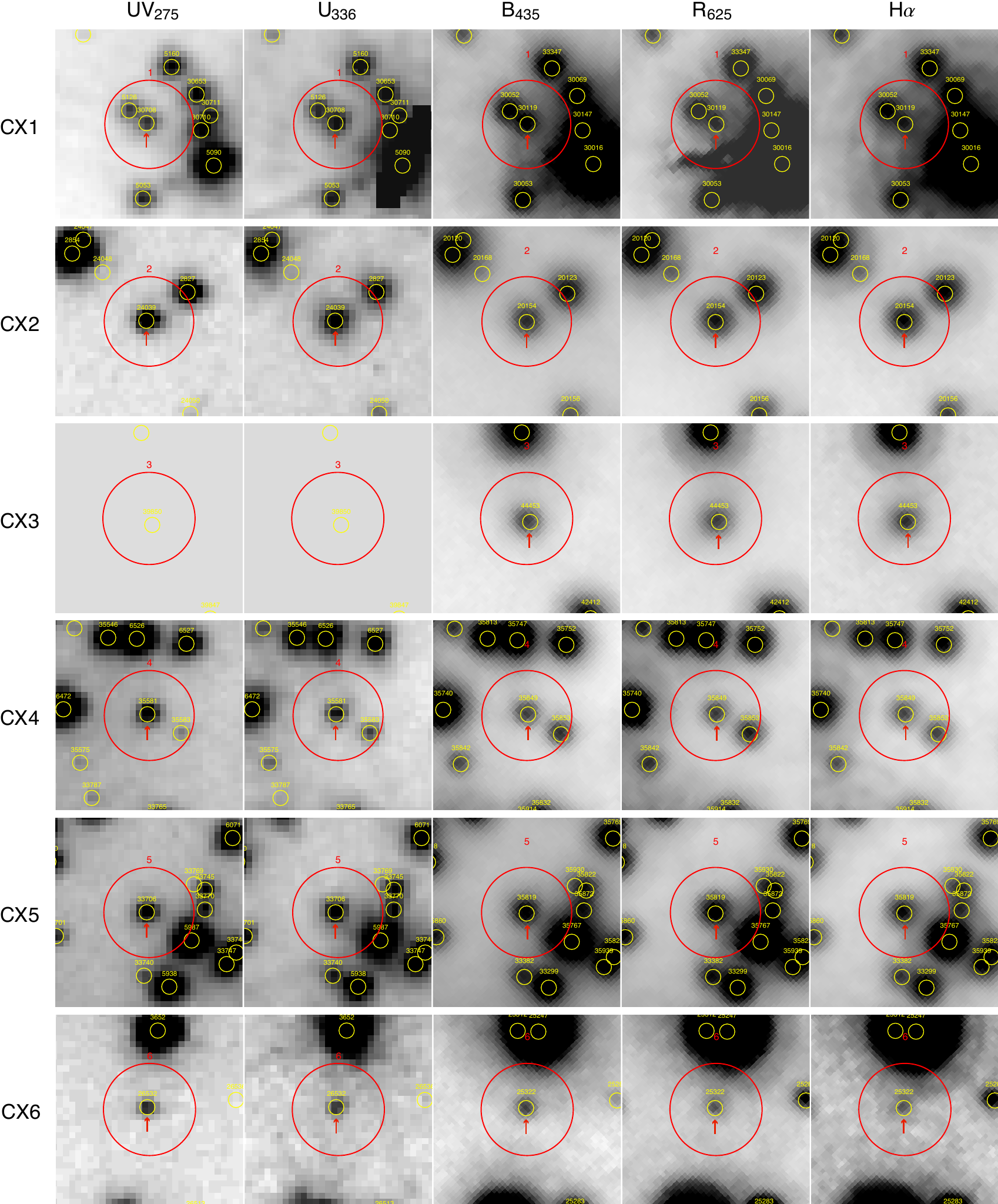}
\caption{Finding charts: CX1 -- CX6. The field size is 1\farcs2 on a side, in this and subsequent figures, unless otherwise noted in the caption. North is up and east is to the left. The star numbers in the HUGS frames are the published sequence numbers in the online HUGS photometry files. The star numbers in the optical frames correspond to the line numbers in the KS2 output files, which also provide a unique star designation.}  \label{f:finding_charts_1-6}
\end{figure*}

\newpage
\begin{figure*}
\includegraphics[width=\textwidth]{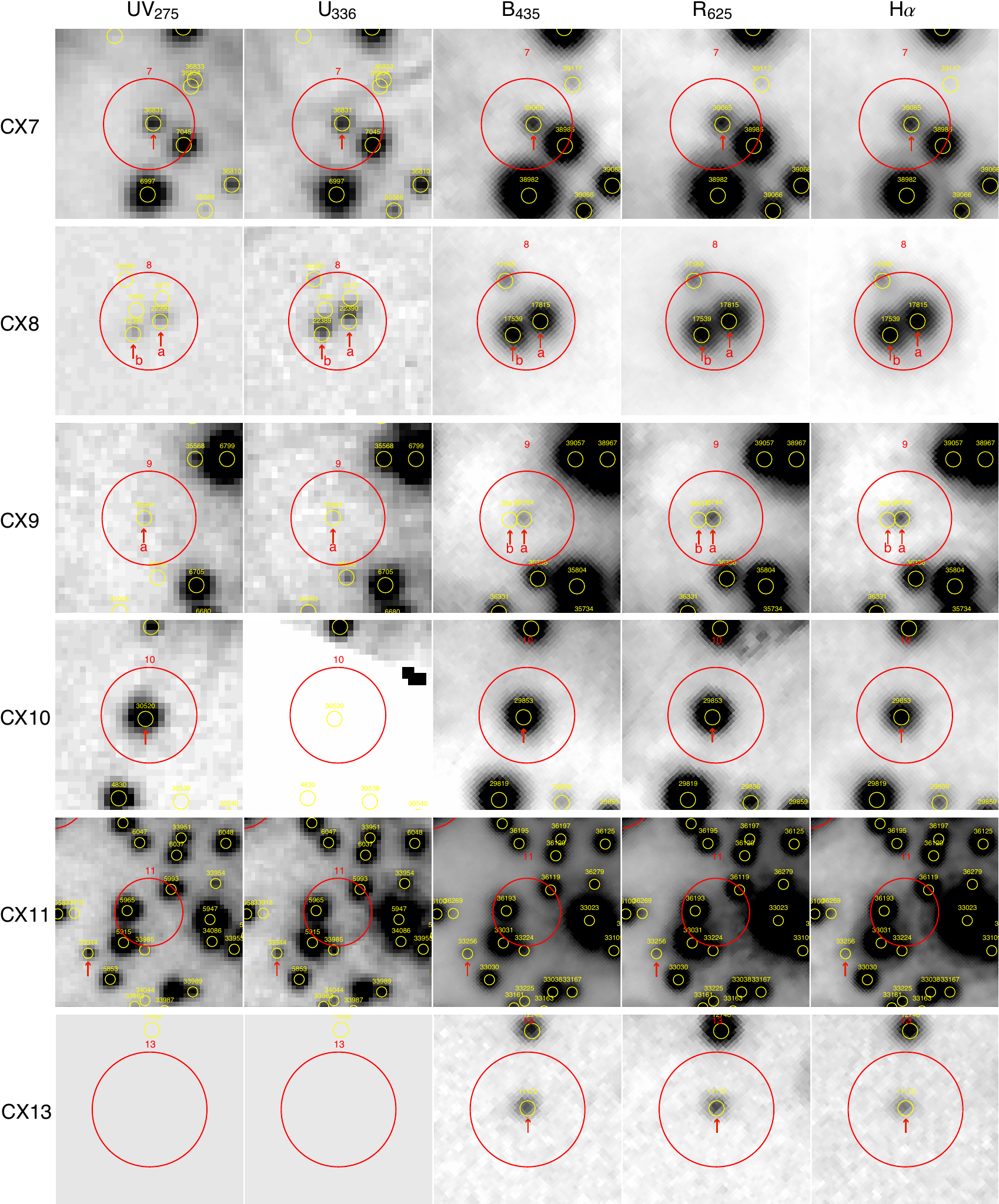}
\caption{Finding charts: CX7 -- CX13. For CX11, the field size is 1\farcs8 on a side.}  \label{f:finding_charts_7-13}
\end{figure*}

\newpage
\begin{figure*}
\includegraphics[width=\textwidth]{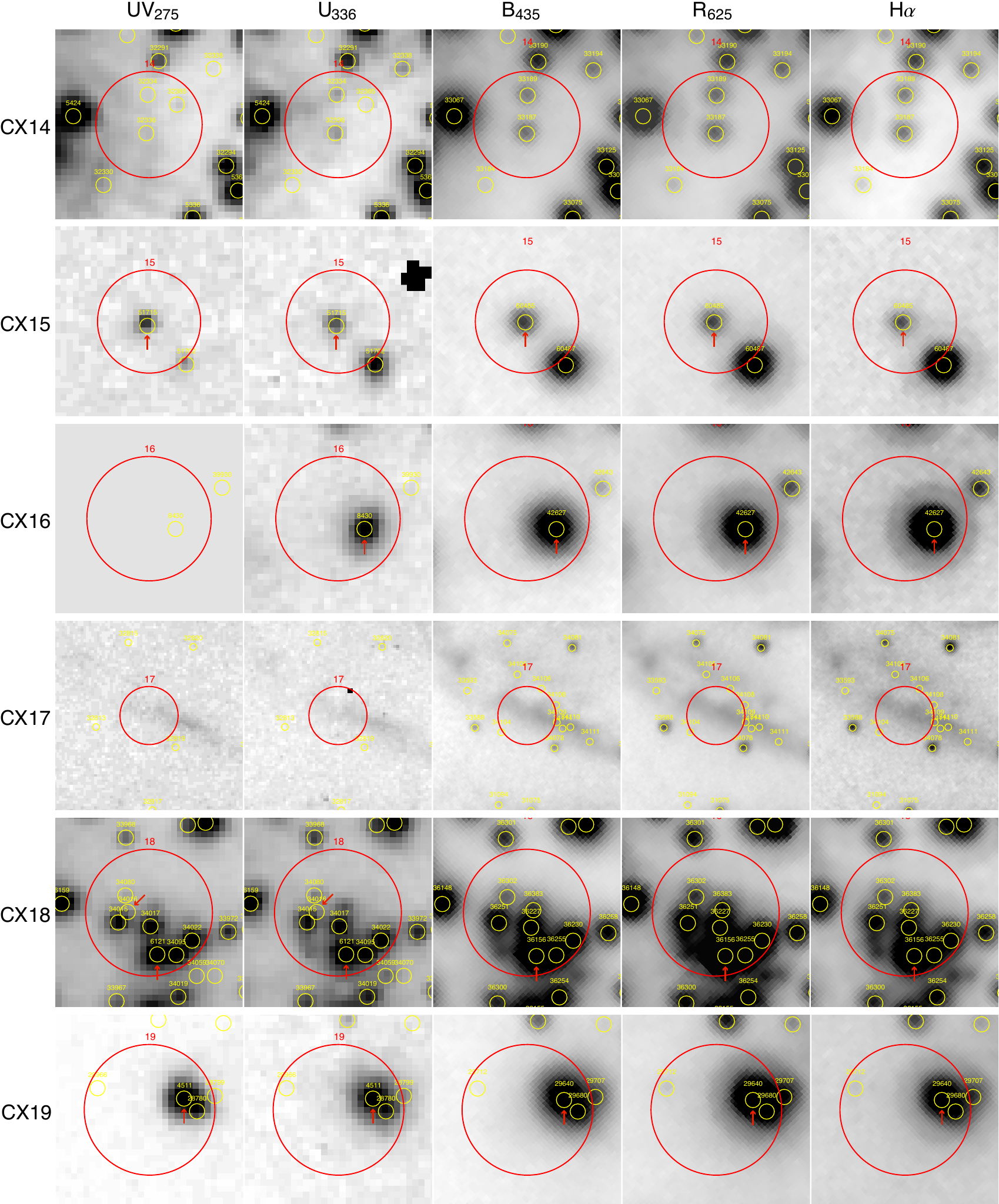}
\caption{Finding charts: CX14 -- CX19. For CX17, the field size is 2\farcs7 on a side. For CX18, the faint candidate counterpart CX18b is located to east of the centre of the error circle.}  \label{f:finding_charts_14-19}
\end{figure*}

\newpage
\begin{figure*}
\includegraphics[width=\textwidth]{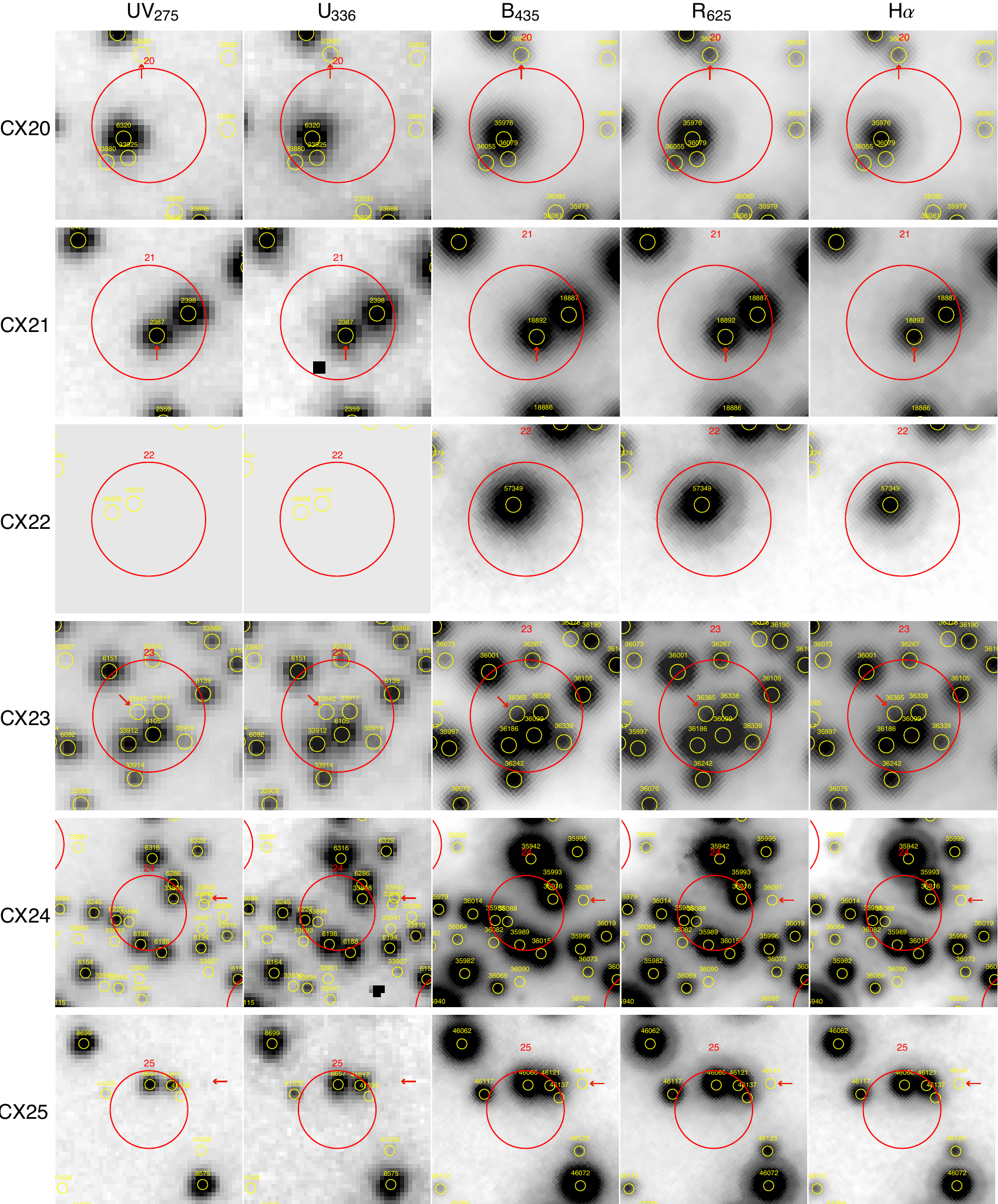}
\caption{Finding charts: CX20 -- CX25. For CX24 and CX25, the field size is 1\farcs8 on a side.}  \label{f:finding_charts_20-25}
\end{figure*}

\newpage
\begin{figure*}
\includegraphics[width=\textwidth]{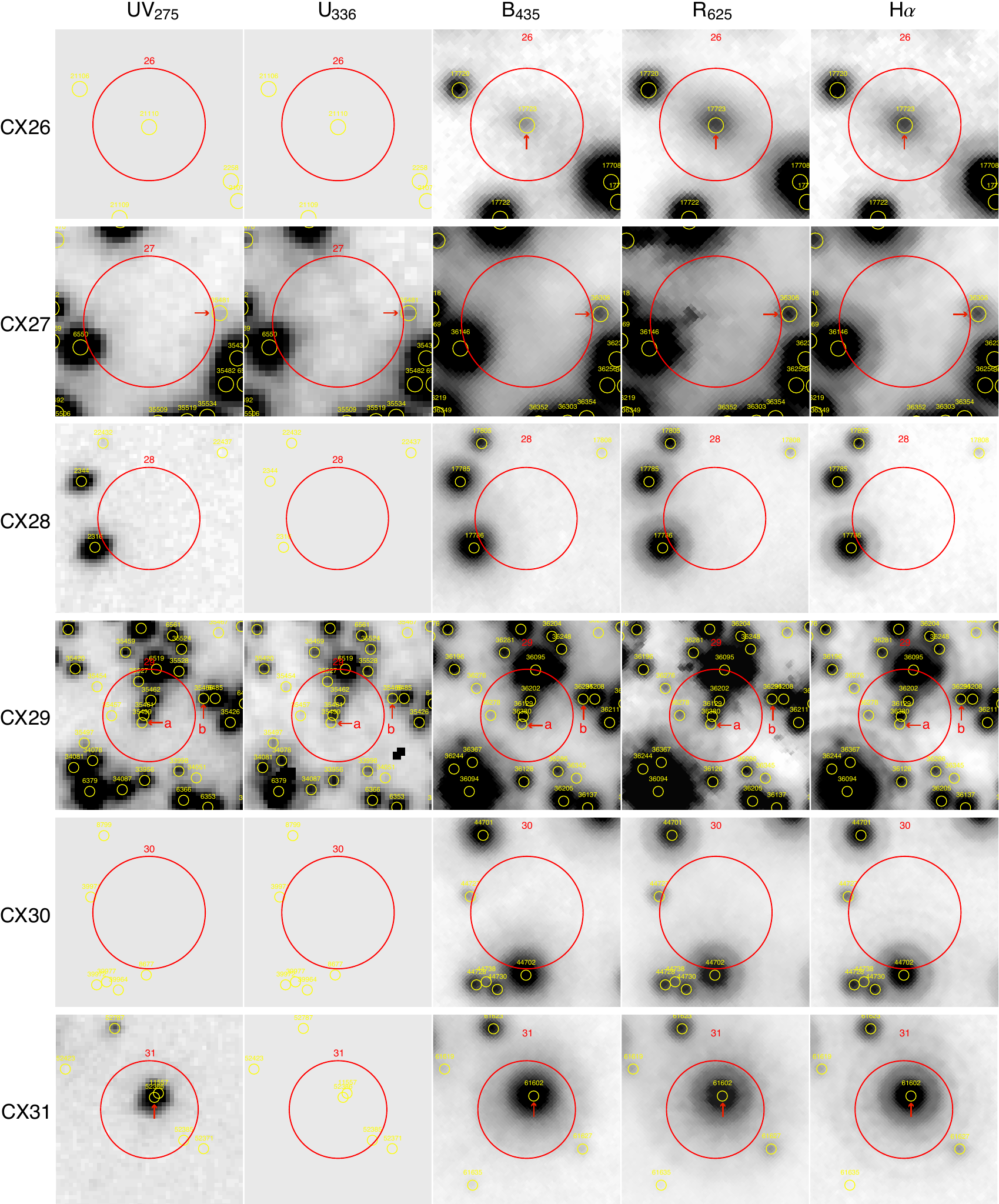}
\caption{Finding charts: CX26 -- CX31. For CX28-CX31, the field size is 1\farcs8 on a side.}  \label{f:finding_charts_26-31}
\end{figure*}

\newpage
\begin{figure*}
\includegraphics[width=\textwidth]{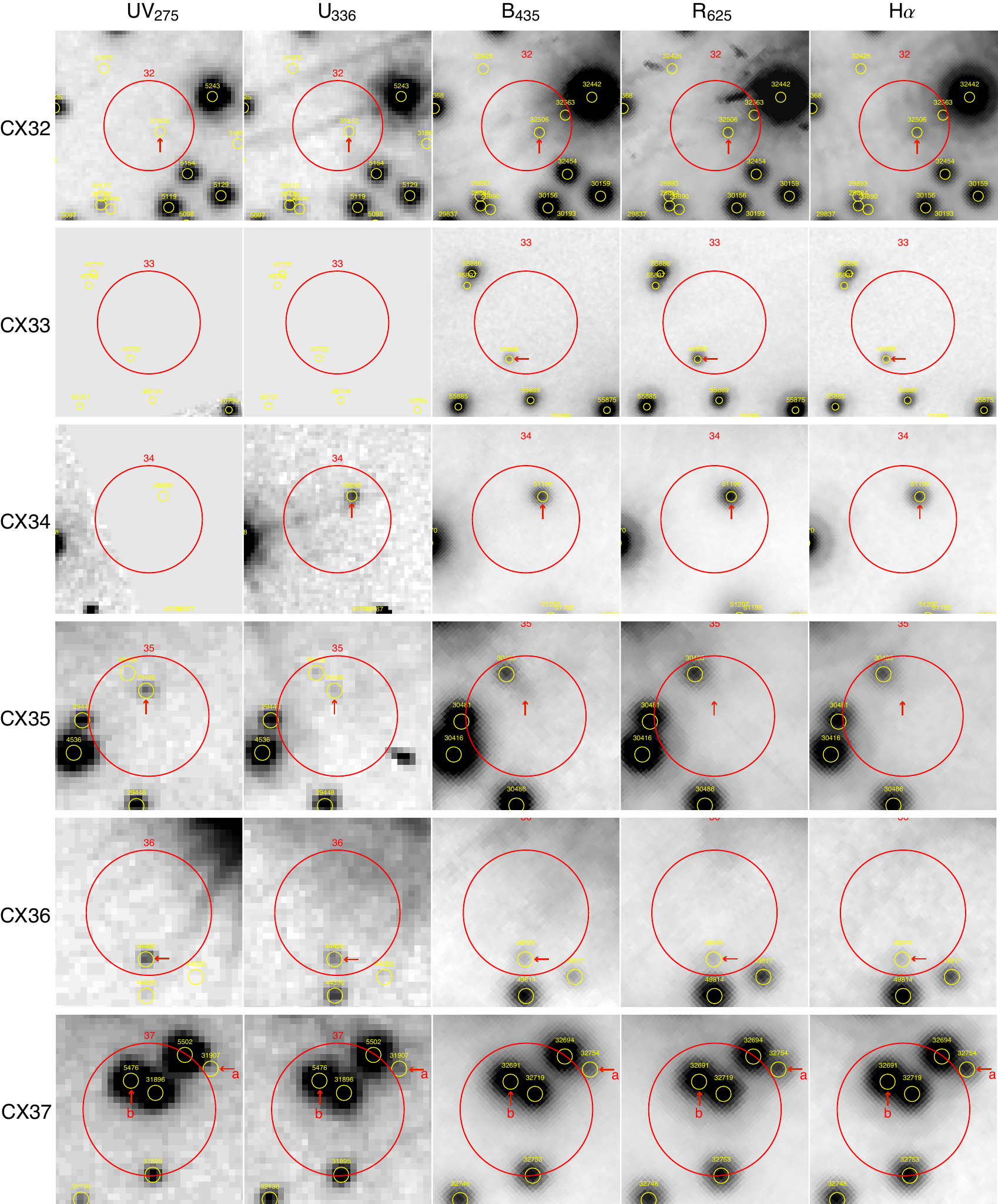}
\caption{Finding charts: CX32 -- CX37. For CX32 and CX34, the field size is 1\farcs8 on a side. For CX33, the field size is 2\farcs7 on a side.}  \label{f:finding_charts_32-37}
\end{figure*}

\newpage
\begin{figure*}
\includegraphics[width=\textwidth]{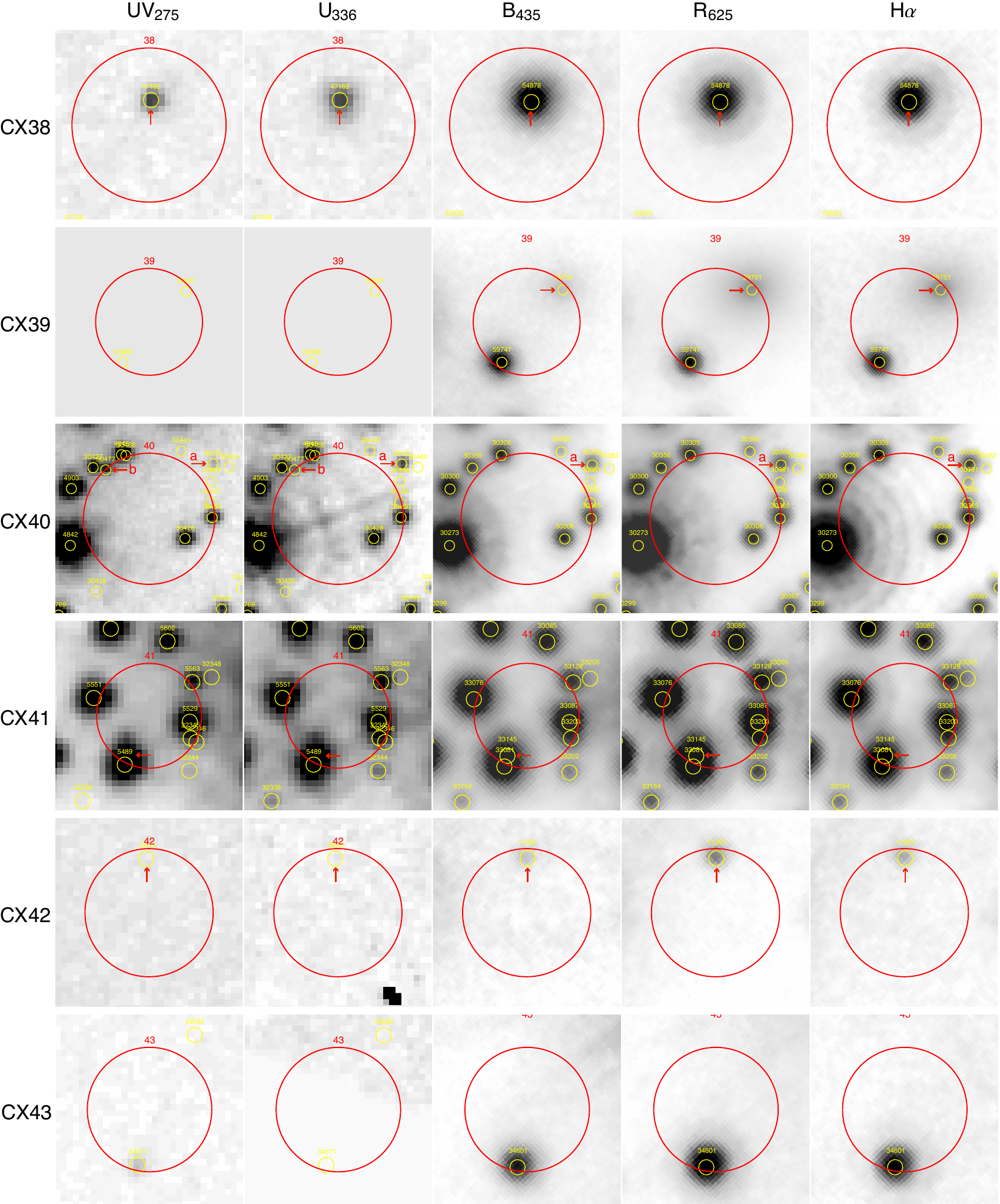}
\caption{Finding charts: CX38 -- CX43. For CX39 and CX40, the field size is 1\farcs8 on a side.}  \label{f:finding_charts_38-43}
\end{figure*}

\newpage
\begin{figure*}
\includegraphics[width=\textwidth]{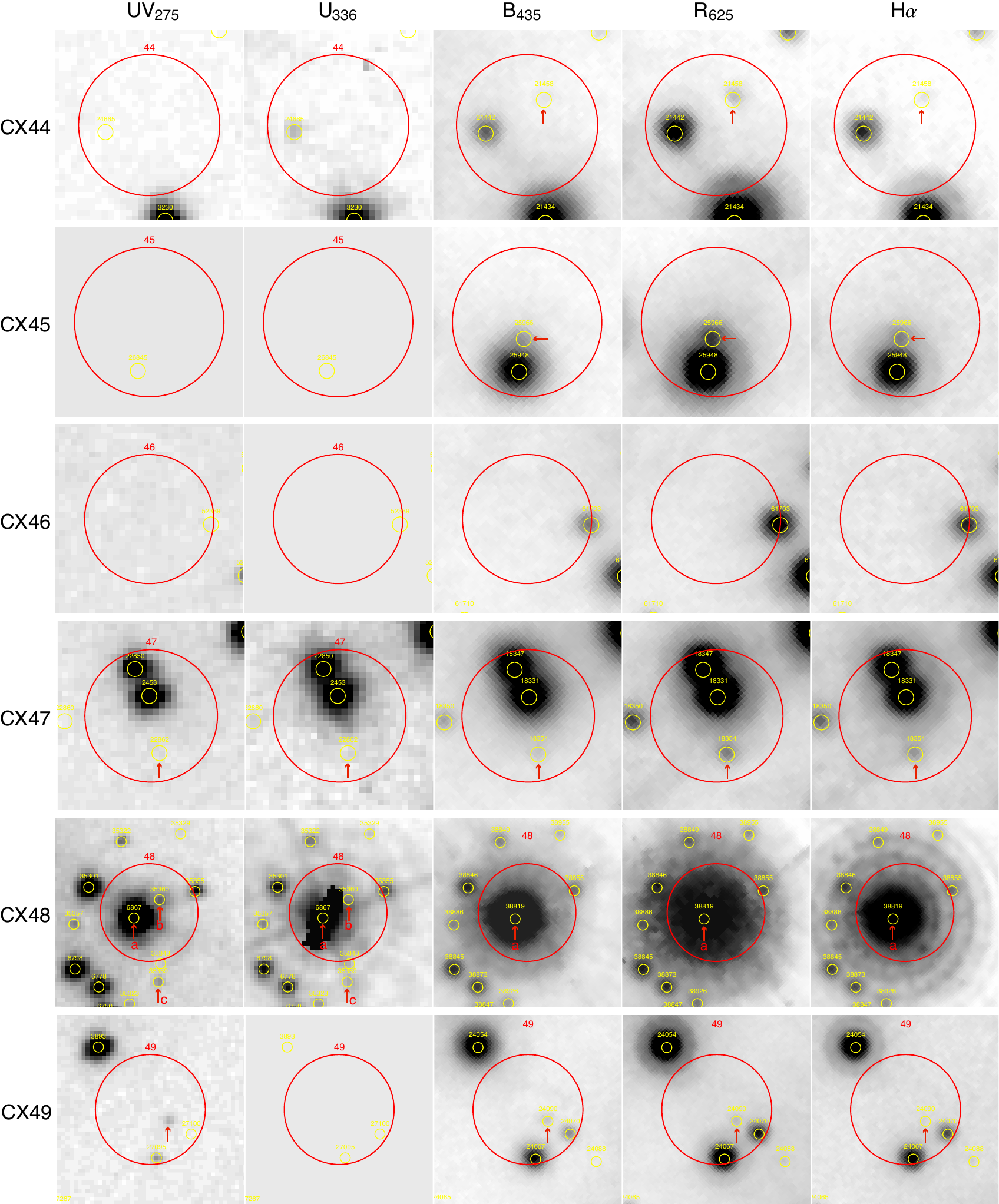}
\caption{Finding charts: CX44 -- CX49. For CX48 and CX49, the field size is 1\farcs8 on a side.}  \label{f:finding_charts_44-49}
\end{figure*}

\newpage
\begin{figure*}
\includegraphics[width=\textwidth]{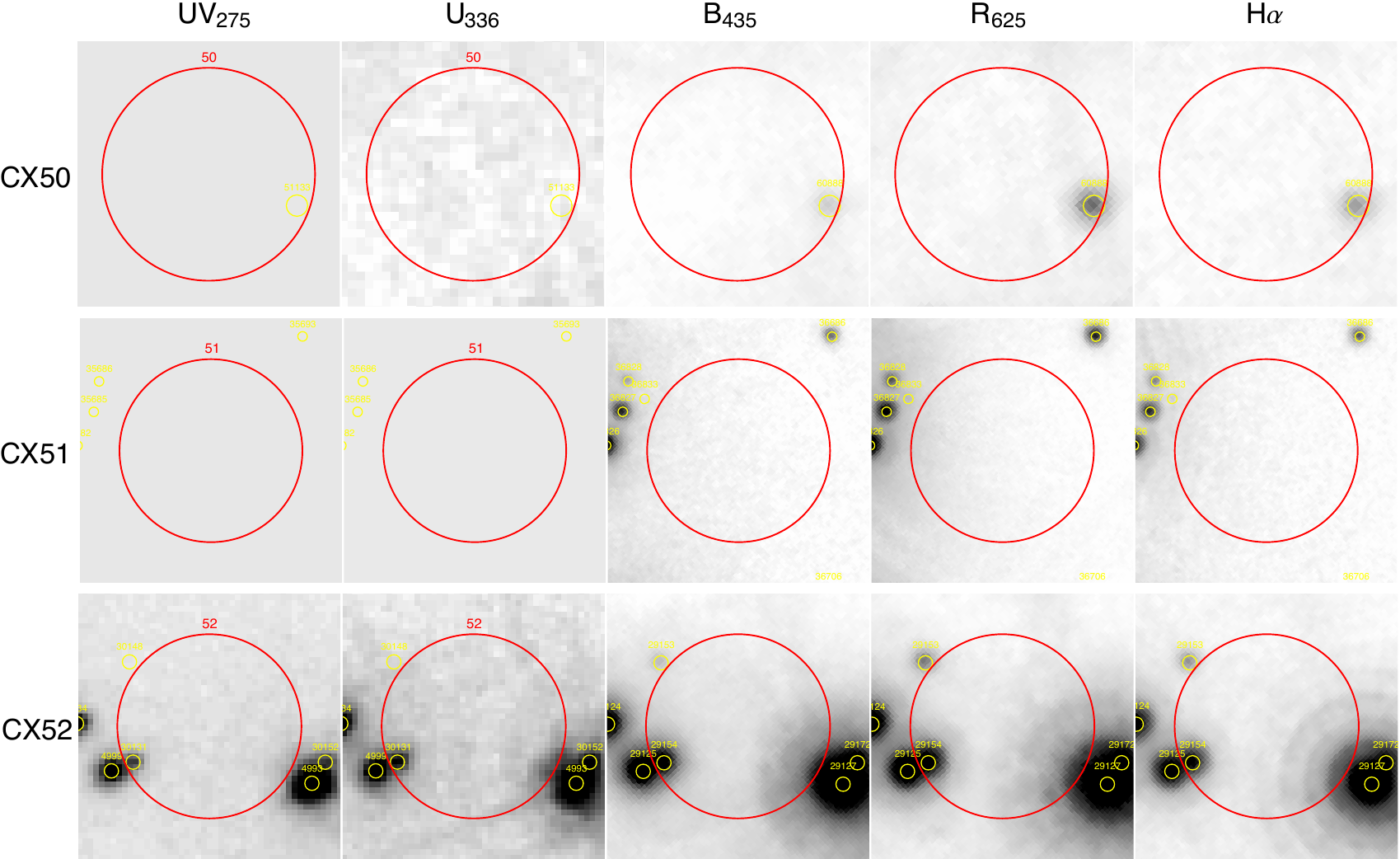}
\caption{Finding charts: CX50 -- CX52. For CX51, the field size is 2\farcs7 on a side. For CX52, the field size is 1\farcs8 on a side.}  \label{f:finding_charts_50-52}
\end{figure*}

\bsp	
\label{lastpage}
\end{document}